\documentclass[fleqn,usenatbib]{mnras}
\pdfoutput=1

\usepackage{amsmath}	

\usepackage[T1]{fontenc}
\usepackage{ae,aecompl}

\usepackage{graphicx}	

\newcommand{\csixty}{C$_{60}$}
\newcommand{\cseventy}{C$_{70}$}

\title[Characterisation of Tc~1]{Characterisation of the Planetary Nebula Tc~1 Based on VLT X-Shooter Observations}

\author[I. Aleman et al.]
{Isabel~Aleman$^{1,2,3}$\thanks{E-mail: bebel.aleman@gmail.com.},
Marcelo~L.~Leal-Ferreira$^{3,4}$\thanks{CNPq Fellow (248503/2013-8)},
Jan~Cami$^{5,6,7}$,
Stavros~Akras$^{8,9}$\thanks{CNPq Fellow (PDI-DA 300336/2016-0)},
\newauthor{Bram~Ochsendorf~$^{10}$,
Roger~Wesson$^{11}$,
Christophe Morisset$^{12}$, Nick~L.J.~Cox$^{13}$,}
\newauthor{Jeronimo~Bernard-Salas$^{13}$,
Carlos~E.~Paladini$^{2}$,
Els~Peeters$^{5,6,7}$,
David~J.~Stock$^{5}$,}
\newauthor{Hektor Monteiro$^{1}$,
Alexander~G.~G.~M.~Tielens$^{3}$}
\\
$^{1}$Universidade Federal de Itajub\'a, Instituto de F\'isica e Qu\'imica, Av. BPS 1303 Pinheirinho, 37500-903 Itajub\'{a}, MG, Brazil\\
$^{2}$IAG-USP, Universidade de S\~{a}o Paulo, Rua do Mat\~{a}o 1226, Cidade Universit\'{a}ria, 05508-090, S\~{a}o Paulo, SP, Brazil\\
$^{3}$Leiden Observatory, University of Leiden, PO Box 9513, 2300 RA, Leiden, The Netherlands\\
$^{4}$Argelander-Institut f\"ur Astronomie, Universit\"at Bonn, Auf dem H\"ugel 71, 53121 Bonn, Germany\\
$^{5}$Department of Physics and Astronomy, The University of Western Ontario, London, ON N6A 3K7, Canada\\
$^{6}$Institute for Earth and Space Exploration, The University of Western Ontario, London, ON N6A 3K7, Canada\\
$^{7}$SETI Institute, 189 Bernardo Ave, Suite 100, Mountain View, CA 94043, USA\\
$^{8}$Observat\'{o}rio Nacional/MCTIC, Rua Gen. Jos\'{e} Cristino, 77, 20921-400, Rio de Janeiro, RJ, Brazil\\
$^{9}$Instituto de Matem\'{a}tica, Estat\'{i}stica e F\'{i}sica, Universidade Federal do Rio Grande, Rio Grande 96203-900, Brazil\\
$^{10}$Space Telescope Science Institute, 3700 San Martin Drive,
Baltimore, MD 21218, USA\\
$^{11}$Department of Physics and Astronomy, University College London, Gower Street, London WC1E 6BT, UK\\
$^{12}$Instituto de Astronomia, Universidad Nacional Autonoma de Mexico, Apartado postal 106, C.P. 22800 Ensenada, Baja California,\\ M\'exico.\\
$^{13}$ACRI-ST, 260 Route du Pin Montard, Sophia-Antipolis, France}

\date{Accepted XXX. Received YYY; in original form ZZZ}

\pubyear{2019}

\begin{document}
\label{firstpage}
\pagerange{\pageref{firstpage}--\pageref{lastpage}}
\maketitle

\begin{abstract}
We present a detailed analysis of deep VLT/X-Shooter observations of the planetary nebula Tc~1. We calculate gas temperature, density, extinction, and abundances for several species from the empirical analysis of the total line fluxes. In addition, a spatially resolved analysis of the most intense lines provides the distribution of such quantities across the nebula. The new data reveal that several lines exhibit a double peak spectral profile consistent with the blue- and red-shifted components of an expanding spherical shell. The study of such components allowed us to construct for the first time a three-dimensional morphological model, which reveals that Tc~1 is a slightly elongated spheroid with an equatorial density enhancement seen almost pole on. A few bright lines present extended wings (with velocities up to a few hundred km~s$^{-1}$), but the mechanism producing them is not clear. We constructed photoionization models for the main shell of Tc~1. The models predict the central star temperature and luminosity, as well as the nebular density and abundances similar to previous studies. Our models indicate that Tc~1 is located at a distance of approximately 2~kpc. We report the first detection of the [Kr~{\sc iii}]~6825~\AA\ emission line, from which we determine the Krypton abundance. Our model indicates that the main shell of Tc~1 is matter bounded; leaking H ionizing photons may explain the ionization of its faint AGB-remnant halo. 
\end{abstract}



\section{Introduction}

\begin{figure}
   \includegraphics[width=7.7cm]{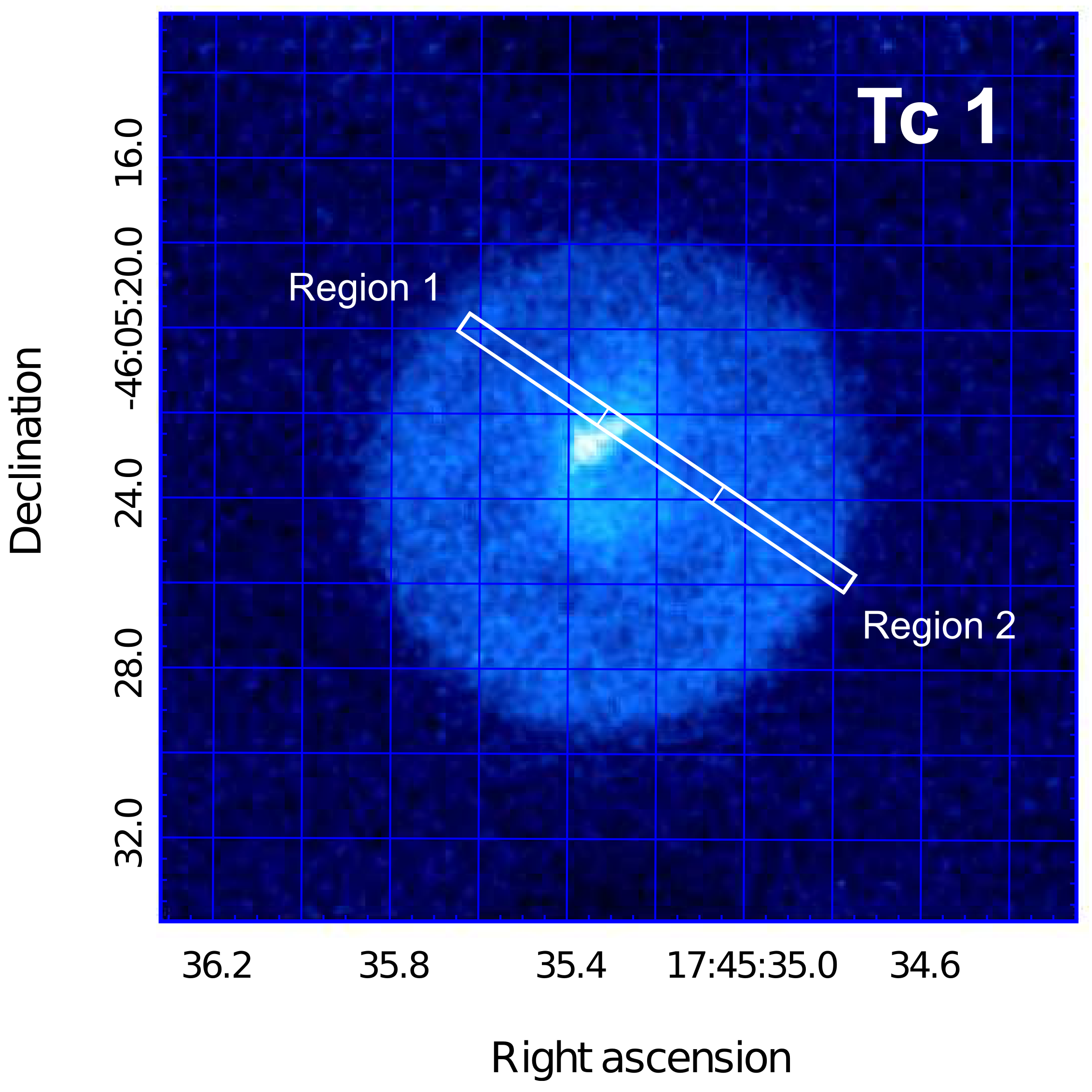}
   \caption{The orientation of the X-Shooter slit position. The image is a narrow-band Qa-band (18.3$\mu$m) Gemini/T-ReCS image of Tc~1 \citep{2018Galax...6..101C}. Regions 1 and 2 are the slit extractions discussed in Sect.~\ref{sect_diagnostic}. The central star is not visible in this image.}
   \label{slitpos}
\end{figure}

Tc~1 (also known as IC~1266) is a young, low-excitation Galactic planetary nebula (PN; \citealt{Pottasch_etal_2011}). Optical images of Tc~1 reveal that the bulk of its nebular emission originates from a fairly round nebula approximately 12~arcsec in diameter \citep{1992A&AS...96...23S,2003MNRAS.340..417C}. This nebula has a bright core of around 3~arcsec in diameter in its centre. Mid infrared (IR) images show a horseshoe-shaped central structure of also approximately 3~arcsec in diameter (Fig.~\ref{slitpos}; \citealt{2018Galax...6..101C}), which is produced by warm dust. This structure, which surrounds the bright core mentioned above, is not apparent in the optical images. Optical images also shows that Tc~1 has a much fainter external shell of approximately 52~arcsec in diameter, whose centre is slightly offset compared to the central nebula. According to \citet{2003MNRAS.340..417C}, this halo is likely a remnant of the progenitor asymptotic giant branch (AGB) star mass loss.

Tc~1 is a C-rich nebula with abundances typical for a low-mass PN progenitor \citep[1.5--2.5~$M_{\sun}$;][]{Pottasch_etal_2011,2014MNRAS.437.2577O}. The Tc~1 nebular emission is photoionized by a central source with effective temperatures ($T_{eff}$) reported in the literature in the range of 28-35~kK and total luminosities ($L_{\star}$) in the range of 1\,400-13\,000~$L_{\sun}$ \citep[e.g.,][]{Gorny_etal_1997,Pauldrach_etal_2004,Gesicki_Zijlstra_2007,Pottasch_etal_2011,2014MNRAS.437.2577O}. The large uncertainty in the luminosity is connected to its still poorly known distance. In the literature we find distances ranging from 0.6 to 4.1~kpc \citep[e.g.,][]{1971ApJS...22..319C,1982ApJ...260..612D,Mendez1988,1995ApJ...452..286S,Gorny_etal_1997,1998AJ....115.1989T,2008ApJ...689..194S}.

Although apparently a fairly normal PN, Tc~1 is now best known by an unusual characteristic: strong fullerene emission bands in its mid-IR spectrum. Fullerenes are a class of large carbonaceous molecules in the shape of an ellipsoidal or spheroidal cage \citep{1985Natur.318..162K} and are a relatively recent addition to the molecular inventory in C-rich PNe. Tc~1 was the first PN in which we detected \csixty, and to date the only object known to also show emission features due to \cseventy\ \citep{2010Sci...329.1180C}. 
Such emission is rare; so far only 24 PNe in the Milky Way and the Magellanic Clouds are known to exhibit the fullerene emission bands \citep{2019MNRAS.482.2354O}.  While several formation mechanisms for \csixty\ in PNe and reflection nebulae have been proposed
\citep[see e.g.][]{2012PNAS..109..401B,Jero:C60excitation,Elisabetta:arophatics}
and some experimentally tested 
\citep{Zhen:formation_lab}, it is not clear what the C$_{60}$ formation route is in PNe \citep{2018Galax...6..101C}.

A study of the physical properties of \csixty-PNe shows that they are all young, low-excitation objects with similar IR spectra, but that have otherwise no unusual properties that sets them apart from their non-\csixty\ containing counterparts \citep{2014MNRAS.437.2577O}. It is thus not clear precisely what physical or chemical conditions result in the formation of these stable species in the circumstellar environments of evolved stars. A key difficulty in making progress is the lack of detailed knowledge about the physical conditions (in particular spatially resolved studies) and the structure of such nebulae.

In this paper we present a detailed analysis of spatially resolved optical/UV VLT/X-Shooter observations of Tc~1. Our aim is to expand the knowledge about this PN, by extracting from this spectrum information of the physical conditions, elemental abundances, kinematics, and morphology of the main shell. Such information will assist in future studies on the formation and survival mechanisms of fullerenes in Tc~1.

This paper is organised as follows: Sect.~\ref{sect_observations} describes the observations and data reduction, Sect.~\ref{sect_diagnostic} presents the nebular line diagnostics, Sect.~\ref{sect_kinem} discusses the nebular kinematics derived from bright lines, Sect.~\ref{sect_morphology} presents our reconstruction of the three-dimensional morphology of Tc~1, Sect. \ref{sect_models} shows the results of the photoionization model we construct for Tc~1, Sect.~\ref{kripton} discuss the detection of a krypton line and its implications, and Sect.~\ref{sect_conclusions} presents our conclusions.

\section{Observations and Data Reduction} \label{sect_observations}

The observations discussed here were obtained as part of Program 385.C-0720 (P.I. N.L.J.~Cox) using the X-Shooter spectrograph \citep{2011A&A...536A.105V} mounted on the Very Large Telescope (VLT). Tc~1 was first observed on July 2, 2010 (night 1), but the data were only of moderate quality due to the mid-way interruption of the observation by a target-of-opportunity triggered observation; it was therefore re-observed on July 5, 2010 (night 2). For the remainder of this paper, we will use the data for night 2.

The object was observed in stare mode with the narrowest slit settings of 11$\times$0.5~arcsec (UVB) and 11$\times$0.4~arcsec (VIS and NIR), resulting in resolving powers of $\sim$9\,100, 17\,000 and 10\,000 for the UVB, VIS, and NIR arms, respectively. The slit orientation is depicted in Fig. \ref{slitpos}. A summary of the observation data is given in Table \ref{Tab_Obs_Log}. In this work, we analyse the observations from the UVB and VIS arms, which covers the spectral range from 3\,200 to 10\,100~\AA.


\begin{table}
  \centering
  \caption{Tc~1 Observation Information}
  \label{Tab_Obs_Log}
  \begin{tabular}{lcc}
	\hline
	Parameter & Value \\
	\hline
	 Instrument & ESO VLT UT2 X-Shooter\\
	 Program & 385.C-0720\\
	 Date & 2010 July 5 \\
	 R.A. (J2000) & 17$^\textrm{h}$~45$^\textrm{m}$~35$\fs$2\\
	 Dec. (J2000) & -46$\degr$~05$\arcmin$~22$\farcs$9\\
     Position Angle & -56.011$\degr$\\
     Air Mass & 1.18 -- 1.22\\
    \hline
  \end{tabular}
\end{table} 

We reduced the data using the ESO X-Shooter pipeline (version 2.5.2) and we used observations of the standard LTT7987 to perform flux calibration. Because the size of Tc~1 is larger than the X-Shooter slit length (11~arcsec), there was unfortunately no part of the slit that was free of nebular emission lines, and therefore we could not perform a satisfactory sky subtraction. Rather than introducing artefacts, we therefore opted to not use the sky-subtracted spectra. Instead, we will only include nebular emission lines in our analysis that are not significantly contaminated by sky emission.

\section{Nebular Line Diagnostics} \label{sect_diagnostic}

We performed the nebular line diagnostics analysis of the Tc~1 X-Shooter spectrum using two different approaches: (i)~integrating the spectra along two regions of the slit and (ii)~spatially resolved (i.e. pixel by pixel).

\begin{figure*}
   \centering
   \includegraphics[width=\textwidth]{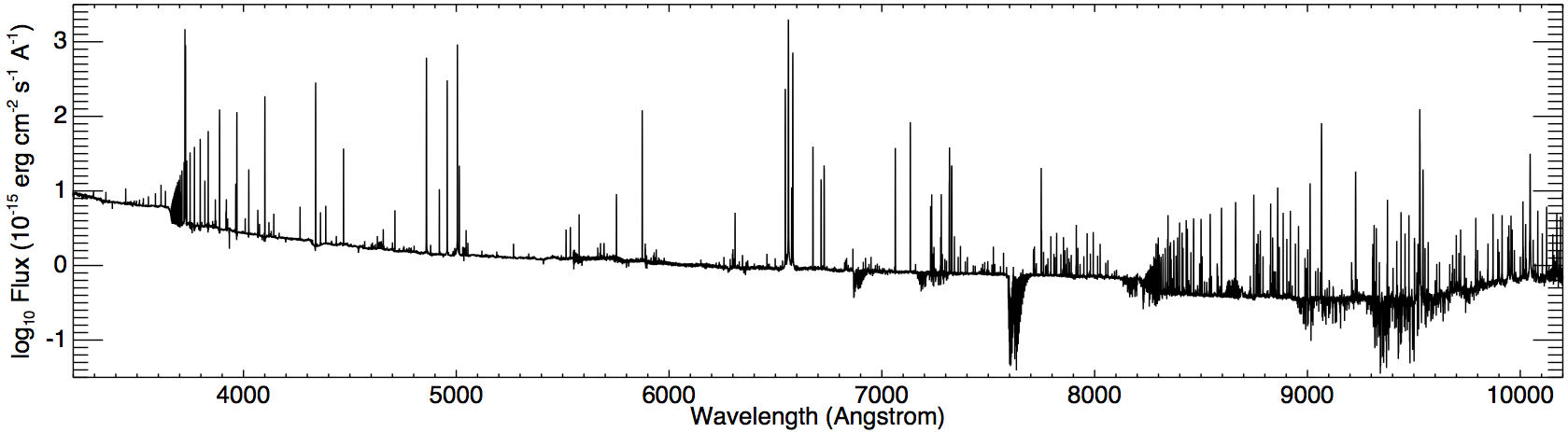}
   \caption{Tc~1 X-Shooter UVB and VIS arms spectra integrated in Regions 1 and 2 (see their positions in Fig.~\ref{slitpos}).} 
   \label{spectrum}
\end{figure*}

The first approach is the most commonly performed in the literature. By integrating the Tc~1 spectrum over a range of spatial pixels we lose spatial information but, on the other hand, we increase the signal-to-noise ratio of all emission lines. This procedure allows weak lines to become detectable, increasing the sample of lines we can detect and analyse. As we only want to analyse the nebular emission, the central region of the slit, where the contamination due to the central star emission is significant, was removed. The central three arcseconds show a mixture of nebular and stellar emission. The fluxes of the two remaining regions, which correspond to the two extremes of the slit (Regions 1 and 2 in Fig.~\ref{slitpos}), were summed. The Tc~1 X-Shooter integrated spectra for both UVB and VIS arms are shown in Fig.~\ref{spectrum}. The rich spectra exhibit a few hundred atomic lines. In total, we identified and measured 230 different lines from 14 elements (20 ions) in the Tc~1 integrated spectrum. The measurements and identification of the numerous lines in the Tc~1 X-Shooter spectrum were performed with the codes {\sc alfa} \citep[Automated Line Fitting Algorithm;][]{2016MNRAS.456.3774W} and {\sc herfit}\footnote{The code was developed by I. Aleman and is available upon request.}, as well as information from the literature. A detailed description of the measurement procedure and the list of the detected nebular lines and their respective fluxes are provided in Appendix~\ref{ap_linetable}\footnote{Supplementary Appendices are only available online.}.

Using the line fluxes as inputs to the Nebular Empirical Analysis Tool ({\sc neat} version 1.7; \citealt{2012MNRAS.422.3516W}), we derived the Tc~1 extinction, physical conditions, and abundances. We present our results in the following subsections, where our results are compared to previous nebular analysis based on data acquired with other instruments and in different positions of the nebula. The results are also used for comparison to our spatially resolved analysis.

In addition to the integrated analysis, we performed spatially resolved diagnostic analysis, to gain insight in the radial distribution of physical parameters and abundances, as the slit was positioned across the nebula (Fig.~\ref{slitpos}). In this case, the analysis was limited to diagnostics involving bright emission lines, as those have sufficient signal-to-noise ratio to be detected in a pixel-by-pixel analysis. For this analysis, we integrated fluxes over two neighbouring spatial pixels to increase the signal-to-noise ratio. Line fluxes were measured using {\sc splot} (\textit{iraf.noao.twodspec.splot}) from {\sc iraf}\footnote{{\sc iraf} is distributed by the National Optical Astronomy Observatories, which are operated by the Association of Universities for Research in Astronomy, Inc., under cooperative agreement with the National Science Foundation.}. The resulting fluxes were used as input parameters for the {\sc 2d\_neb} code \citep{2011MNRAS.411.1395L, 2013A&A...560A.102M}, which performs two--dimensional nebular diagnostics. We used the same atomic data and method for evaluation in both the integrated and spatially resolved analysis. The atomic data used in the code are listed in Appendix~\ref{ap_atomicdata}. The results of the spatially resolved analysis are also presented in the following subsections.

\subsection{Correction for Extinction} \label{sect_extinction}

The line fluxes were corrected for extinction by comparison of the observed and theoretical hydrogen Balmer series emission line ratios (see, e.g., \citealt{2006agna.book.....O}). Both {\sc neat} and {\sc 2d\_neb} follow this traditional method of de-reddening. We use the Galactic extinction law from \citet{1989ApJ...345..245C}\footnote{The Galactic extinction law used by {\sc 2d\_neb} has been updated with the correction suggested by \citet{1994ApJ...422..158O}. Although {\sc neat} 1.7 does not include this correction, the very small difference will not affect the line fluxes significantly and will not change our conclusions.}, assuming the ratio of total-to-selective extinction $R_V =$~3.1. In our calculations, the extinction coefficient $c$(H$\beta$) was derived from the H$\gamma$/H$\beta$ ratio. H$\alpha$ is saturated and for consistency with the spatially resolved analysis, we did not include H$\delta$ (see discussion below).

We found $c$(H$\beta$) equal to 0.40, which is similar to the measurements found in the literature. \citet{Cahn_etal_1992}, \citet{1994MNRAS.271..257K},  \citet{2008ApJ...677.1100W}, and \citet{2013MNRAS.431....2F} derived 0.28, 0.44, 0.33, and 0.43 respectively. For all these measurements, the extinction was determined from the Balmer decrement. From He optical/UV lines, \citet{1994MNRAS.271..257K} calculated a value of 0.30 for $c$(H$\beta$). \citet{Cahn_etal_1992}, \citet{1994MNRAS.271..257K}, and \citet{Pottasch_etal_2011} determined $c$(H$\beta$) using the comparison between the measured H$\beta$ and radio continuum flux. The values they obtained are 0.40, 0.40, and 0.36, respectively. Note, however, that the region probed here is not the same as the region studied in these other works.

\begin{figure}
\includegraphics[width=7.5cm]{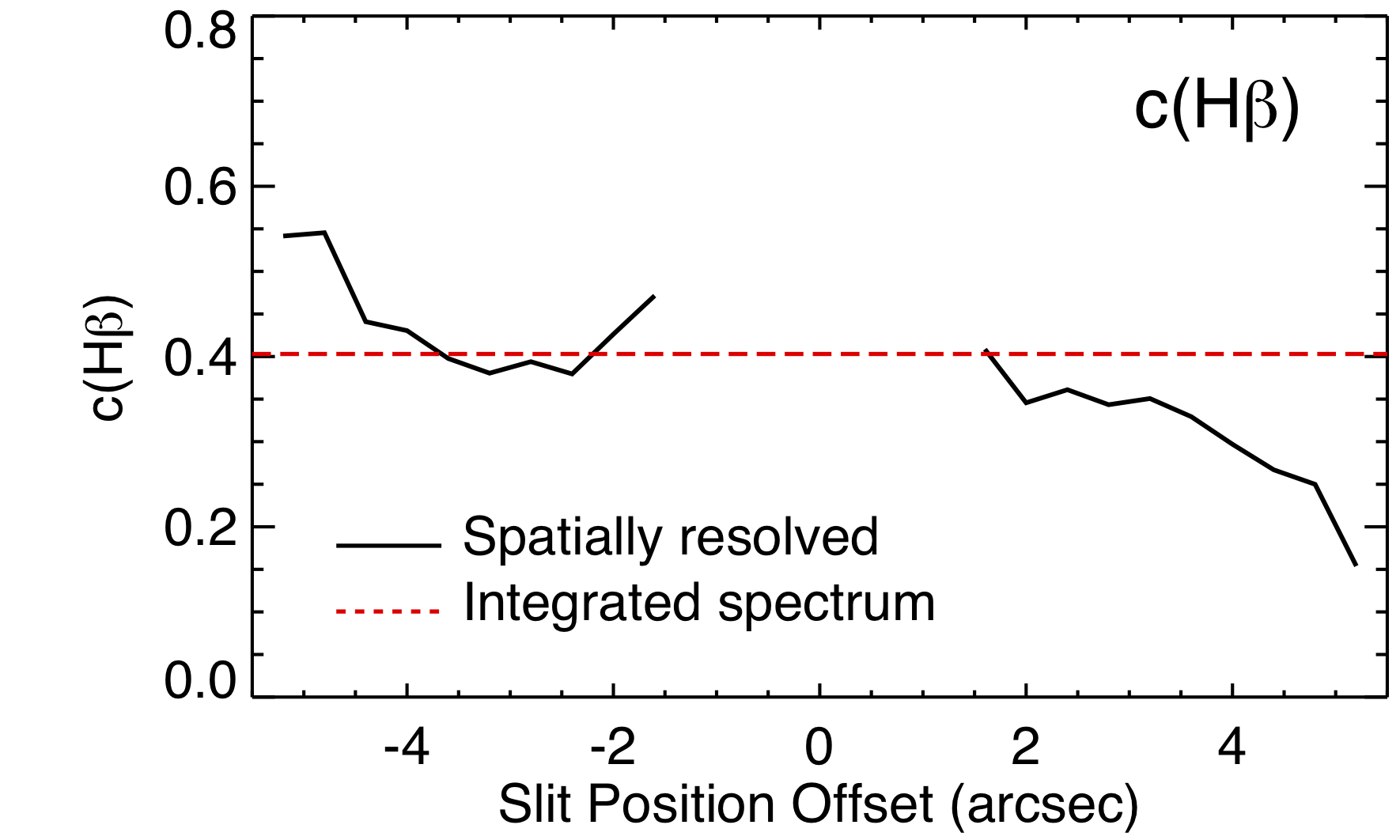}
\caption{Spatial variation of extinction across the slit. c(H$\beta$) was derived from the H$\gamma$/H$\beta$ ratio. The red dashed line indicates the c(H$\beta$) value obtained from the integrated spectrum.}   
\label{fig_extinction}
\end{figure}

We also determined the spatial variation of the extinction coefficient using the H$\gamma$/H$\beta$ ratio. Using only the most intense lines in the UVB arm was necessary to accurately determine this quantity in our pixel-by-pixel study. As it can be seen in Fig.~\ref{fig_extinction}, $c$(H$\beta$) varies from 0.15 to 0.55 along the nebular region. Such a variation appears to be typical for PNe as shown in the $c$(H$\beta$) maps derived by \citet{2008MNRAS.386...22T}, \citet{2011MNRAS.411.1395L}, and  \citet{2016A&A...588A.106W} for a few PNe.

The colour excess and extinction in magnitudes derived from the $c$(H$\beta$) integrated value are $E(B-V)$~=~0.31~mag and $A_\textrm{V}$~=~0.96~mag \citep[here we also assumed the same R values and extinction law mentioned above;][]{2006agna.book.....O}. The $c$(H$\beta$) variation along the slit seen in Fig.~\ref{fig_spatial_phcond} converts to a range of $A_\textrm{V}$ from 0.36 to 1.31 mag. Part of this extinction can be internal to the nebulae. According to \citet{2015ApJ...810...25G} the foreground extinction towards Tc~1 is $E(B-V)$~=~0.22~$\pm$~0.03, corresponding to $A_\textrm{V}$~=~0.68~$\pm$~0.09. The $c$(H$\beta$) maps of NGC 7009 derived by \citet{2016A&A...588A.106W} show in great detail that $c$(H$\beta$) traces some of the structures of the nebulae. Their analysis indicates that part of the extinction is internal to the nebula. This is also observed in the PNe extinction maps derived by \citet{2008MNRAS.386...22T} and by \citet{2011MNRAS.411.1395L}.

\begin{figure*}
\begin{center}

   \includegraphics[width=8.5cm,trim=0.3cm 1.3cm 0.4cm 0.5cm,clip]{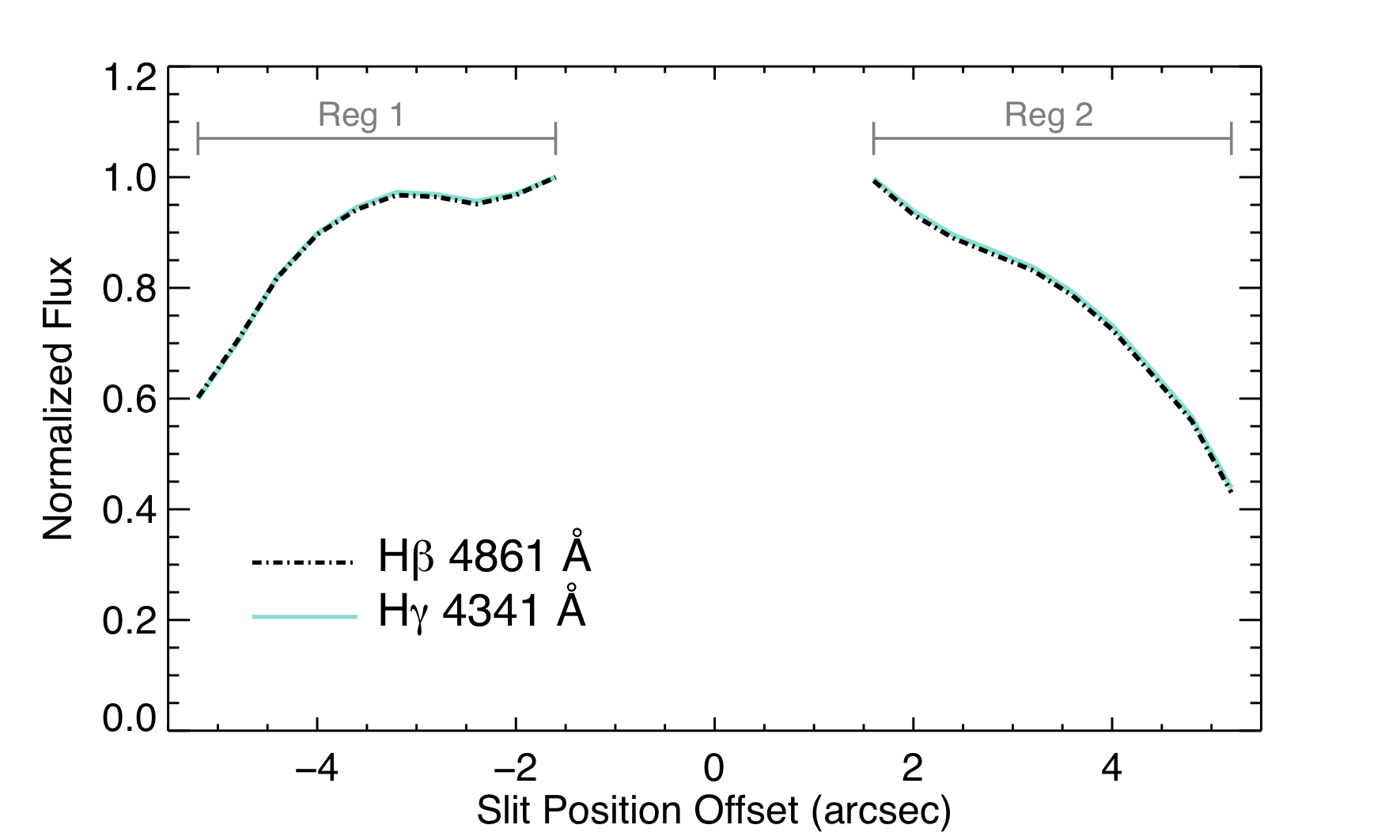} 
   \includegraphics[width=8.5cm,trim=0.3cm 1.3cm 0.4cm 0.5cm,clip]{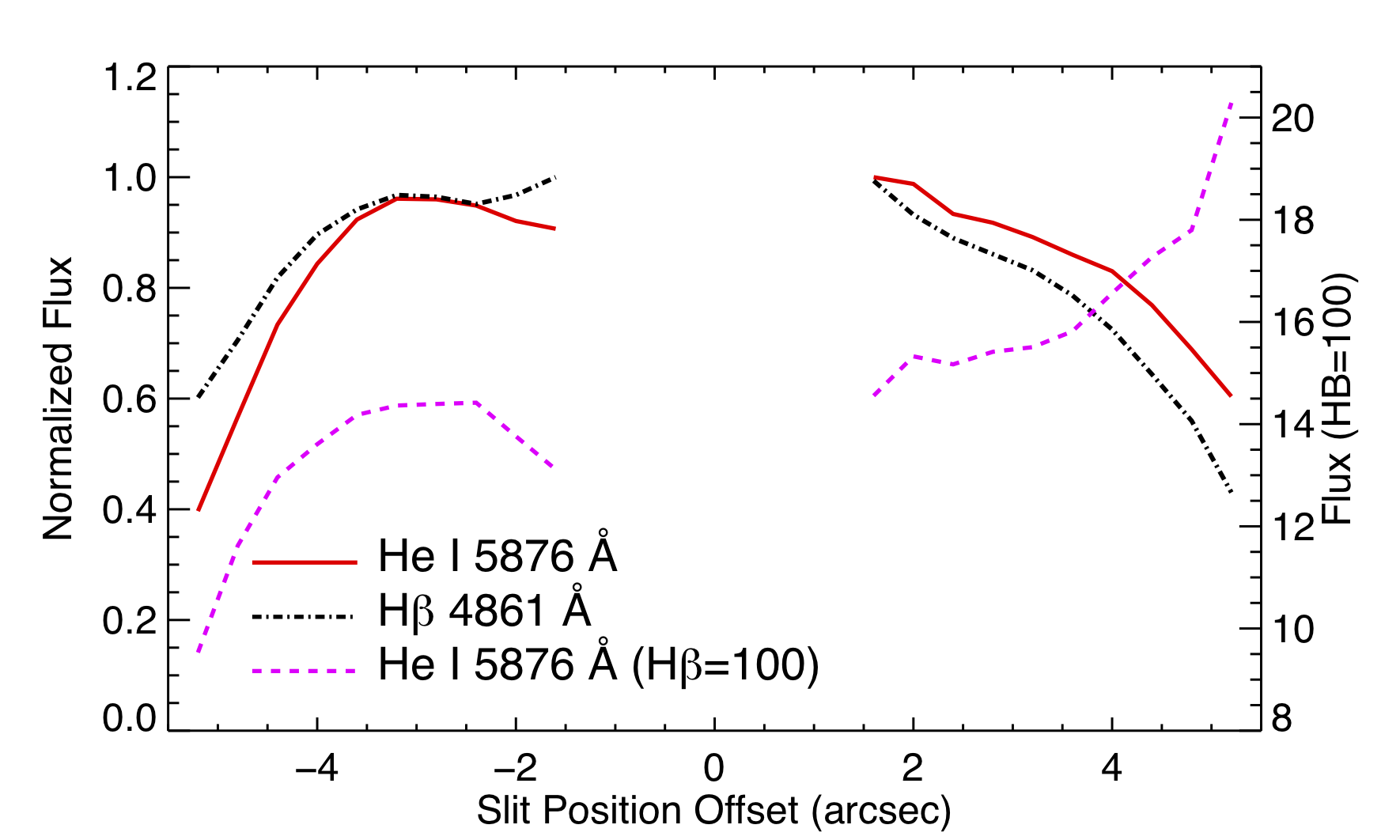}
   \includegraphics[width=8.5cm,trim=0.3cm 1.3cm 0.4cm 0.5cm,clip]{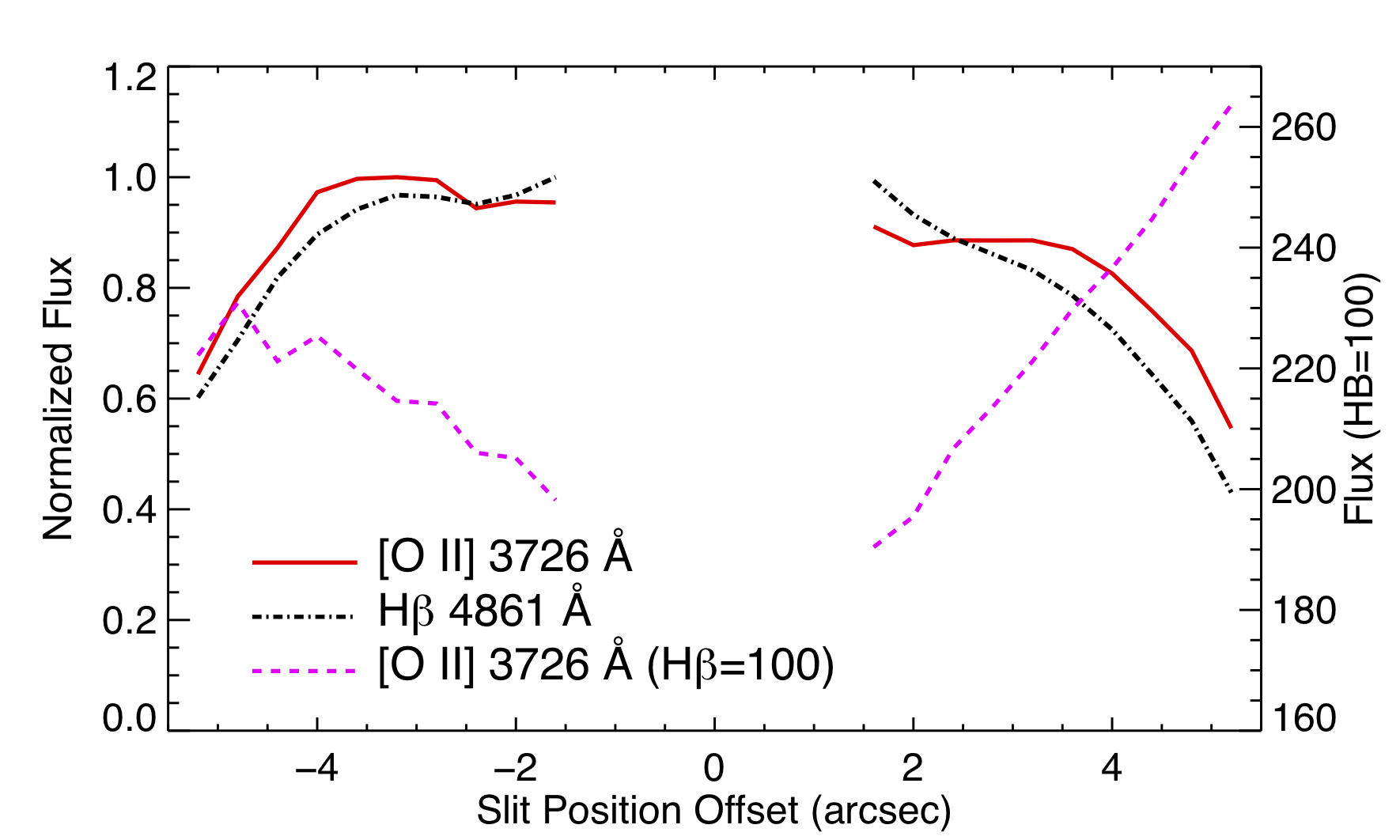} 
   \includegraphics[width=8.5cm,trim=0.3cm 1.3cm 0.4cm 0.5cm,clip]{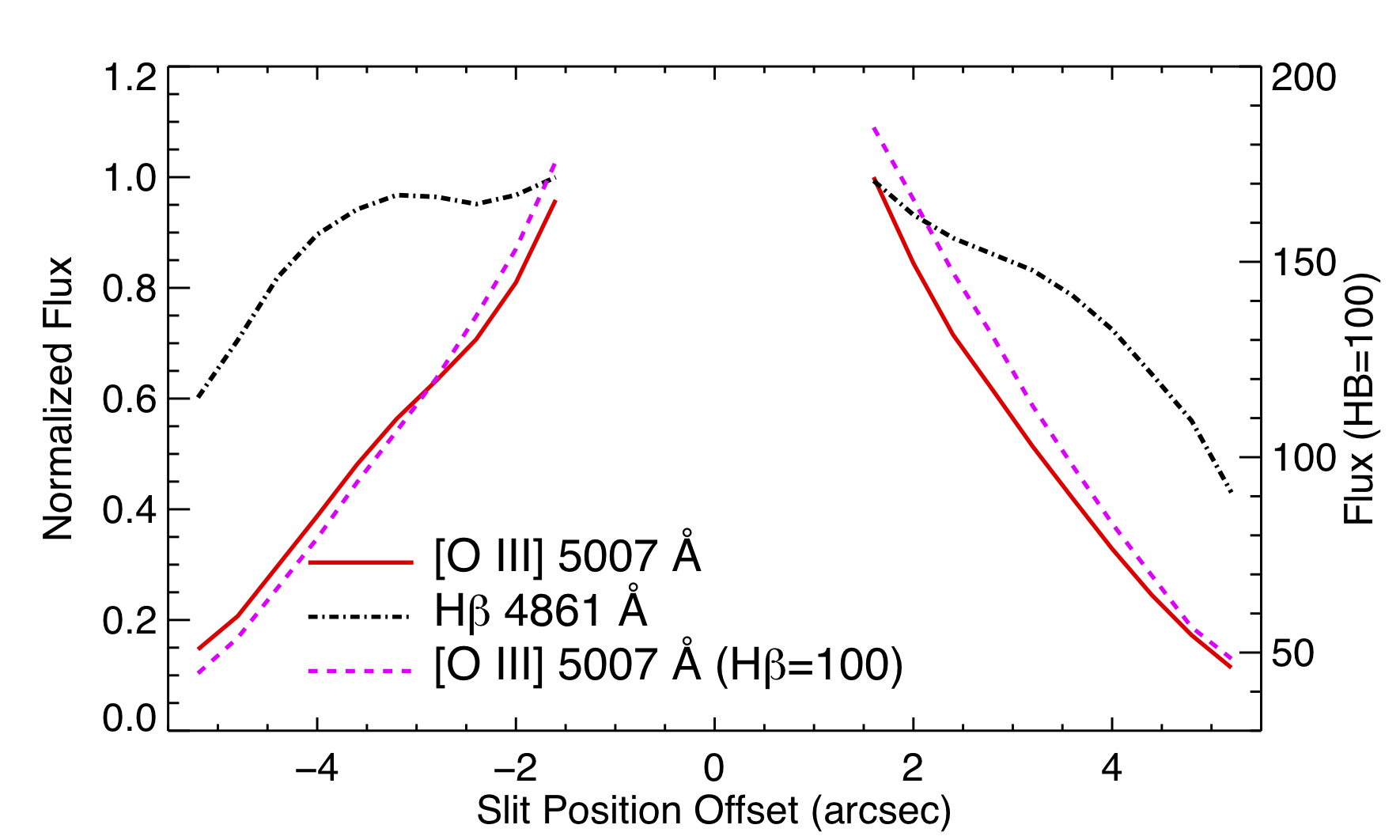} 
   \includegraphics[width=8.5cm,trim=0.3cm 0.3cm 0.4cm 0.5cm,clip]{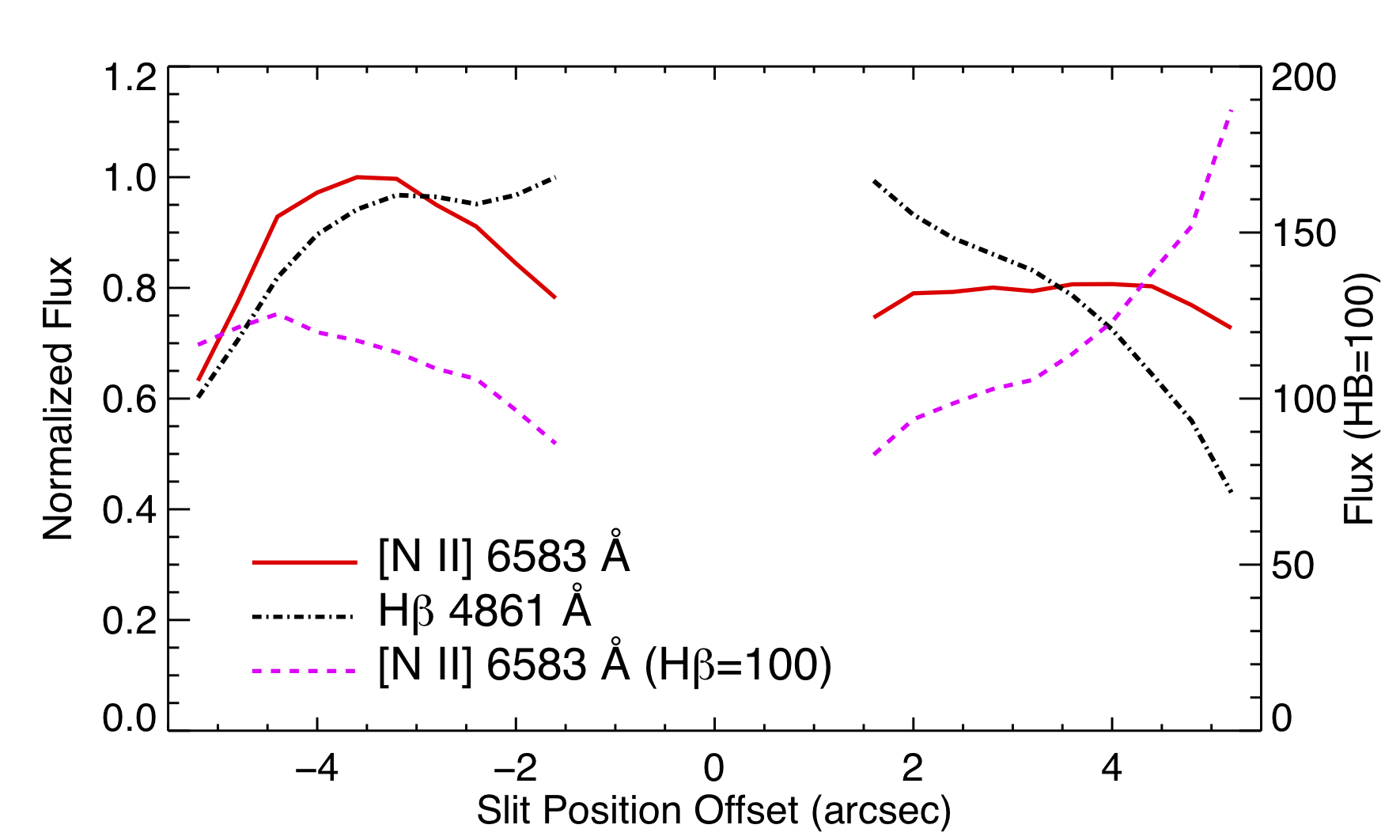}
   \includegraphics[width=8.5cm,trim=0.3cm 0.3cm 0.4cm 0.5cm,clip]{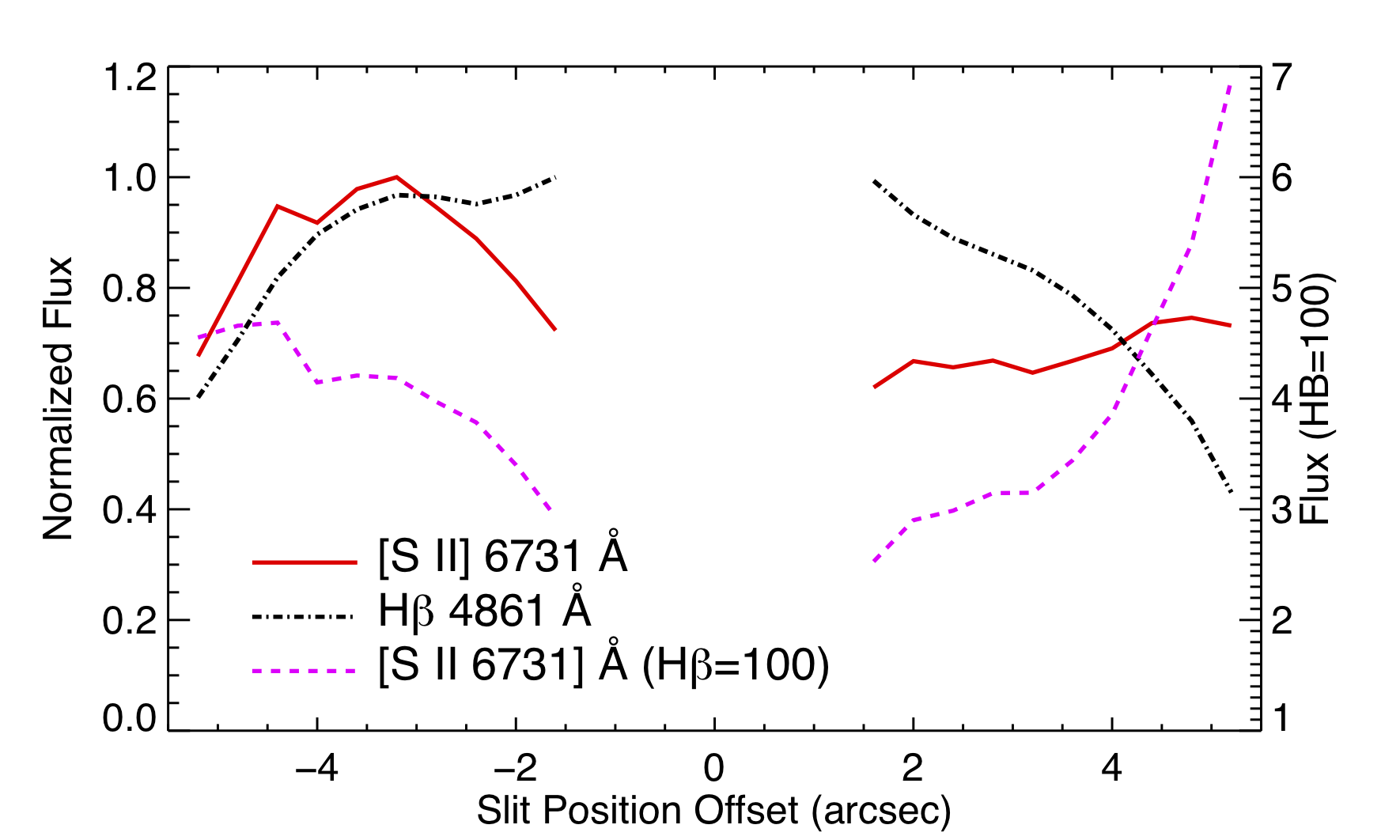}
   
   \caption{Reddening corrected line fluxes of intense lines across the slit. The pixels corresponding to the central region of the slit, where the contamination from the central star is significant, have been excluded. Regions 1 and 2 discussed in the text are identified in the top left plot. In that plot, each profile is normalised to its own peak value. Note that H$\beta$ and H$\gamma$ are superimposed (natural outcome of the extinction correction process; see text). In the other plots, the spatial profile of the line normalised to its own peak is shown in solid red, the profile normalised to H$\beta$~=~100 in dashed magenta, and the H$\beta$ profile normalised to its peak value in dot-dashed black.} 
   \label{lineprof}
\end{center}
\end{figure*}

Integrated line fluxes corrected for extinction (using $c$(H$\beta$) = 0.40) are presented in Appendix~\ref{ap_linetable}. For the spatially resolved analysis, we used the $c$(H$\beta$) values in Fig.~\ref{fig_extinction} to correct the fluxes in the corresponding position along the slit. Figure~\ref{lineprof} shows the radial profiles of the brightest lines corrected for extinction. H$\alpha$ is not presented because it was saturated. In the figure, Region~1 (Reg~1) is on the left and Region~2 (Reg~2) is on the right. As for the integrated analysis, the central region was eliminated to avoid contamination by the central star emission in the analysis and will not be shown in the plots. We show the H$\beta$ radial profile together with each one of the other radial profiles for comparison.

In Figure~\ref{lineprof}, we also include the line ratios to H$\beta$, which are used in the empirical analysis and can be used for comparison to the ratios from the integrated spectrum available in Appendix \ref{ap_linetable}. Note that although some emission lines are brighter towards the centre of the nebula, their fluxes relative to H$\beta $=~100 increase outwards (e.g., He~{\sc i}~5876~\AA\ and [O~{\sc ii}]~3726~\AA). This is a consequence of the lower H$\beta$ emission in the outer region of the source.

There is a discernible difference in the radial flux of the [N~{\sc ii}] 6583~\AA~ and [S~{\sc ii}] 6731~\AA~ lines between region 1 (on the left) and region 2 (on the right). Part of the difference is due to the slit position centre being offset with relation to the nebula centre. The slit position, however, does not explain all the differences. A flux enhancement is likely present. This will be further discussed in Sect.~\ref{sect_morphology}.

\subsection{Electronic Densities and Temperatures} \label{temdensec}

\subsubsection{Forbidden Lines Diagnostics}

\begin{table}
\centering
\caption{Electronic densities and temperatures obtained from the integrated spectrum analysis (Region~1 + Region~2).}
\label{temden_new}
\begin{tabular}{lclc}
\hline
\multicolumn{2}{c}{Density (cm$^{-3}$)} & \multicolumn{2}{c}{Temperature (K)}\\
\hline
\multicolumn{4}{c}{Low ionization}\\
$n_\mathrm{e}$[O~{\sc ii}]  & 1\,859 & $T_\mathrm{e}$[O~{\sc ii}]  & 7\,668\\ 
$n_\mathrm{e}$[S~{\sc ii}]  & 2\,188 & $T_\mathrm{e}$[S~{\sc ii}]  & 9\,542\\
$n_\mathrm{e}$(Low)         & 2\,024 & $T_\mathrm{e}$[N~{\sc ii}]  & 8\,697\\
                   &        & $T_\mathrm{e}$(Low)         & 8\,671\\
\hline
\multicolumn{4}{c}{Medium ionization}\\
$n_\mathrm{e}$(Medium) & $= n_\mathrm{e}$(Low) & $T_\mathrm{e}$[O~{\sc iii}] & 8\,905 \\
              &              & $T_\mathrm{e}$[Ar~{\sc iii}] & 7\,916 \\
              &              & $T_\mathrm{e}$(Medium) & 8\,577\\
              
\hline
\multicolumn{4}{c}{Balmer decrement}\\
$n_\mathrm{e}$[Bal] & $5.7 \times 10^7$ &  & \\

\hline 
\end{tabular}
\end{table}

Values of $n_\mathrm{e}$ and $T_\mathrm{e}$ obtained from the Tc~1 integrated spectrum are summarised in Table~\ref{temden_new}. {\sc neat} uses typical line ratio diagnostics\footnote{In Appendix~\ref{ap_diag}, we provide a list of the line ratios and the corresponding input $n_\mathrm{e}$/$T_\mathrm{e}$ used for the each diagnostic.} and iterative calculations. The quantities $n_\mathrm{e}$(Low), $T_\mathrm{e}$(Low), and $T_\mathrm{e}$(Medium) in the table refer to averages of the corresponding results for different ions, weighted as described by \cite{2012MNRAS.422.3516W}.

\begin{figure}
    \centering
    \includegraphics[width=\columnwidth]{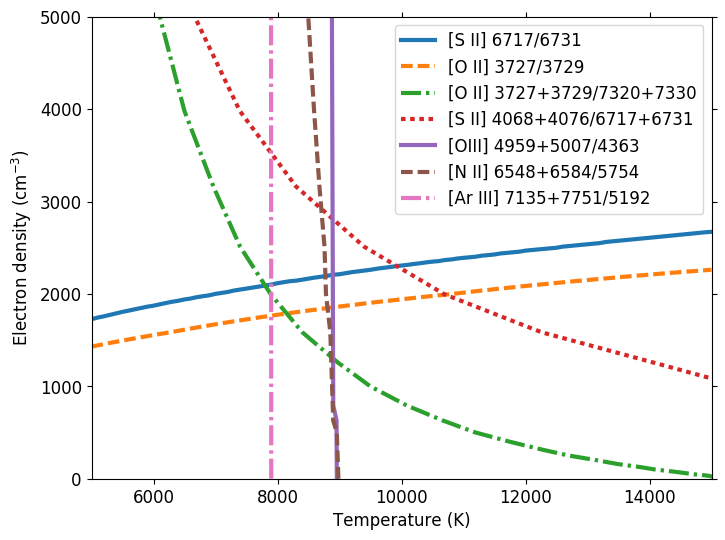}
    \caption{Plasma diagnostic diagram showing the range of values of electron temperature and density permitted by the observed diagnostic line ratios in the integrated spectrum.}
    \label{fig:diagnosticdiagram}
\end{figure}

Figure~\ref{fig:diagnosticdiagram} shows the values of T$_e$ and n$_e$ constrained by the observed line ratios. The densities calculated from the forbidden line diagnostics have typical values of $\sim$2\,000~cm$^{-3}$, while the temperatures are in the range 7\,700--9\,500~K. These are typical values obtained from these diagnostic lines and are in agreement with the values determined by \citet{2008ApJ...677.1100W} and \citet{Pottasch_etal_2011}. The figure clearly shows that [S~{\sc ii}] (4068~\AA+4076~\AA)/(6717~\AA+6731~\AA) and [O~{\sc ii}] (3727~\AA+3729~\AA)/(7320~\AA+7330~\AA) are not straightforward diagnostics of either temperature or density, having a significant dependence on both; {\sc neat} uses them as temperature diagnostics, by first calculating the density in the low ionization zone. As described in \citet{2012MNRAS.422.3516W}, they receive a low weighting in the calculation of the average T$_e$ for the low ionization zone, both being weighted 1.0, while T$_e$[N~{\sc ii}] is given a weight of 5.0. The locus of the [S~{\sc ii}] (4068~\AA+4076~\AA)/(6717~\AA+6731~\AA) line is slightly displaced from the region where the other diagnostic lines converged. This is probably due to the blue line pair being blended with O~{\sc ii} recombination lines; a $\sim$20\% reduction in their line fluxes would bring the locus into close agreement with the rest.

\begin{figure*}
   \begin{center}
   \includegraphics[width=5.8cm]{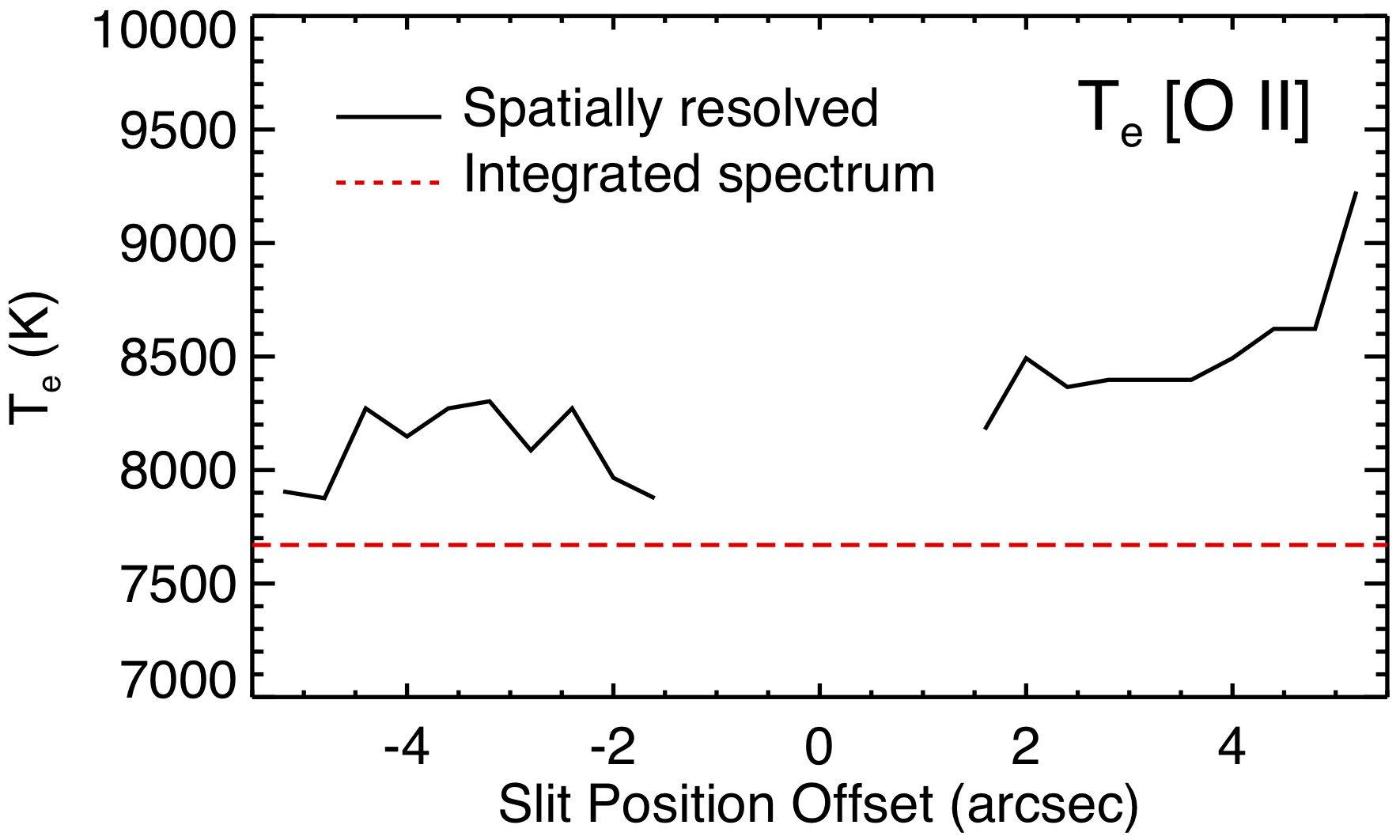}
   \includegraphics[width=5.8cm]{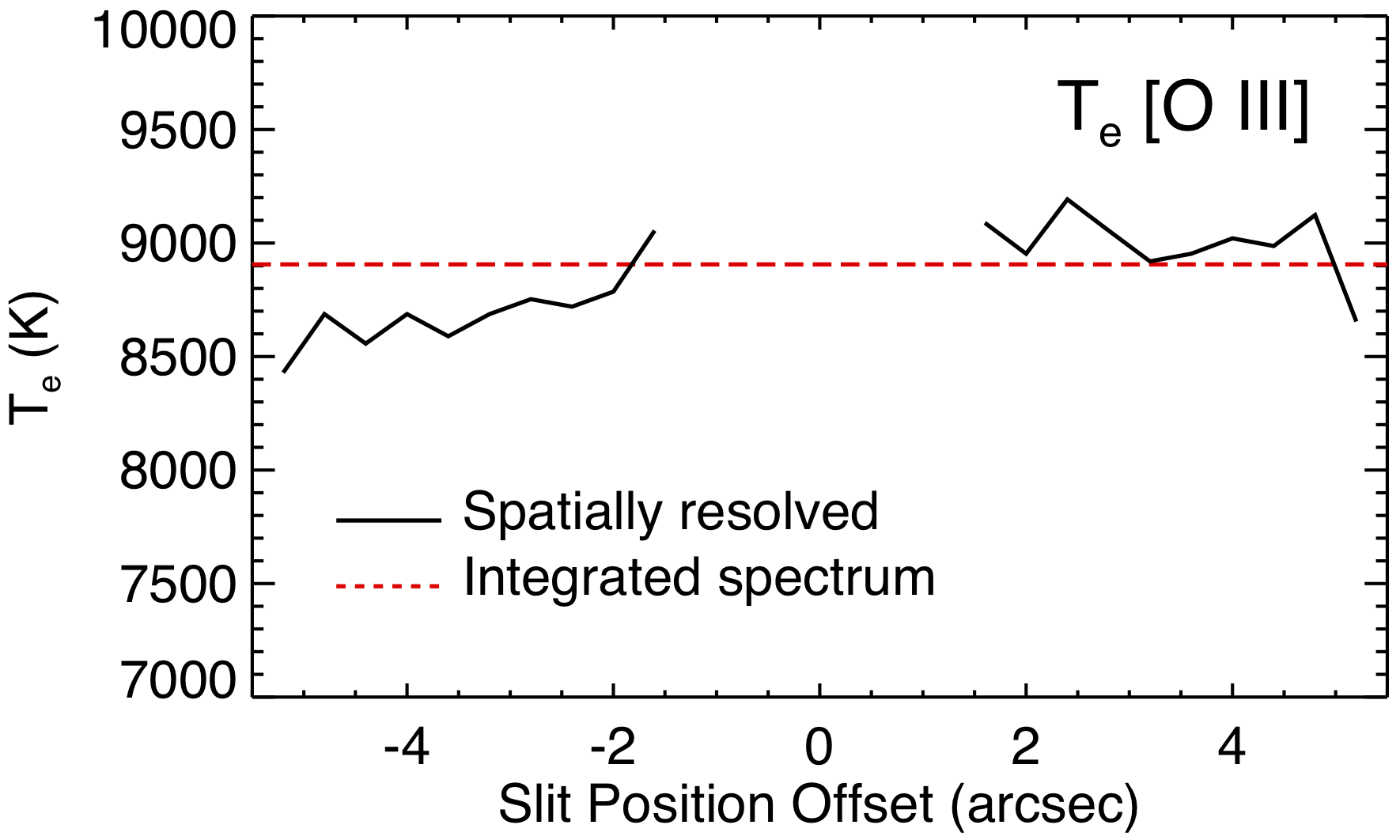}
   \includegraphics[width=5.8cm]{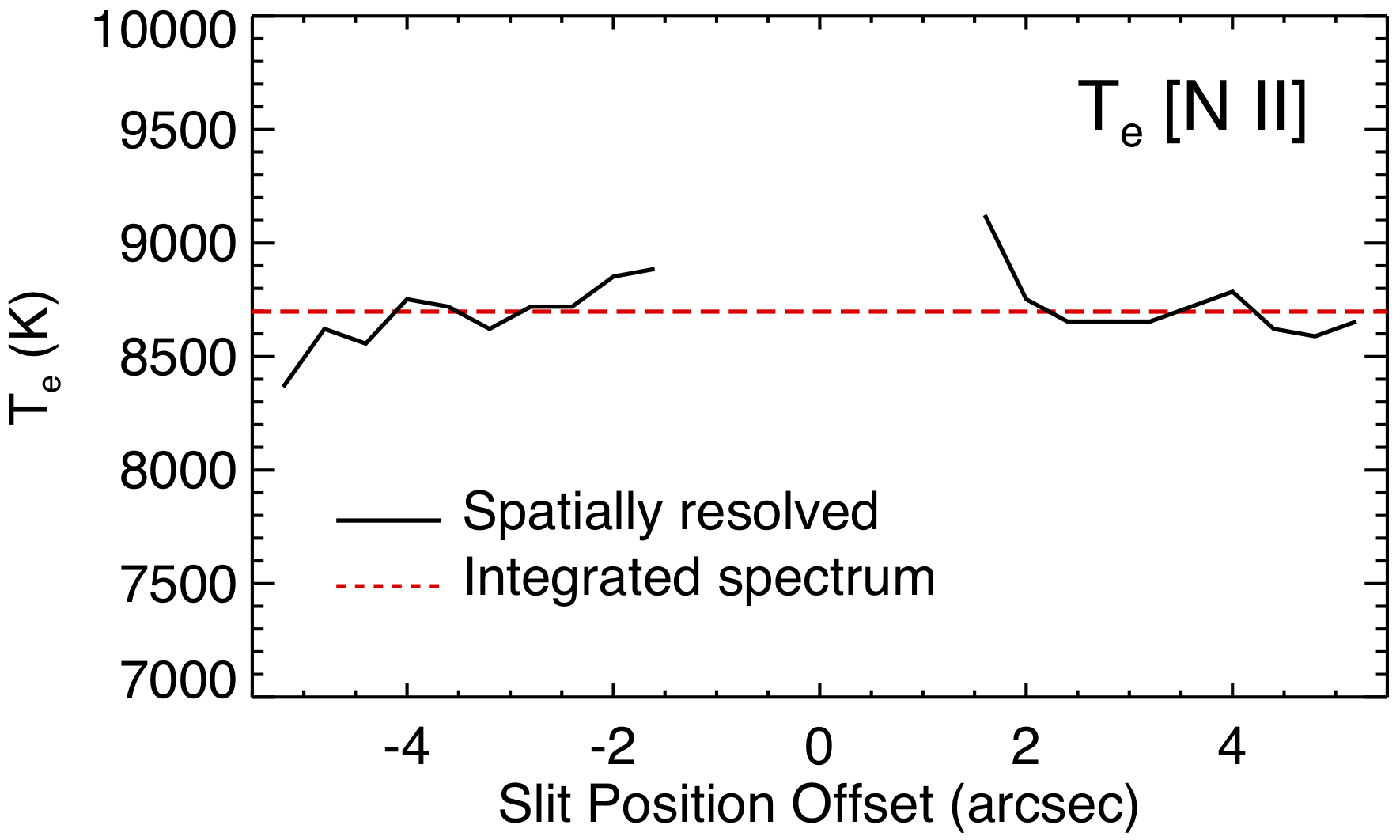}
   \includegraphics[width=5.8cm]{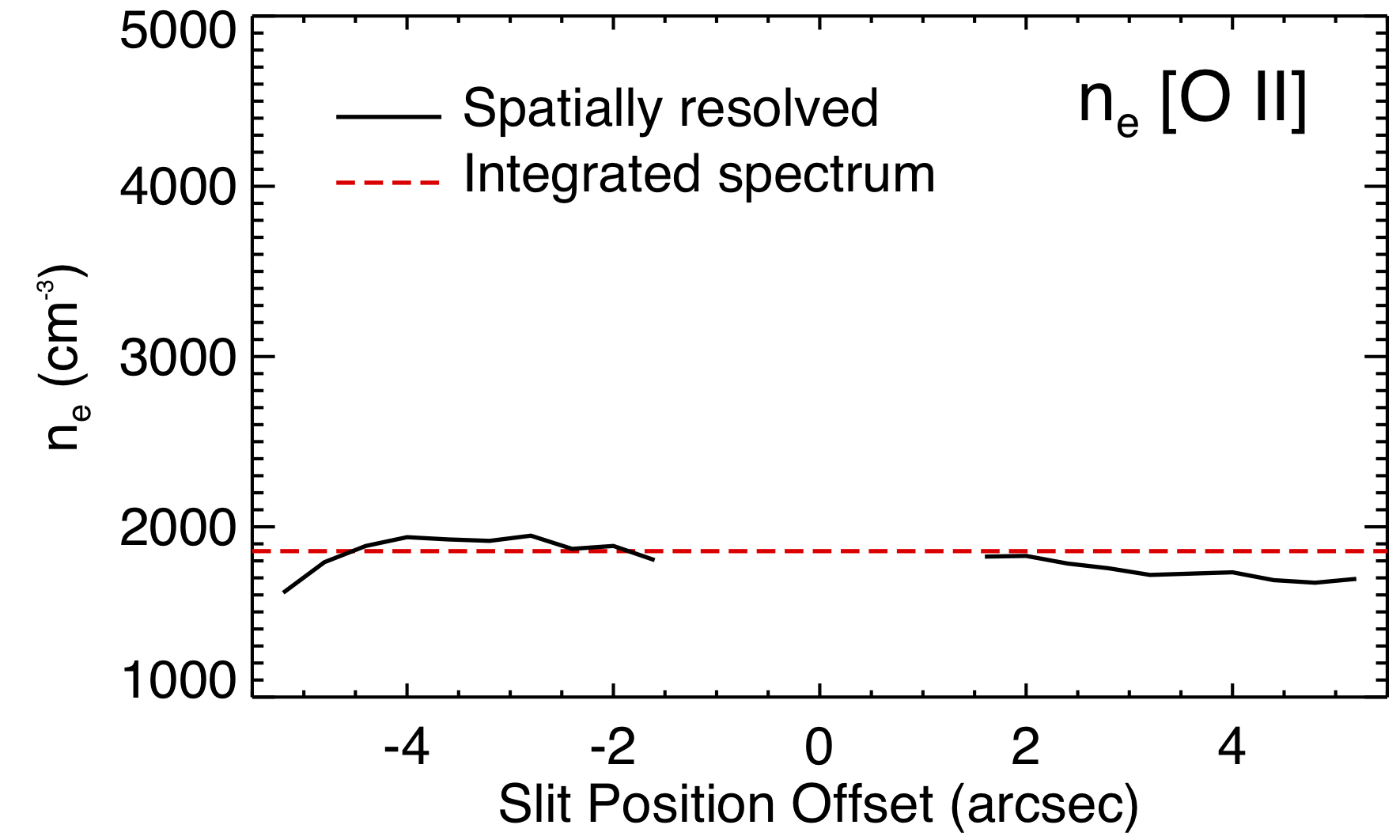}
   \includegraphics[width=5.8cm]{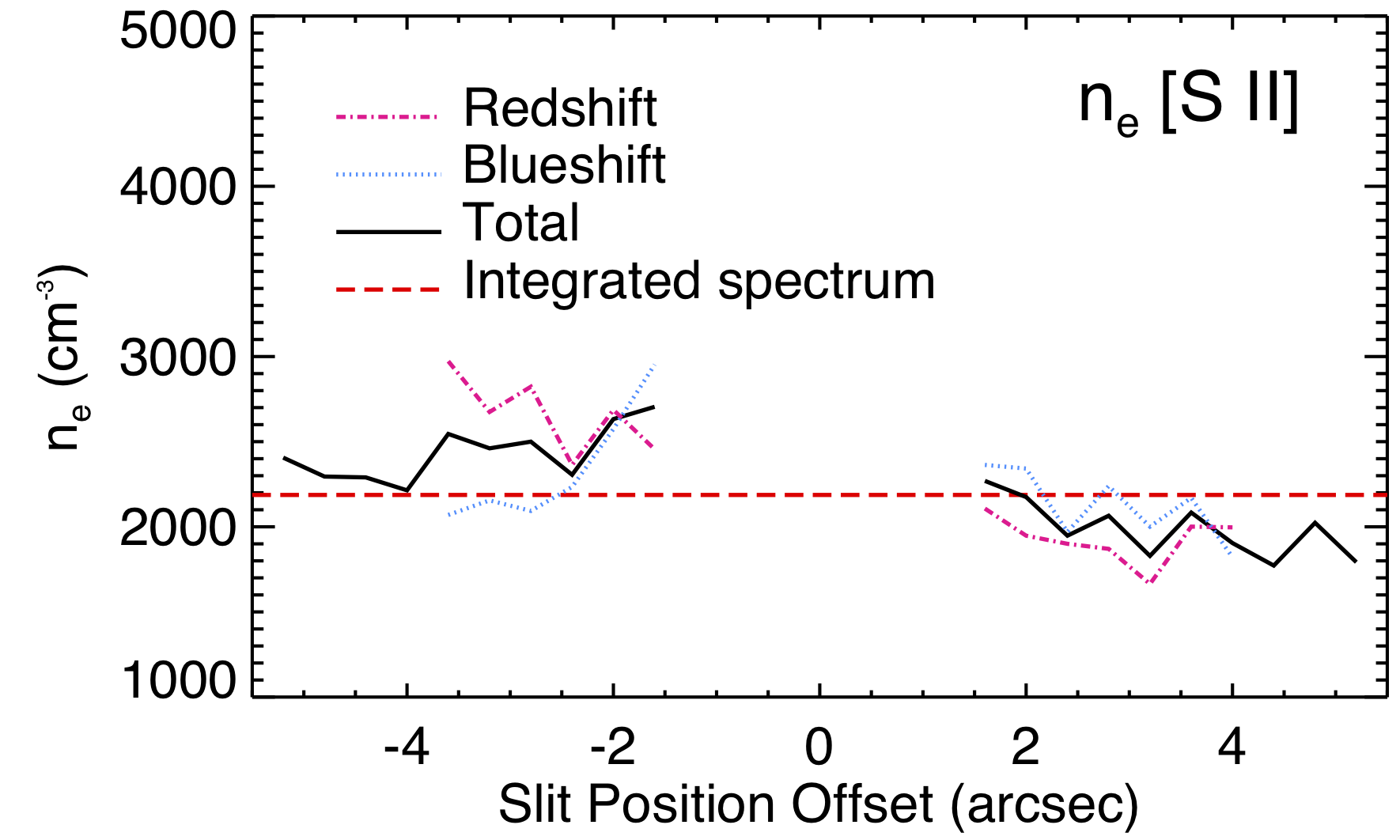}
   \caption{Spatial variation of the physical conditions across the slit. $n_\mathrm{e}$[S~{\sc ii}] was derived from the red- and blue-shifted components, as well as from the total fluxes of the [S~{\sc ii}] lines. In all panels, the red dashed line indicates the corresponding value obtained from the integrated spectrum.}   
   \label{fig_spatial_phcond}
   \end{center}
\end{figure*}

Figure~\ref{fig_spatial_phcond} shows the spatial variation of $n_\mathrm{e}$ and $T_\mathrm{e}$ along the slit, derived by {\sc 2d\_neb} from strong emission lines traditionally used for such diagnostics. The values of $n_\mathrm{e}$[O~{\sc ii}] and $n_\mathrm{e}$[S~{\sc ii}] (total) are similar and their values along the slit does not differ much from the integrated value. Both densities seems to be marginally higher in Reg~1.

The [S~{\sc ii}] emission lines at 6717~\AA\ and 6731~\AA\ show interesting features. The high spectral resolution of our data allowed us to detect a blue- and a red-shifted component in these lines (see Sections~\ref{sect_kinem} and \ref{sect_morphology} for more details). We derived the total $n_\mathrm{e}$[S~{\sc ii}] from the flux ratios of the combined blue- and red-shifted components of each line and individual $n_\mathrm{e}$ from the blue and from the red-shifted components. This analysis was only possible because we did not integrate the flux over a large spatial region of the slit (otherwise, the two components blend together). In Figure~\ref{fig_spatial_phcond}, we show the results from the blue and for the red-shifted components. There is a spatial region in Reg 1 where the density derived from the red-shifted component is significantly higher than that from the blue-shifted component. In Reg 2, the blue-shifted component is slightly higher through (almost) the whole slit. The difference in values between the blue- and red-shifted components in Reg 2 is small and might not be significant. The total $n_\mathrm{e}$[S~{\sc ii}] profile shows that the density in Reg 1 is slightly higher than in Reg 2, and that the density is marginally higher towards the centre of the slit (which corresponds to a region closer to the centre of the source).

$T_\mathrm{e}$[N~{\sc ii}] and $T_\mathrm{e}$[O~{\sc iii}] do not show significant variation along the slit, apart from a slightly increase towards the centre of the nebula. Both curves are compatible to the corresponding values from the integrated analysis. The spatial distribution of $T_\mathrm{e}$[O~{\sc ii}] exhibits a similar variation, but the values are higher than the temperature obtained from the integrated spectrum (7\,700~K). Such difference is caused by small differences in the input $n_\mathrm{e}$ used for the derivation (see Appendix~\ref{ap_diag}), which can produce the observed difference in $T_\mathrm{e}$[O~{\sc ii}].

\subsubsection{Hydrogen Recombination Lines}\label{sect_Hdecrement}

\begin{figure}
\begin{center}
\includegraphics[width=8cm,trim=0.0cm 1.0cm 0.0cm 0.0cm,clip]{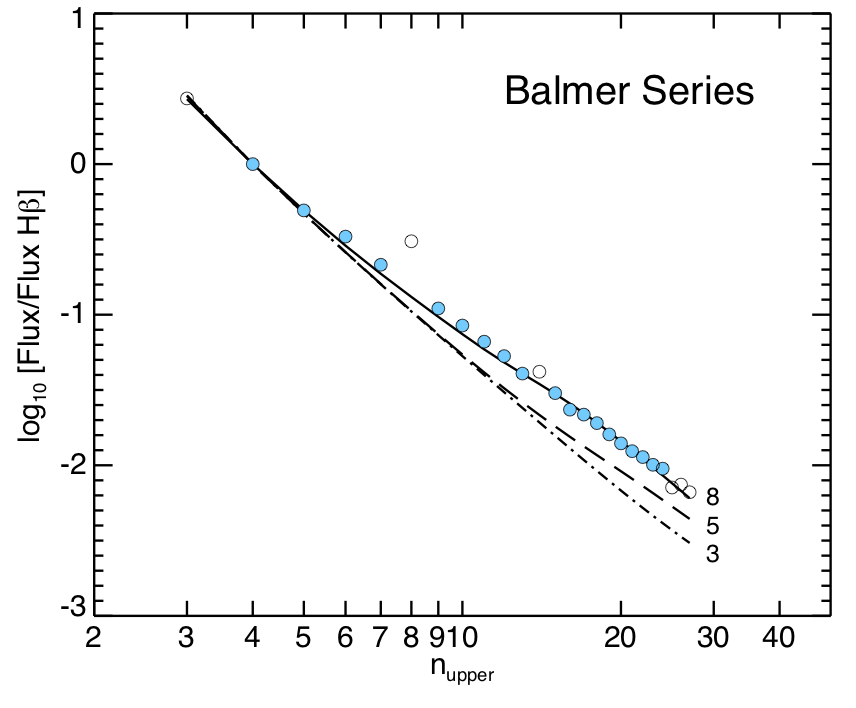}
\includegraphics[width=8cm,trim=0.0cm 0.4cm 0.0cm 0.0cm,clip]{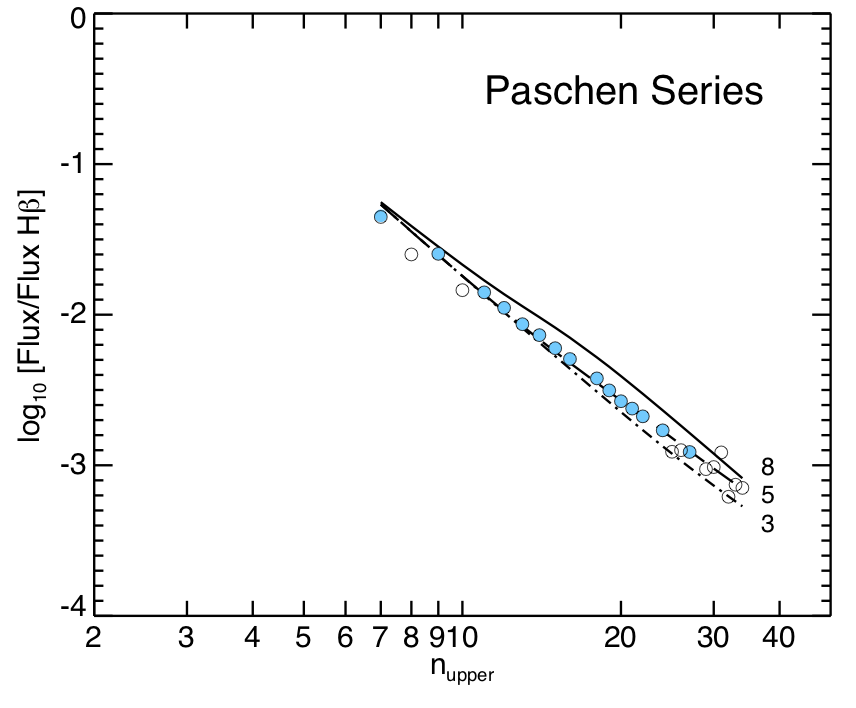}
\caption{Balmer and Paschen hydrogen series decrements. Dots are values from observation; open dots represents strongly blended or contaminated lines not considered in the fit.} The best fit for the Balmer series observed curve corresponds to theoretical ratios for [$T_e$,$n_e$] = [10\,000~K,10$^8$~cm$^{-3}$], while for Paschen the best fit corresponds to [$T_e$,$n_e$] = [12\,500~K,10$^5$~cm$^{-3}$]. Two other curves for different $n_e$ (but the same $T_e$) are added in each plot for comparison. The logarithm of the density is shown close to the corresponding curve.
\label{decrement}
\end{center}
\end{figure}

The Balmer decrement density calculated with {\sc neat} is 5.7$\times10^{7}$~cm$^{-3}$, which is very high for the typical diffuse nebular gas of PNe. Such high densities are expected to be found inside dense clumps such as cometary knots or in discs produced in a binary central stellar system. 
We explore this topic a little further in Fig.~\ref{decrement}, where we show the observed (de-reddened) line fluxes relative to H$\beta$ for all the lines we measured from the Balmer series (i.e. the Balmer decrement). The equivalent plot for the Paschen line series is also shown. \citet{Storey_Hummer_1995} provide recombination coefficients for such lines for a wide range of physical conditions. After exploring the whole range provided, we determined the best fit to our observations. Figure \ref{decrement} compares the observed and theoretical ratios. The best fit to the observed Balmer decrement is found for [$T_\mathrm{e}$,$n_\mathrm{e}$]~=~[10\,000~K,10$^8$~cm$^{-3}$]. For the Paschen series, we found [$T_\mathrm{e}$,$n_\mathrm{e}$] = [12\,500~K,10$^5$~cm$^{-3}$]. The theoretical line ratios for such condition are shown in the corresponding panels in Fig.~\ref{decrement}. Two additional curves for the same $T_\mathrm{e}$, but different $n_\mathrm{e}$ are also displayed in each plot for comparison. Although showing in the plot, we eliminated from the fit lines that could have been significantly contaminated by other atomic or sky lines, as well as lines affected by blends or saturation. For the Balmer series decrement fitting, we considered lines from H$\beta$ up to H24, except for H$\alpha$ (saturated), H8 and H14. For the Paschen series, lines from P7 to P27 were considered, except P8, P10, P23, and P25. For lines above H24 and P28 the continuum determination is quite uncertain and line intensities can be significantly affected.

Reasonable fits can also be obtained for temperatures in the range $T_\mathrm{e} =$~5\,000--12\,500~K for Balmer and $T_\mathrm{e} =$~7\,500--20\,000~K for Paschen and for densities in the range $n_\mathrm{e} =$~10$^6$--10$^{10}$~cm$^{-3}$ for Balmer and $n_\mathrm{e} =$~10$^4$--10$^6$~cm$^{-3}$ for Paschen. The hydrogen line ratios are not very sensitive to the density or gas temperature, but it is clear that high densities are needed to explain the Balmer decrement for the lines produced in upper levels.

\citet{2004MNRAS.351..935Z} performed a systematic study of temperatures and densities obtained from hydrogen recombination lines for a sample of PNe. Two objects of their sample, the PNe M2-24 and IC~4997, show values of density obtained from forbidden line indicators and from the Balmer decrement with similar discrepancies as ours. For M2-24, the most extreme case, log($n_\mathrm{e}$[O~{\sc ii}])~=~4.0~$\pm$~0.2 and log($n_\mathrm{e}$[S~{\sc ii}])~=~3.2~$\pm$~0.1, while log($n_\mathrm{e}$[Bal])~=~7.0~$\pm$~0.7 ($n_\mathrm{e}$ in cm$^{-3}$). \citet{2004MNRAS.351..935Z} suggested that the presence of high density regions embedded in the ionized gas could explain such discrepancy.

Hydrogen recombination lines may reveal dense gas regions, which cannot be probed by forbidden lines due to their typically lower critical densities. Paschen \citep{tg1997} and H~{\sc i} $\alpha$ \citep{strel1996,aleman2018} high-n lines, for example, have been used to measure electron densities in high density stellar envelopes and even inferring the presence of a circumstellar disc.

\subsection{Ionic Abundances}

The $n_\mathrm{e}$ and $T_\mathrm{e}$ we determined (Sect.~\ref{temdensec}) were used as input parameters to derive ionic abundances in the nebula. The specific $n_\mathrm{e}$/$T_\mathrm{e}$ pair used to derive the ionic abundance of a given species is listed in Appendix~\ref{ap_diag}. For ionic species with more than one emission line observed, we derived their abundances from each of the lines independently, obtained the final result by taking a flux-weighted average of these results.

We present the ionic abundances obtained from the collisionally excited lines (CELs) and from optical recombination lines (ORLs) for the integrated spectrum analysis in Table~\ref{ionic_reg}. Our ionic abundances are similar to the values calculated by \citet{Pottasch_etal_2011} (typically within 0.5~dex).

\begin{table}
\centering
\caption{Ionic abundances derived from the Tc~1 integrated spectrum (Region 1 + Region 2).}
\label{ionic_reg}
\begin{tabular}{lclc}
\hline
Ratio & log       & Ratio & log\\
      & Abundance &       & Abundance\\
\hline
\multicolumn{4}{c}{CEL Abundances}\\ 
\hline
N$^{+}$/H$^+$  & -4.54 & S$^{+}$/H$^+$   & -6.46\\
O$^{0}$/H$^+$  & -6.44 & Ne$^{2+}$/H$^+$ & -5.90\\
O$^{+}$/H$^+$  & -3.44 & Ar$^{2+}$/H$^+$ & -5.96\\
O$^{2+}$/H$^+$ & -4.20 & &\\

\hline
\multicolumn{4}{c}{ORL Abundances}\\ 
\hline
He$^{+}$/H$^+$       & -0.98 & O$^{2+}$/H$^+$ (V2)    & -3.86\\
C$^{2+}$/H$^+$       & -3.22 & O$^{2+}$/H$^+$ (V5)    & -3.05\\
N$^{2+}$/H$^+$ (V3)  & -3.40 & O$^{2+}$/H$^+$ (V10)   & -3.90\\
N$^{2+}$/H$^+$ (V5)  & -3.00 & O$^{2+}$/H$^+$ (V19)   & -3.59\\ 
N$^{2+}$/H$^+$ (V20) & -2.94 & O$^{2+}$/H$^+$ (V25)   & -2.71\\
N$^{2+}$/H$^+$ (V28) & -3.12 & O$^{2+}$/H$^+$ (V28)   & -3.43\\
N$^{2+}$/H$^+$       & -3.40 & O$^{2+}$/H$^+$ (3d-4f) & -2.66\\
O$^{2+}$/H$^+$ (V1)  & -3.92 & O$^{2+}$/H$^+$         & -3.86\\
\hline 
\end{tabular}
\end{table}

\begin{figure*}
\begin{center}
\includegraphics[width=5.8cm]{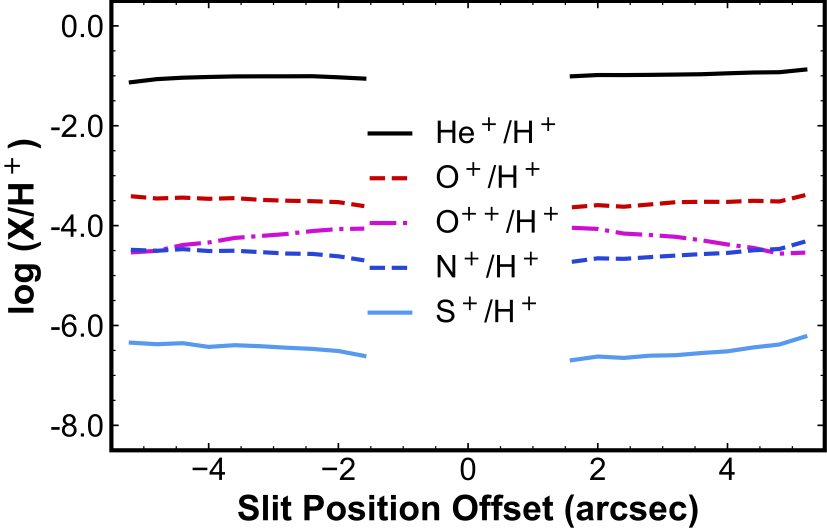}
\includegraphics[width=5.8cm]{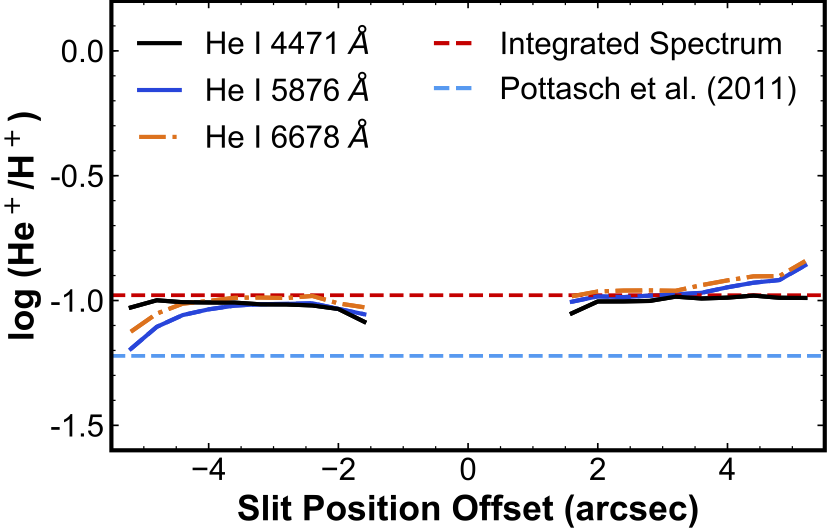}
\includegraphics[width=5.8cm]{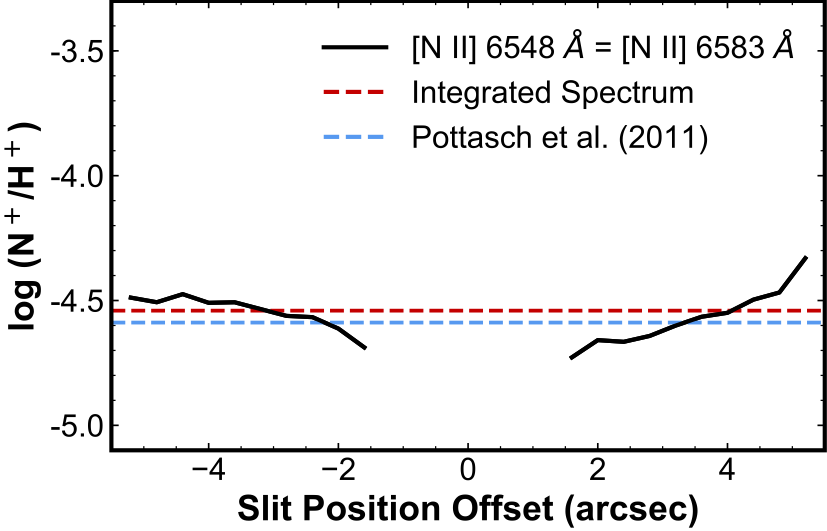}
\includegraphics[width=5.8cm]{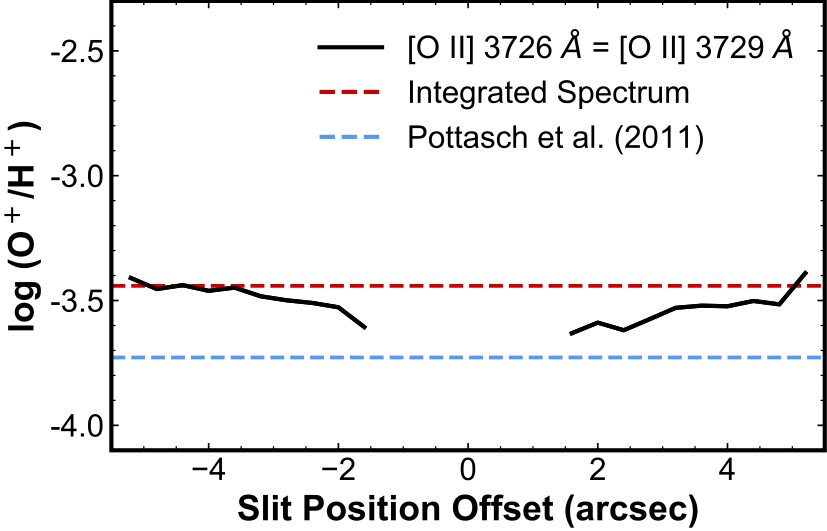}
\includegraphics[width=5.8cm]{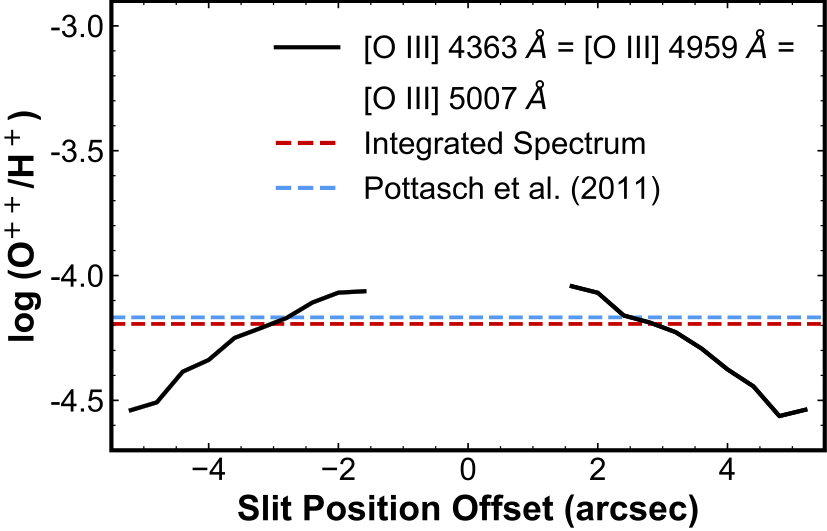}
\includegraphics[width=5.8cm]{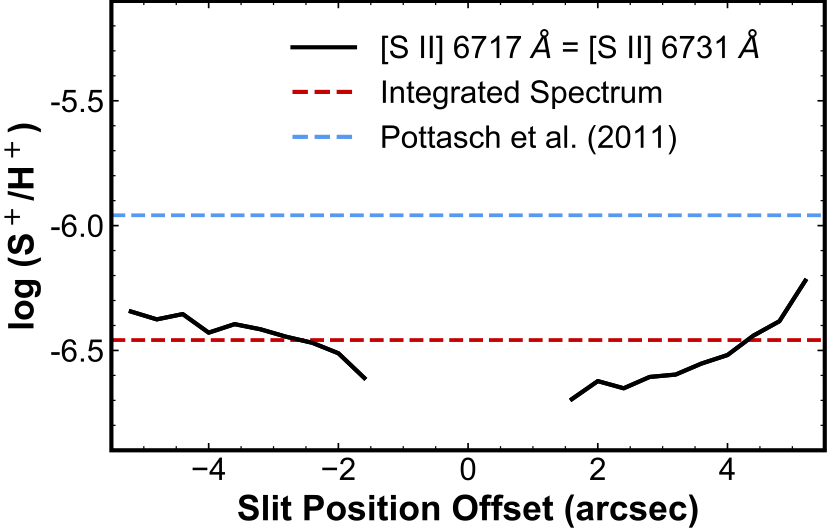}
\caption{Tc~1 ionic abundances spatial variation across the slit. The first plot shows the weighted average result for each ion, whiles the reminder plot show results for different lines for individual ions. For comparison, values from the literature and from this work obtained from the integrated analysis are also shown (horizontal lines).}
\label{ionic}
\end{center}
\end{figure*}

The spatial variation of the ionic abundances for He$^+$, N$^+$, O$^+$, O$^{2+}$, and S$^+$ obtained from the most intense emission lines is shown in Fig.~\ref{ionic}. In the top left plot, the weighted average abundance of each ion is shown. The expected change between dominant oxygen ion due to recombination can be seen: the O$^{2+}$ abundance decreases towards the outer parts of the nebula, while the opposite is seen for O$^{+}$. It is also possible to see the increase of N$^{+}$ and S$^{+}$ towards the outer edge of the nebula. However, one should be careful when reading such increase towards the outer zones, as these are relative to H$^+$. Note that the abundance shape of the O$^{+}$, N$^{+}$ and S$^{+}$ abundance profiles are very similar. For example, all of them increase abruptly around +2 to +5~arcsec. An inspection of Fig.~\ref{lineprof} reveals that the intrinsic fluxes of these emission lines (red curves) do not rise significantly between +2 and +5~arcsec. As the H$\beta$ profile is not constant and decreases towards the edges of the slit, the pixel-by-pixel normalisation of a given line to H$\beta =$~100 causes the line ratio to increase where H$\beta$ is lower. Therefore, the strong variation observed in the profiles of the ionic abundances does not reflect an absolute increase of concentration of the ionic species. It reflects its growth in respect to the (lower) absolute H$^+$ abundance.

In the other panels, we show the results for individual ions obtained from individual emission lines and the averaged value. We also add the integrated spectrum and literature values for comparison. The abundances obtained from individual lines are very similar and, for most of the ions (except for He$^+$), the curves coincide in the scale shown.

\subsection{Total Abundances} \label{sect_tot_abundances}

From the ionic abundances, we derived the total elemental abundances, using the Ionization Correction Factors (ICFs) determined by \citet{1994MNRAS.271..257K} to correct for unobserved ions. For He, we assume no correction as recommended by \citet{2014MNRAS.440..536D}.


\begin{table*}
\centering
\caption{Tc~1 Parameters from the literature$^{a}$ and from our work.}

\label{abund_reg}
\label{Tab_Tc1_Literature}
\label{tab:models}

\begin{tabular}{lcccccccccc}
\hline
Parameter & \multicolumn{2}{c}{Po11}  & Ot14 & \multicolumn{4}{c}{This Work} \\

& Empirical$^{b}$ & Model$^{c}$ & Model$^{c}$ & \multicolumn{2}{c}{Empirical$^{d}$} & LCO Model$^{c}$ & XS Model$^{c}$ \\

\hline

Distance (kpc)  &  & 1.8 & 3.0 &  &  & 2.2 & 2.1 \\
Inner Radius (arcsec) &  & 0.4 & -- &  &  & 0.3 & 0.3 \\
Outer Radius (arcsec) &  & 6.25 & -- &  &  & 6 & 6 \\
Filling Factor &  & 1 & -- &  &  & 1 & 1 \\

\hline

\multicolumn{8}{l}{Central Star:} \\

$T_{eff}$ (K)  &  & 34\,700  & 34\,060 &  &  & 30\,000  & 32\,000  \\
$L_{\star} (L_{\sun})$  &  & 1\,480  & 2\,512 &  &  & 3768 & 2026 \\

\hline

\multicolumn{8}{l}{Nebula:} \\

$n_\textrm{H}$ (cm$^{-3}$) &  & 2850 & --  &  &  & 2500 & 2500 \\
Dust Grains &  & Graphite & --  &  &  & Graphite & Graphite \\
Grain Distribution &  & Single Size & --  &  &  & MRN$^{e}$ & MRN$^{e}$ \\
Grain sizes ($\mu$m) &  & 1.0 & --  &  &  & 0.0005 to 1.5 & 0.0005 to 1.5 \\
Dust-to-gas Ratio &  & --  & --  &  &  & 2.8$\times$10$^{-3}$ & 3.0$\times$10$^{-3}$ \\

\hline

Gas Abundances: & Empirical$^{f2}$ & Model & Model & CEL & ORL  & Model & Model \\
log He/H  & $\geq$-1.22 & -1.08 & -- & -- & -0.98 & -0.98 & -0.98 \\
log C/H   & -3.44 & -3.33 & -3.29 & --    & -2.91 & -3.25 & -3.10 \\
log N/H   & -4.44 & -4.41 & -4.17 & -4.47 & -2.60 & -4.25 & -4.25 \\
log O/H   & -3.59 & -3.57 & -3.39 & -3.37 & -3.06 & -3.60 & -3.25 \\
log Ne/H  & -4.20 & -4.52 & -4.02 & -5.08 & --    & -4.40 & -4.40 \\
log Mg/H  & --    & -6.50 & --    & --    & --    & -5.10 & -5.10 \\
log Si/H  & --    & -6.22 & --    & --    & --    & -6.10 & -6.10 \\
log S/H   & -5.55 & -5.80 & -5.42 & -6.05 & --    & -5.65 & -5.80 \\
log Cl/H  & -7.03 & -7.04 & --    & --    & --    & -7.10 & -6.95 \\
log Ar/H  & -5.29 & -5.52 & -5.93 & -5.69 & --    & -6.00 & -5.60 \\
log Fe/H  & -6.81 & --    & --    & --    & --    & -6.40 & -6.40 \\
log Kr/H   & --   & --   & --   & --   & --   & --  &  -7.90$^{g2}$\\
C/O       & 1.4   & 1.7   & 1.3   & --    & --    & 2.2   & 1.4  \\

\hline

\multicolumn{8}{l}{Ionization Correction Factors (ICFs):}\\

He & --  & -- & -- & --  & 1.0 & 1.3$^{h}$ & 2.2$^{h}$ \\ %
C  & 1.0 & -- & -- & --  & 2.1 & 1.6  & 2.8 \\ %
N  & 1.4 & -- & -- & 1.2 & 6.3 & 1.9/2.1$^{i}$ & 1.3/4.1$^{i}$ \\ 
O  & 1.0 & -- & -- & 1.0 & 6.3 & 1.3/4.3$^{j}$ & 1.1/8.6$^{j}$ \\ %
Ne & 1.0 & -- & -- & 6.6 & --  & 14.3 & 25.4 \\ %
S  & 1.0 & -- & -- & 2.6 & --  & 6.7  & 3.3 \\ 
Ar & 1.0 & -- & -- & 1.9 & --  & 1.4  & 2.4 \\ %

\hline

\multicolumn{8}{l}{$^{a}$References: \citet[][Po11]{Pottasch_etal_2011} and \citet[][Ot14]{2014MNRAS.437.2577O}. $^{b}$Values from classical analysis.}\\

\multicolumn{8}{l}{$^{c}$Values from photoionization models. $^{d}$Elemental abundances obtained from the Tc 1 integrated spectrum}\\ 

\multicolumn{8}{l}{(Reg 1 + Reg 2). $^{e}$Power law distribution typical for the interstellar medium; see \citet{1977ApJ...217..425M}.}\\

\multicolumn{8}{l}{$^{f}$Abundances determined from CELs, except for He. $^{g}$The Kr abundance was determined separated from}\\

\multicolumn{8}{l}{the other abundances. See Sect.~\ref{kripton} for details. $^{h}$As the He~{\sc i} lines are not well reproduced by the models, the}\\

\multicolumn{8}{l}{resulting ICF(He$^{+}$) should not be used (see text). $^{i}$The two values correspond to ICF(N$^{+}$) and ICF(N$^{2+}$),}\\

\multicolumn{8}{l}{respectively. $^{j}$The two values correspond to ICF(O$^{+}$) and ICF(O$^{2+}$), respectively.}

\end{tabular}
\end{table*}

From the integrated spectrum, we derived the total elemental abundances for N, O, Ne, Ar, and S from CELs, and for He, C, N, and O from ORLs. The results are presented in Table~\ref{abund_reg}. The calculated ICFs used to derive each abundance are also shown in the table.

For heavy elements, ORLs are detected from C$^{2+}$, O$^{2+}$ and N$^{2+}$. Nineteen O~{\sc ii} lines are detected, and for the final O$^{2+}$ abundance, we take a flux-weighted average of lines in multiplets V1, V2, V10 and V19; the multiplets all give abundances consistent with each other. Weak lines from other multiplets give higher values. 

Four recombination lines of C$^{2+}$ are detected. The strong 4267~\AA\ line has a flux of 0.55 when normalised to H$\beta$=100. Additionally, we detect three lines from the recombination ladder which populates the upper level from which the 4267~\AA\ line emission arises; their abundances are in excellent agreement with that from 4267~\AA. The adopted carbon abundance in Table \ref{abund_reg} is from 4267~\AA\ alone, as it is by far the strongest line.

The abundances we found are reasonably similar to the values found by \citet{Pottasch_etal_2011} and \citet{2014MNRAS.437.2577O} (see Table~\ref{Tab_Tc1_Literature}), with the exception of Ne. \citet{Pottasch_etal_2011} and \citet{2014MNRAS.437.2577O} calculate similar higher Ne abundances when including the mid-IR Ne$^{+}$ lines (there are no [Ne~{\sc ii}] optical lines). The difference observed in our Ne abundance seems then to be caused by ICF underestimation. Photoionization models indicate that the Ne$^{2+}$ ICF is above 14 (see Sect.~\ref{sec:bestmodel}).

Abundances from ORLs always exceed those from CELs. The mechanism behind such difference is not completely understood yet, but the largest discrepancies occur in nebulae with short-period binary central stars (\citealt{corradi2015}, \citealt{jones2016}, \citealt{wesson2018}). The abundance discrepancy factor (ADF) can be determined most accurately for O$^{2+}$, for which both types of line are seen in optical spectra. We derive the O$^{2+}$ ORL abundance by first calculating a flux-weighted mean abundance for each detected multiplet, and then averaging the abundances for the well detected V1, V2, V10 and V19 multiplets. V5 lines are also well detected but give a much higher abundances, as has been observed in many other planetary nebulae (e.g. \citealt{wesson2005}). Lines from other multiplets are detected but with higher measurement uncertainty. The derived value of O$^{2+}$/H$^{+}$ from ORLs exceeds the CELs value by a factor of 2.2, which is equal to the median value of over 200 measurements in the literature (\citealt{wesson2018}). 
ADF(N/H) can also be measured, but its value relies on the ICF for nitrogen, which is extremely uncertain as CELs trace only N$^{+}$, while RLs are available only for N$^{2+}$. The ICF derived from CELs is 1.18, which indicates that N$^{+}$ is the dominant ionization stage. The ICF to determine N/H from the N$^{2+}$ RL abundance would then be 6.30, which gives an extremely high abundance value of -2.60 (in log) and an ADF(N/H) of 75. This could be attributed to one or more of the following: N/H from RLs is overestimated; N/H from CELs is underestimated; the ICF scheme does not well represent the ionization structure of this low excitation object; or that applying an ICF calculated from CELs to RL abundances is not valid. Photoionization models presented in Sec.~\ref{sect_models} provide ICFs between 1.3 and 1.9 for N$^{+}$ and between 2 and 4 for N$^{2+}$. The N$^{2+}$ recombination lines have intensities less than 0.1 per cent of H$\beta$ and are thus subject to relatively large measurement uncertainties. 

An estimation of C/O is only possible with ORLs, as there are no CELs of ionized carbon in the optical. While ORLs always give higher nebular abundances than CELs, ionic ratios from ORLs are generally found to be consistent with those from CELs (e.g. \citealt{2000MNRAS.312..585L}, \citealt{wesson2005}), and so we determine the C/O abundance ratio from ORLs.

The abundance ratio C$^{2+}$/O$^{2+}$ is classically assumed to be similar to C/O, but \citet{2014MNRAS.440..536D} find it to differ by up to an order of magnitude. We use their ICF to correct for the unseen C$^{+}$. This relies on the ratio of O$^{+}$/O$^{2+}$, which is only available from CELs. This gives a C/O ratio of 1.4, which is very close to the value of 1.38 found by \citet{Pottasch_etal_2011}, using UV data to estimate C/H and optical data to estimate O/H, both from CELs. This has the additional systematic uncertainty of comparing different apertures: 4.7$\times$22~arcsec and 11$\times$22~arcsec slits centred on the star in the IR, and two 2$\times$4~arcsec slit extractions at different positions excluding the central star in the optical. Our measurement of C/O from lines of the same type observed at the same time should be subject to smaller systematic uncertainty and confirms Tc~1 to be a C-rich nebula. In the models presented in Sec.~\ref{sect_models} the ICF for C$^{2+}$ is estimated to be between 1.6 and 2.8.

\begin{figure*}
\begin{center}
   \includegraphics[width=6.3cm]{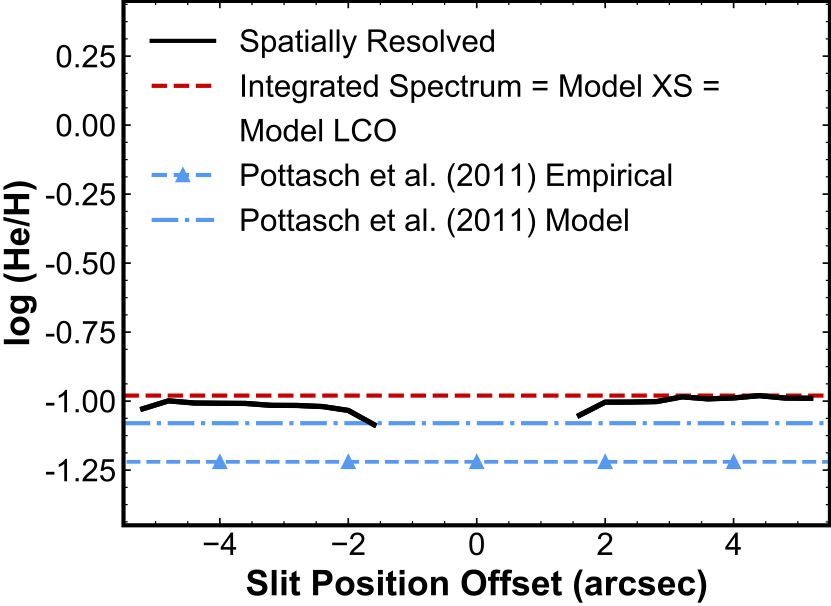}
   \includegraphics[width=6.3cm]{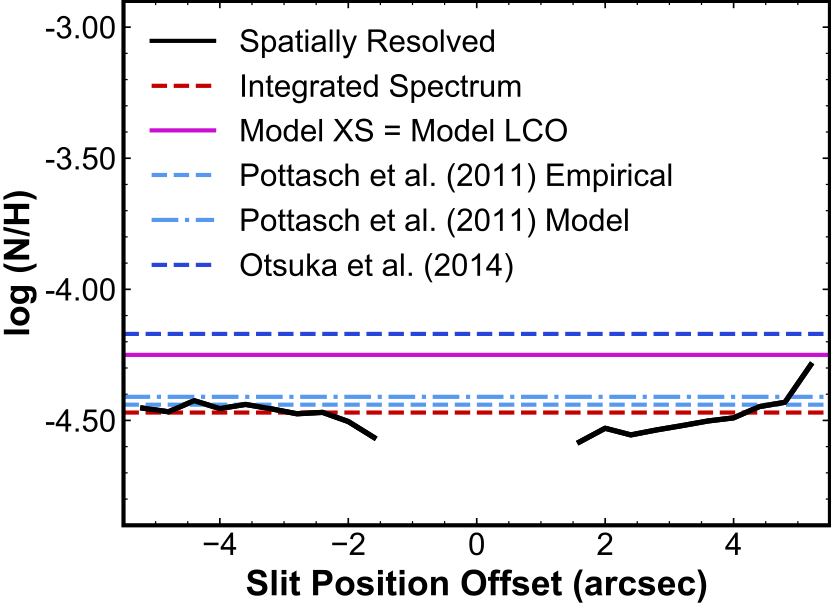}\\
   \includegraphics[width=6.3cm]{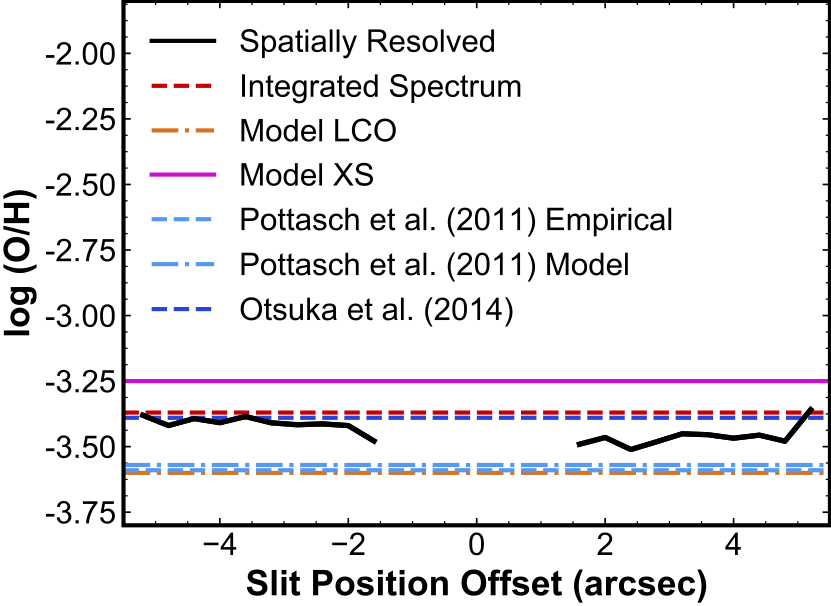}
   \includegraphics[width=6.3cm]{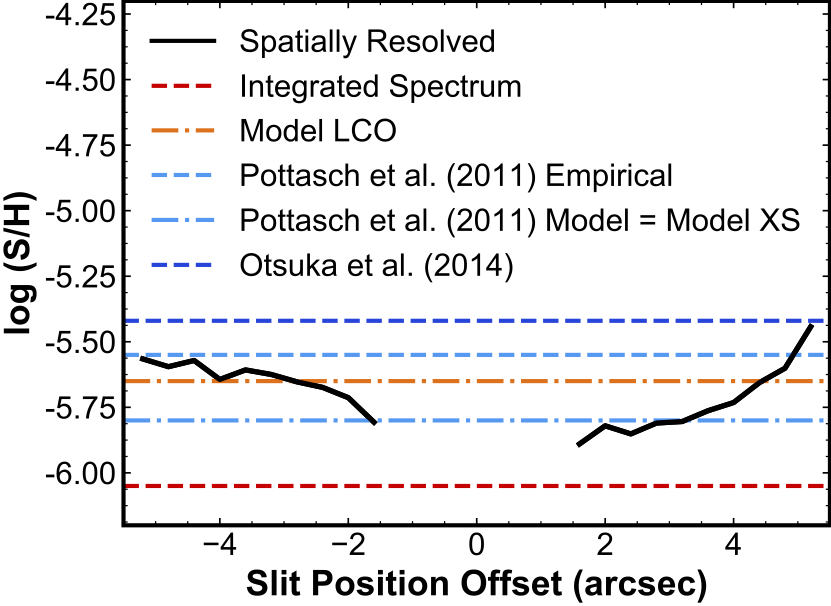}
   \caption{Spatial variation of elemental abundances across the slit on Tc~1 (black solid curves). Empirical abundances from our integrated spectrum, as well as from our models and from the literature are also indicated (horizontal lines). The He abundance determined by \citet{Pottasch_etal_2011} is a lower limit.}
   \label{abund_prof}
   \end{center}
\end{figure*}

The spatial variation of the He, N, O, and S abundances is shown in Fig.~\ref{abund_prof}. No significant gradient was found (less than 0.3~dex variation is seen in all cases). The variation seen is, however, systematic. In all cases the abundances increase toward the outer nebula. This is likely an effect of the lower H$^+$ abundance propagated in the calculations. 

It is not straightforward to determine the uncertainties in the abundances, as they should account for the errors in the observations and in the classical method used to determine the abundances and not all these errors are well known. We estimate the uncertainties are 30-50 per cent (i.e. 0.15-0.30~dex). Values in the largest offset positions have higher uncertainties due to the lower fluxes.

One of the sources of uncertainty is the ICFs. The ICFs are derived from one-dimensional photoionization models. \citet{2012IAUS..283..144G} discuss the limitations of the ICFs derived from one-dimensional models and the ICF dependence on the morphology of the source. As Tc~1 is an almost spherical source, we expect that uncertainties due the use of ICFs derived from one-dimensional models should be low in our study, although still present.

The abundances obtained from integrated spectrum in our work and from the literature are also included in the plots of Fig.~\ref{abund_prof} for comparison. The variation seen can also be seen as an estimation of the uncertainties in the abundances for each element.

\section{Kinematics} \label{sect_kinem}

Most line profiles in our integrated spectra could be well fitted by a single Gaussian profile. Some of the lines, however, deviate from this profile. Below we study these lines and their integrated and spatially-resolved profiles. The study provides information on the kinematics and structure of the nebula. In the tables provided in Appendix~\ref{ap_linetable}, we indicate the lines that deviate significantly from a Gaussian profile and their fluxes. In such cases, we present both the fluxes calculated from a Gaussian fit as well as those obtained by integrating the observed spectral profile.

\subsection{Double-Peaked Lines}  \label{sect_2peak}

\begin{figure*}
\begin{center}
   \includegraphics*[width=11.1cm,trim=4.cm 3.cm 4.5cm 3.5cm,clip]{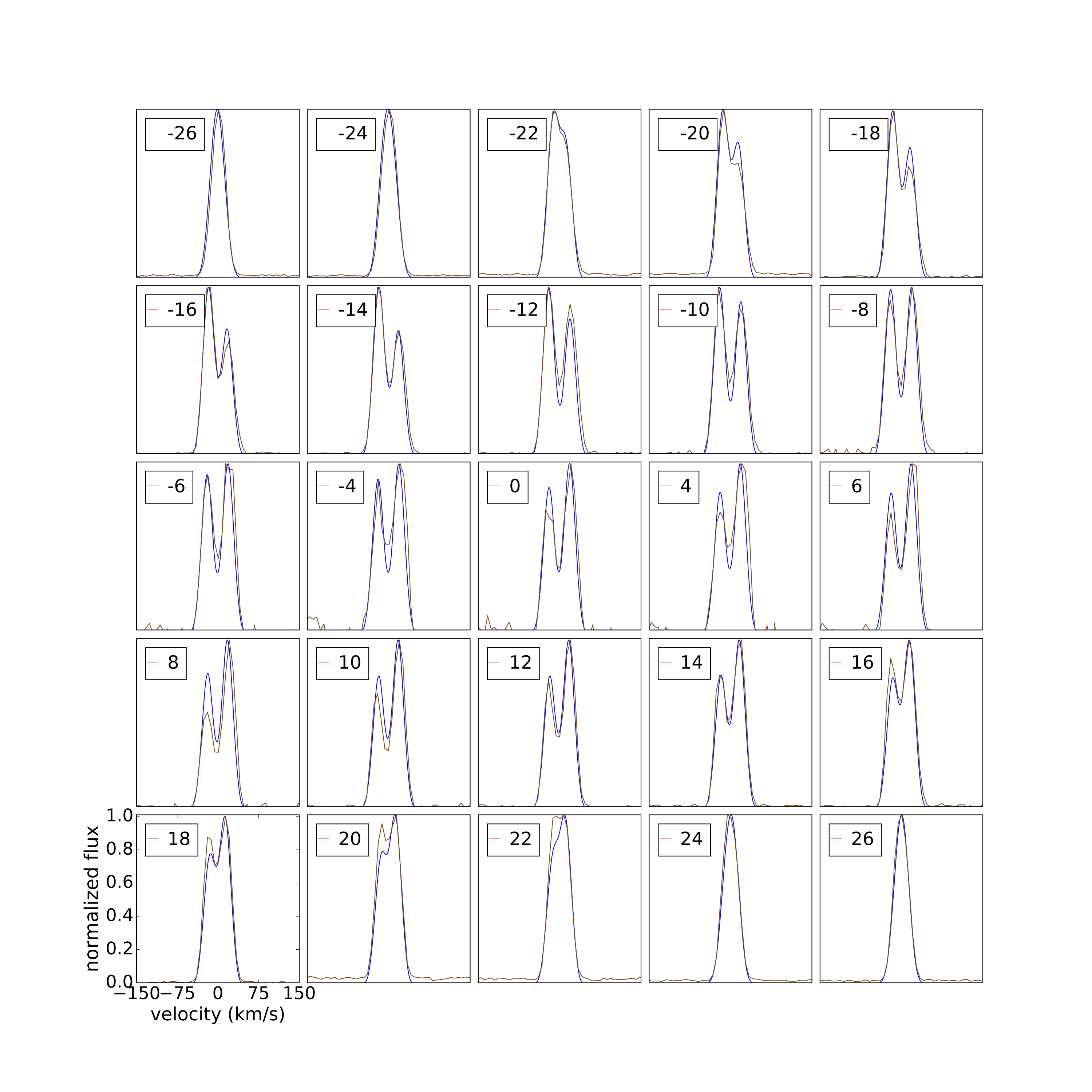}
   \caption{Comparison between the observed (red curve) and modelled (blue) profiles of the [S~{\sc ii}] emission lines in different positions along the slit. The modelled curve was obtained from the morpho-kinematical model simulated with the code {\sc shape}. The number in each plot corresponds to the spatial pixel from which the profile is derived.}
   \label{sii}
   \end{center}
\end{figure*}

The lines [N~{\sc ii}] at 6548 and 6583~\AA, [S~{\sc ii}] at 6717 and 6731~\AA, and [Cl~{\sc ii}] 8579~\AA\ show a double-peaked profile in several spatial positions when they are examined pixel by pixel. The double-peaked structure is clearer in the pixels closer to the PN centre. The separation of the peaks decreases towards the outer regions of Tc~1 and single-peak profiles are seen in the pixels close to the two extremes of the slit. Fig.~\ref{sii} shows the [S~{\sc ii}] line profiles as an example. 

\begin{figure}
\begin{center}
\centering
\includegraphics[width=7.5cm]{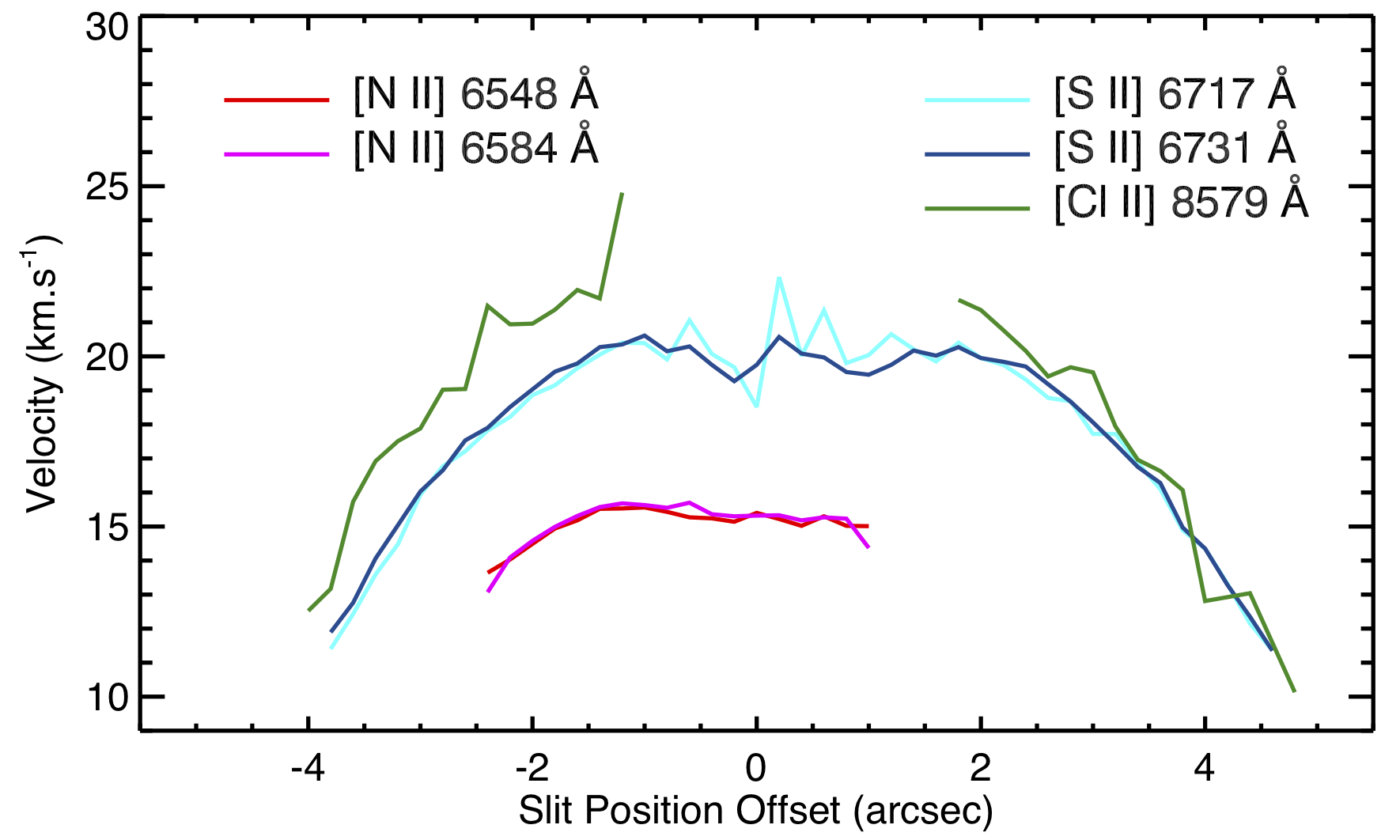} 
\caption{Doppler velocities inferred from the blue- and red-shifted peak separation in the most intense [N~{\sc ii}], [S~{\sc ii}], and [Cl~{\sc ii}] emission lines.}
\label{kinem_profile}
\end{center}
\end{figure}

Such behaviour is characteristic of the blue- and red-shifted components of emission lines in an expanding spherical shell. Therefore, measuring the spectral peak distance of the two components in each of the five lines listed above we derive the expansion velocity component of the [N~{\sc ii}], [S~{\sc ii}], and [Cl~{\sc ii}] shells in the direction of the line of sight. The peak-to-peak distance measured in velocity space should be twice the actual expansion velocity. We perform this calculation in each of the 55 individual spatial pixels to have the maximum possible spatial resolution. Figure~\ref{kinem_profile} shows the results. The expansion velocity is the maximum value in that plot, where the line-of-sight component is parallel to the expansion velocity vector.

Among the five lines mentioned above, the separation of the components is larger in the [S~{\sc ii}] lines and, therefore, the deblending of the components is easier for this line. The [Cl~{\sc ii}] 8579~\AA\ emission line is very weak, and we included in the analysis for completeness. However, given its low signal-to-noise ratio and the consequently large uncertainty on its measurements, the values derived from this line should be taken with caution.

We found expansion velocities of 16~km~s$^{-1}$ for [N~{\sc ii}] and 22~km~s$^{-1}$ for [S~{\sc ii}] lines. Expansion velocities for [N~{\sc ii}] in the literature agree with the values we obtain \citep[e.g.,][]{1989A&AS...78..301W,1992A&A...260..314B}. The general agreement between the results obtained from both [S~{\sc ii}] lines is good. The same can be said for the velocities derived from the two [N~{\sc ii}] lines. For the [Cl~{\sc ii}] line, velocities up to 25~km~s$^{-1}$ are found, but the low signal-to-noise ratio of the line makes the estimates uncertain compared to those from the other lines. Such velocities are typical for PNe.

Only the velocity profiles of a few lines could be well resolved. A good quality spatially resolved position-velocity diagram spectrum can only be obtained for the brightest lines. In addition, the resolving power in the UVB arm is not enough to reliably resolve velocities similar to those obtained above. The instrumental spectral resolution for the UVB arm is 9\,100, which corresponds to 33~km~s$^{-1}$, while the VIS arm has a much higher spectral resolution of $R =$~17\,400, which corresponds to velocities displacements of 17~km~s$^{-1}$. 

In the UVB spectra, the lines of [S~{\sc ii}] 4068 and 4076~\AA, [Fe~{\sc iii}] 4658, 4881, and 5270~\AA, and Si~{\sc ii} 5056~\AA\ show weak evidence for a double peaked structure. The lines of the hydrogen Balmer series and He lines in this arm do not present clear indications of blue- and red-shifted components, despite the deviation from a Gaussian profile seen in some of the brightest H lines. Furthermore, the bright H$\beta$ line, for example, has a high thermal broadening (21~km~s$^{-1}$ at 10\,000~K), which should results in a barely resolved double-peak line profile.

In addition to the VIS arm lines mentioned earlier in this section, hints of double peak structure can also be seen in the [N~{\sc ii}] 5755~\AA, [S~{\sc iii}] 6312 and 9068~\AA, and [O~{\sc i}] 6300~\AA\ lines. As in the UVB arm, the H (Paschen) and He lines in this arm do not show double peaked structure. We remind the reader that H$\alpha$ is saturated.

In a nebula expanding homologously (i.e. $v = K \times r$, where $r$ is the distance from the central star and $K$ is a constant), emission lines from high ionization species (such as [O~{\sc iii}]) emanate from gas that lies closer to the central star and therefore should have lower expansion velocities compared to the emission lines from low ionization species (such as [N~{\sc ii}] and [S~{\sc ii}]). Similarly, since the ionization potential of singly ionized nitrogen is higher than that of sulphur (29.6 and 23.3~eV, respectively) the [N~{\sc ii}] shell is closer to the central star and should thus expand at a lower velocity. Indeed, as mentioned above, the [N~{\sc ii}] shell is moving outward with lower velocities than the [S~{\sc ii}] shell.

Following the same reasoning, singly ionized oxygen lines should also show a double peak profile given that its ionization potential is similar to those of singly ionized N and sulphur. We expect that the expansion velocity for the [O~{\sc ii}] lines should be close to that of [N{\sc ii}] and [S~{\sc ii}], on the order of 15-20~km~s$^{-1}$ (see Figure~\ref{vel_components}). The width of the [O~{\sc ii}] line gives a velocity of the order of 20~km~s$^{-1}$ (see Table~\ref{components}) but due to the lower resolution of the UVB spectrum, the line is not resolved. The [O~{\sc ii}] lines in the VIS spectrum are unfortunately weak and blended. The brighter [O~{\sc ii}] 3726 and 3729~\AA\ lines are in the UVB arm which has an insufficient resolution to adequately resolve such velocity. \citet{1989A&AS...78..301W} published a velocity of 5.5~km~s$^{-1}$ for [O~{\sc ii}] in Tc~1.

For high ionization lines such as the doublet [O~{\sc iii}] 4959 and 5007~\AA\, we expect expansion velocities lower than those found for the [S~{\sc ii}] and [N~{\sc ii}]. Both [O~{\sc iii}] lines have widths that correspond to velocities about 15~km~s$^{-1}$, which shows that the line is not resolved by the instrument. This also implicates that the expansion velocity for this line is of only a few km~s$^{-1}$. A velocity of 4~km~s$^{-1}$ has been previously determined for [O~{\sc iii}] \citep{1989A&AS...78..301W}.

Deviations from homologous expansion laws have been reported in highly collimated outflows due to a poloidal velocity component \citep[][]{2009ApJ...691..696S}. We found no evidence of such collimated outflows in Tc~1 which could significantly alter the global velocity field of the nebula.

\subsection{Lines with Broad Wings}

\begin{figure*}
\begin{center}
  \includegraphics[width=8.7cm]{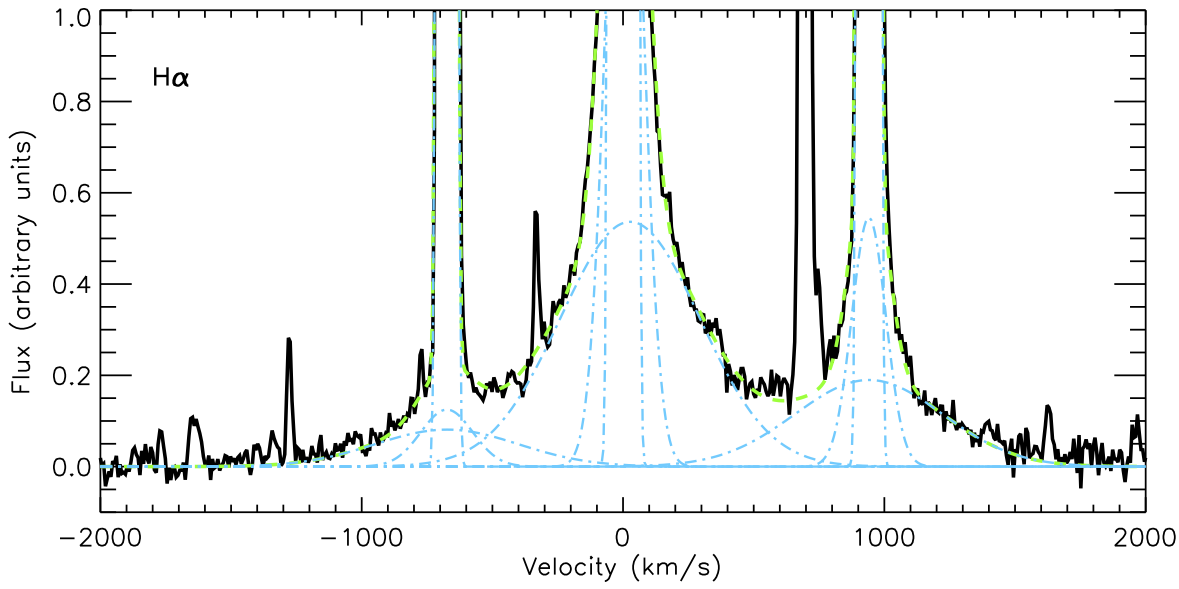}
  \includegraphics[width=8.7cm]{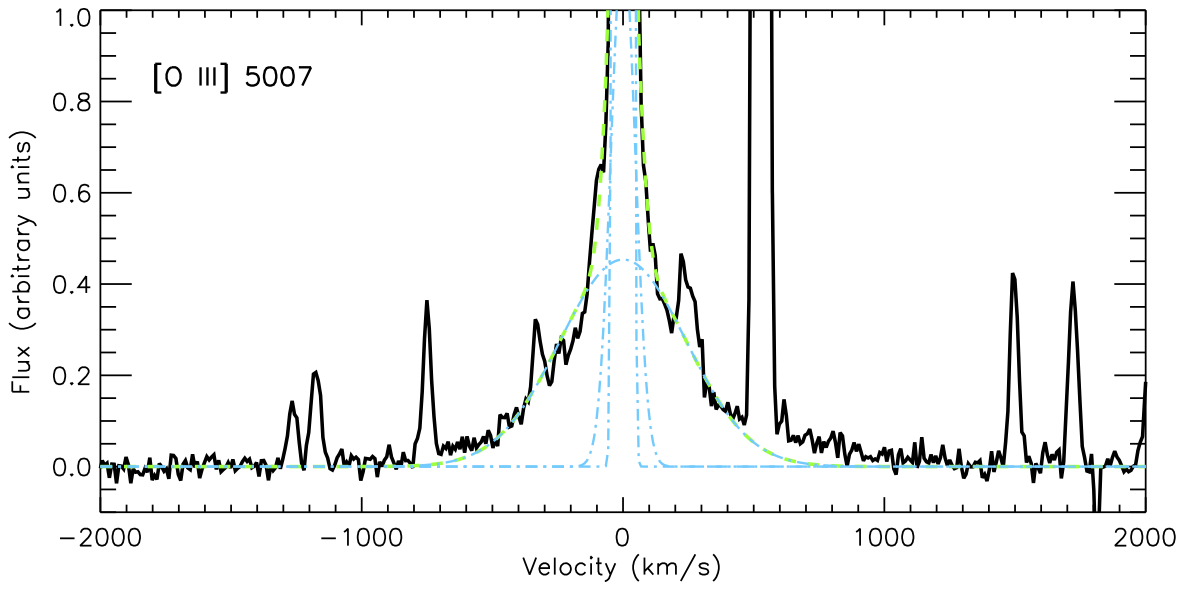}
  \includegraphics[width=8.7cm]{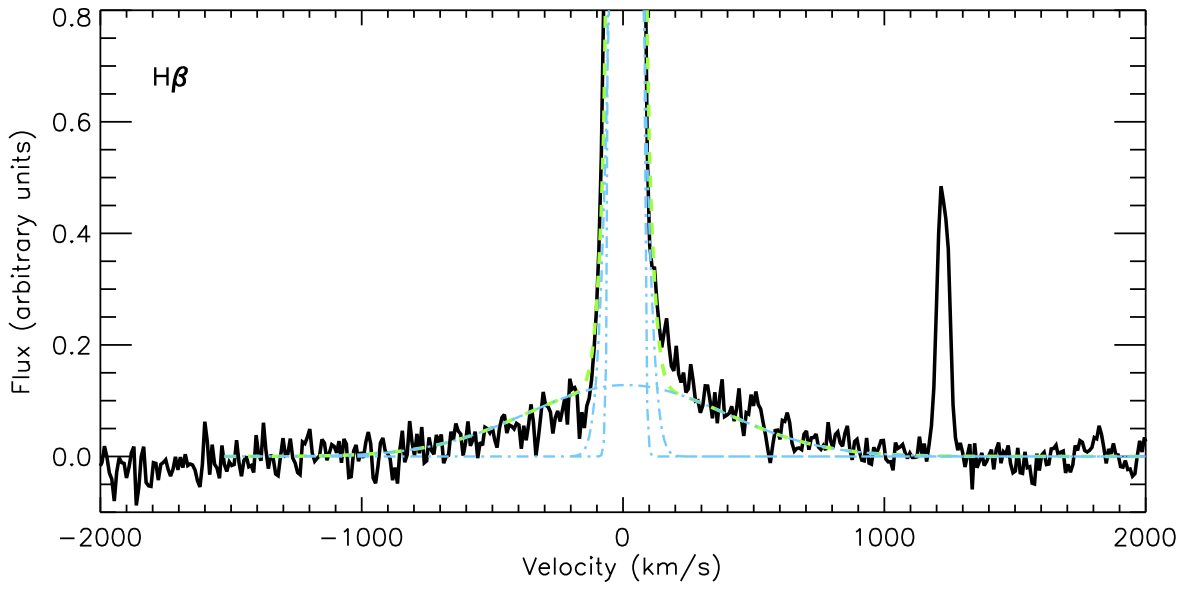}
  \includegraphics[width=8.7cm]{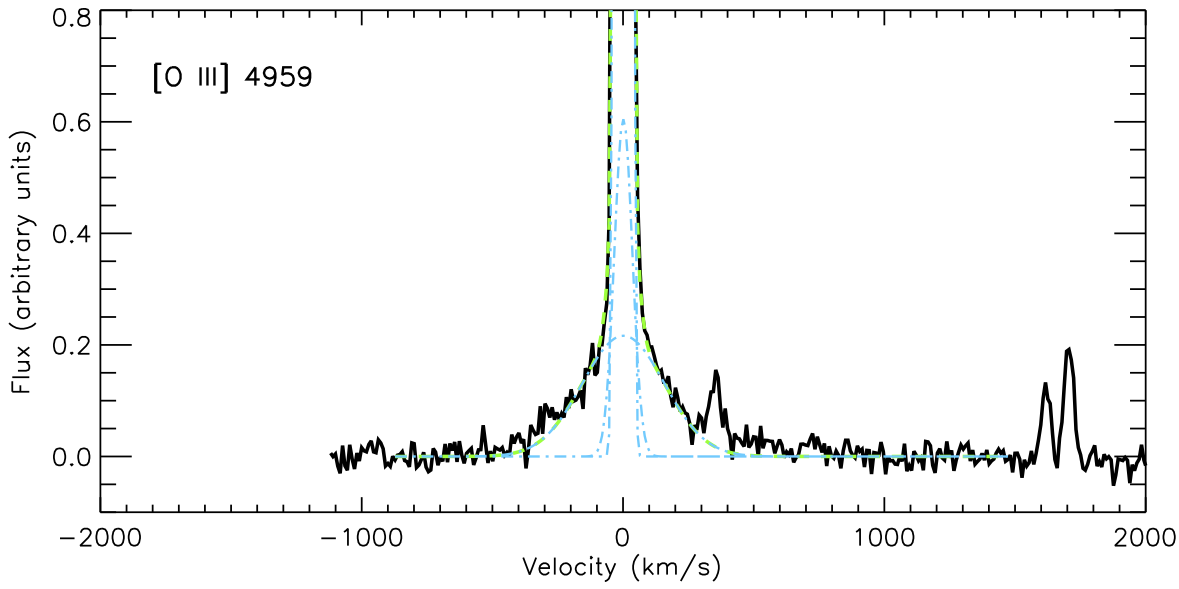}
  \includegraphics[width=8.7cm]{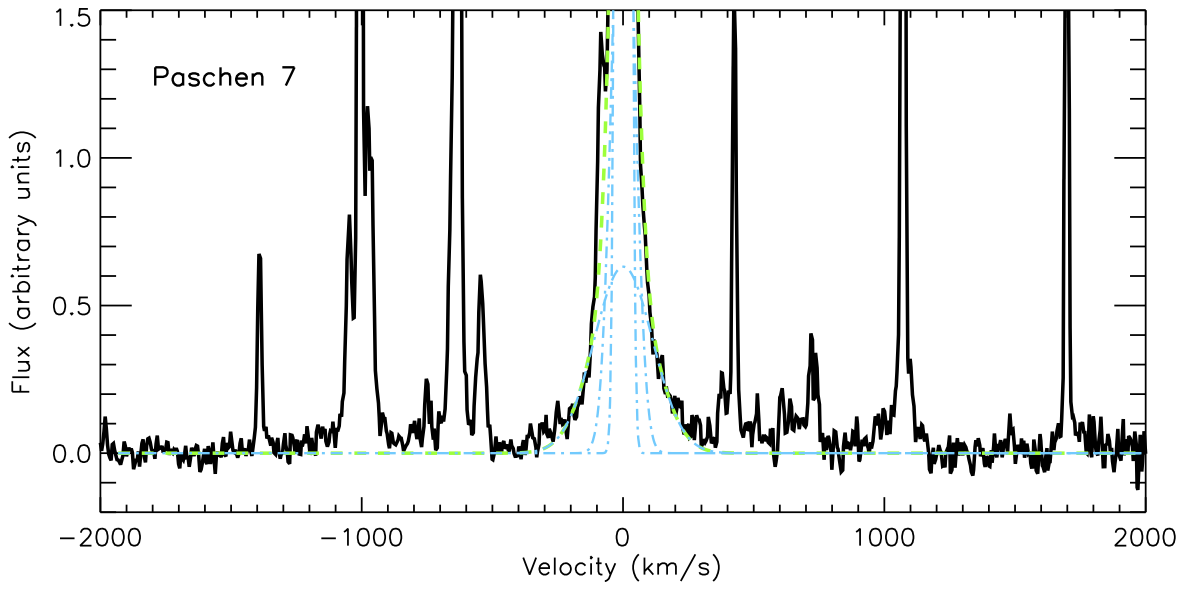}
  \includegraphics[width=8.7cm]{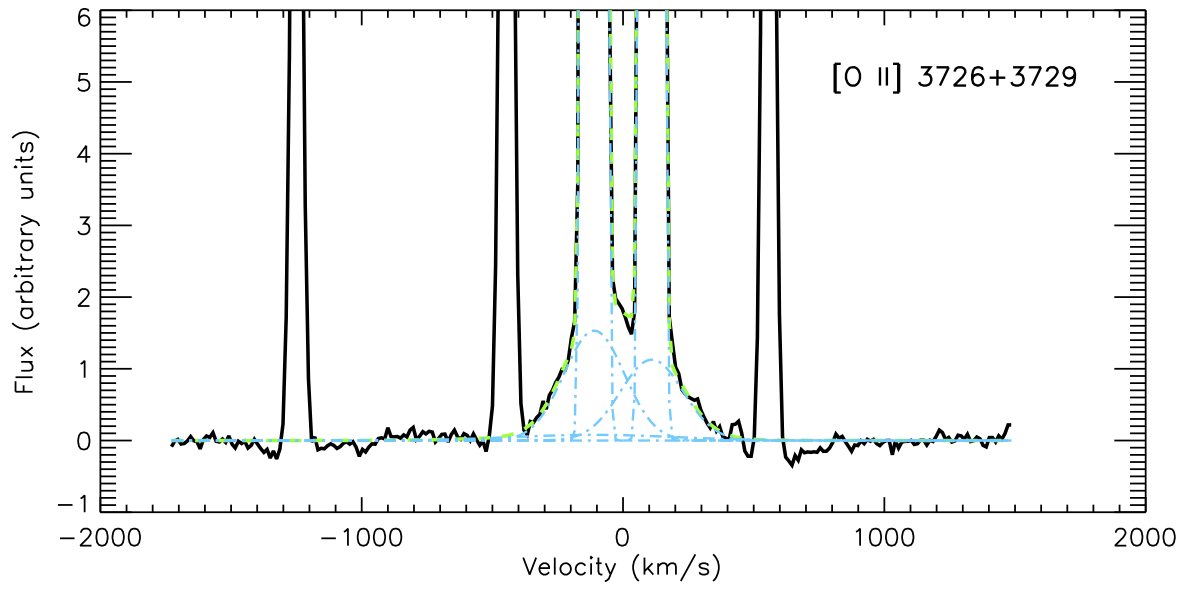}
  \caption{Strong lines with broad wings in the X-Shooter spectrum of Tc~1 (black), Gaussian fits to the individual components (blue dot-dashed line; see Table~\ref{components}) and the overall fit (sum of all components; green). The H$\alpha$ line is affected by saturation and Paschen~7 may be affected by instrumental effects. The broad wings in the remaining lines seem to be produced by a physical mechanism.}
  \label{vel_components}
\end{center}
\end{figure*}

Inspection of the integrated spectra reveals that several strong lines show a faint broad wing component, which may be seen in Fig.~\ref{vel_components}. These components are very faint compared to the total line intensity. We used an IDL routine based on {\sc herfit} to fit multiple Gaussian components to these lines. We also attempt to fit a combination of Gaussian and Lorentz profiles, which did not yield reasonable results.

To perform the fitting, we assumed that the central wavelength is the same for all components. This was helpful to enhance asymmetries in the components (e.g. H$\beta$ and [O~{\sc iii}] 5\,007). For the blends in the H$\alpha$/[N~{\sc ii}] and [O~{\sc ii}] lines, we determined the central wavelength by eye; in the remaining cases, the wavelength was a free parameter of the fit. We kept the number of components to the lowest possible while still ensuring a good fit.

\begin{table}
  \centering
  \caption{Multi-component fit parameters for broad-wing lines.}
  \label{components}
  \begin{tabular}{lrrrr}
  \hline
Line  	&	$\lambda$ & $\sigma_{\rm obs}^{(a)}$	&	$\sigma_{\rm int}^{(b)}$	&	$I/I_{\rm tot}$\\
        & [\AA] & [km~s$^{-1}$] &  [km~s$^{-1}$] & [\%] \\ 
  \hline  
~H$\alpha$      & 6563   &  19  &   13 & 99.26 \\
               &        &  78  &  77  &  0.34 \\
               &        & 314  & 314  &  0.40 \\
~H$\beta$       & 4861   &  22  &  17  & 98.77 \\
               &        &  59  &  57  &  0.82 \\
               &        & 445  & 445  &  0.41 \\
~Pa 7           & 10049  &  19  &  13  & 72.28 \\
               &        &  48  &  46  & 15.97 \\
               &        & 132  & 131  & 11.75 \\
~[O~{\sc iii}] & 4959   &  15  &  --  & 98.71 \\
               &        &  41  &  37  &  0.49 \\
               &        & 185  & 184  &  0.80 \\
~[O~{\sc iii}] & 5007   &  15  &  --  & 98.73 \\
               &        &  54  &  51  &  0.41 \\
               &        & 284  & 284  &  0.85 \\
~[N~{\sc ii}]  & 6548   &  16  &  13  & 99.19 \\
               &        & 116  & 116  &  0.29 \\
               &        & 322  & 322  &  0.52 \\
~[N~{\sc ii}]  & 6583	&  17  &  14  & 99.32 \\
               &        &  77  &  76  &  0.27 \\
               &        & 337  & 337  &  0.41 \\
~[O~{\sc ii}]  & 3727	&  20  &  11  & 99.26 \\
               &        & 136  & 135  &  0.65 \\
               &        & 333  & 333  &  0.09 \\
~[O~{\sc ii}]  & 3730	&  20  &  11  & 99.23 \\
               &        & 141  & 140  &  0.77 \\
               \hline
\multicolumn{5}{l}{$^{(a)}$ Line width obtained from Gaussian fit.}\\
\multicolumn{5}{l}{$^{(b)}$ See Appendix \ref{Ap_intrinsic_widt} for the intrinsic line}\\
\multicolumn{5}{l}{width calculation.}\\

\end{tabular}
\end{table}

The velocities of the resulting components are listed in Table \ref{components}, as well as the fraction of the total line flux that is emitted in each component. Most of the lines were well fitted only by three velocity components, with the exception of [O~{\sc ii}]~3730~\AA. In the case of [O~{\sc ii}]~3730~\AA\ lines, only two components can be well distinguished by the fitting code. This is likely due to the strong blend of the high-velocity components of [O~{\sc ii}]~3730~\AA\ and the nearby [O~{\sc ii}]~3727~\AA\ line.

Table~\ref{components} shows recurring components in different lines. The main velocity component for all lines is in the range 15--22 km~s$^{-1}$, corresponding to the expansion velocities found in the previous sections for the double-peaked lines. Clearly, this indicates that the bulk of the overall emission produced in the nebula is caused by this expanding outflow. However, all strong lines show the clear presence of high-velocity components as well, with velocities up to a few hundred km~s$^{-1}$. While they only represent a generally small contribution to the total line flux, they show up in recombination and forbidden lines alike, are seen in high-, medium- and low-ionization regions (i.e. are distributed all over the nebula). The Pa~7 line is an exception, showing up to 28 per cent of the total line flux in the secondary components. The telluric lines close to Pa~7 also show a large base. In the case of Pa~7, the effect may then be instrumental. In addition, there is a blended telluric line in the left wing of Pa~7 that affects the fitting results. Therefore, the results for Pa~7 in Table~\ref{components} should be taken with caution.

H$\alpha$ shows broad wings with fitting parameters similar to H$\beta$. However we remind that the H$\alpha$ is saturated, which may affect its profile. The fitting results for H$\alpha$ in Table~\ref{components} are only included for completeness as they are need to reproduce the fitting of the [N~{\sc ii}] neighbour lines.

High-velocity components are not unusual in PNe. They have been reported in several PNe, e.g. \citet{1989AJ.....97..476B}, \citet{2000ApJ...530L..49L}, and \citet{2003ApJS..147...97A}.
Such authors explored a few mechanisms responsible for the broad wings seen in H$\alpha$ emission in PNe (e.g. stellar winds and Rayleigh-Raman scattering).

A more detailed analysis of the Tc~1 broad wings component and the mechanism behind them will be done in a future paper.

\section{Morphology} \label{sect_morphology}

In Section~\ref{sect_2peak}, we presented the detection of the blue- and red-shifted components in the [N~{\sc ii}] and [S~{\sc ii}] emission lines at 6548~\AA, 6583~\AA, 6717~\AA, and 6731~\AA. Here, we explore which morphologies could lead to these observed profiles. We focus on the shapes of the two [S~{\sc ii}] lines in this analysis, as the blue- and red-shifted components are better defined in these lines.

Assuming that the shape of the double-peaked lines can be explained by components from the projection effect of the three-dimensional morphology of the source, we used the computational tool {\sc Shape} \citep{2006RMxAA..42...99S,2011ITVCG..17..454S} to model a number of different morphologies and investigate which ones could reproduce the observed shapes of the lines. In this analysis we aimed to reproduce not only the variation of the intensity of the blue- and red-shifted components, but also the depth of the valley between the two components (see Fig.~\ref{sii}). 

In optical [S~{\sc ii}] emission Tc~1 seems to be spheroidal. Therefore, a simple model for the nebula can be constructed starting from a spherical structure with a constant density distribution and a homologous expansion law. This simple model reproduces the main characteristics of the Tc~1 position-velocity (PV) diagram, i.e. the blue and the red-shifted components \citep[see also][]{2012MNRAS.425.2197A}. However it does not reproduce the intensity difference between the two components. The next step was to insert and explore a more complex density profile such as an equatorial density enhancement (a maximum density at equator, decreasing density polewards, with a minimum at the poles). This model can adequately reproduce the observed line profile. However, in order to be consistent with the expansion velocities of the two components and the depth of the valley between the peaks, an inclination angle should be also applied to the model. This inclination angle is found to be 5-7 degrees with respect to the line of sight.

Usually, PNe with equatorial density enhancements have an axisymmetric structure rather than a spherical one. We therefore ran a number of models assuming an elliptical structure with the major axis in the line-of-sight direction, and varying the coefficient ($K$) of the expansion law \citep[e.g.][]{2015A&A...582A..60C} until a good match with the observed line profiles was obtained. But it was not possible to constrain the model with the data we have. Generally, the size of a nebula in the line-of-sight direction cannot be constrained if it is seen nearly pole-on \citep[low inclination angle; see][]{2012MNRAS.425.2197A}. In Fig. \ref{sii}, we compare the observed and modelled lines profiles in several positions along slit from the north-east to south-west direction.

\begin{figure}
\begin{center}
  \includegraphics[width=7.5cm,trim=0.5cm 0.cm 1.4cm 0cm,clip]{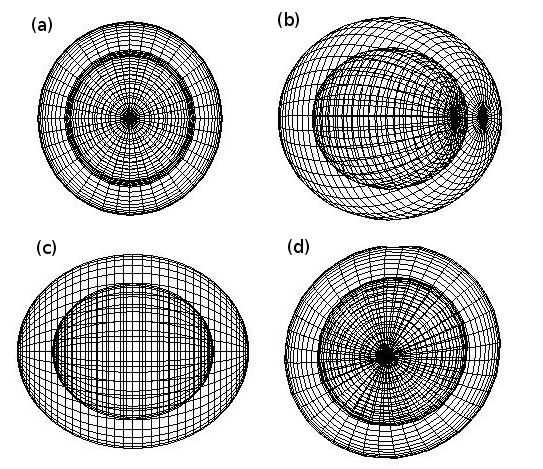}
  \caption{3D mesh of the morphology that best reproduces our Tc~1 [S~{\sc ii}] observations. Panels (a) to (d) shows our SHAPE model from different angles. In (a) the ellipsoid is seen pole on; in (b) and (c) the ellipsoid is turned to the right by 50 and 90 degrees angles, respectively; (d) is the observer's view, i.e., the ellipsoid originally pole-on (a) is turned down by 6 degrees from the light of sight and clockwise 50 degrees.}
\label{Tc1_Shape}
\end{center}
\end{figure}

Finally, from the models we tested, we could eliminate an open-end hour-glass-like morphology. In conclusion, Tc~1 very likely has an axisymmetric/elliptical structure seen almost pole on, with an inclination of 5-7 degrees and a position angle of approximately -50 degrees. The exact size of Tc~1 along the major axis cannot be constrained with the current information. We also find evidence of an equatorial density enhancement. The [S~{\sc ii}] emitting region seems to be a factor of $\sim$3-8 denser in the equator than in the rest of the nebula. The 3D mesh of the model that best reproduces the observed data is presented in Fig.~\ref{Tc1_Shape}.

\begin{figure}
\begin{center}  
  \includegraphics[width=8cm]{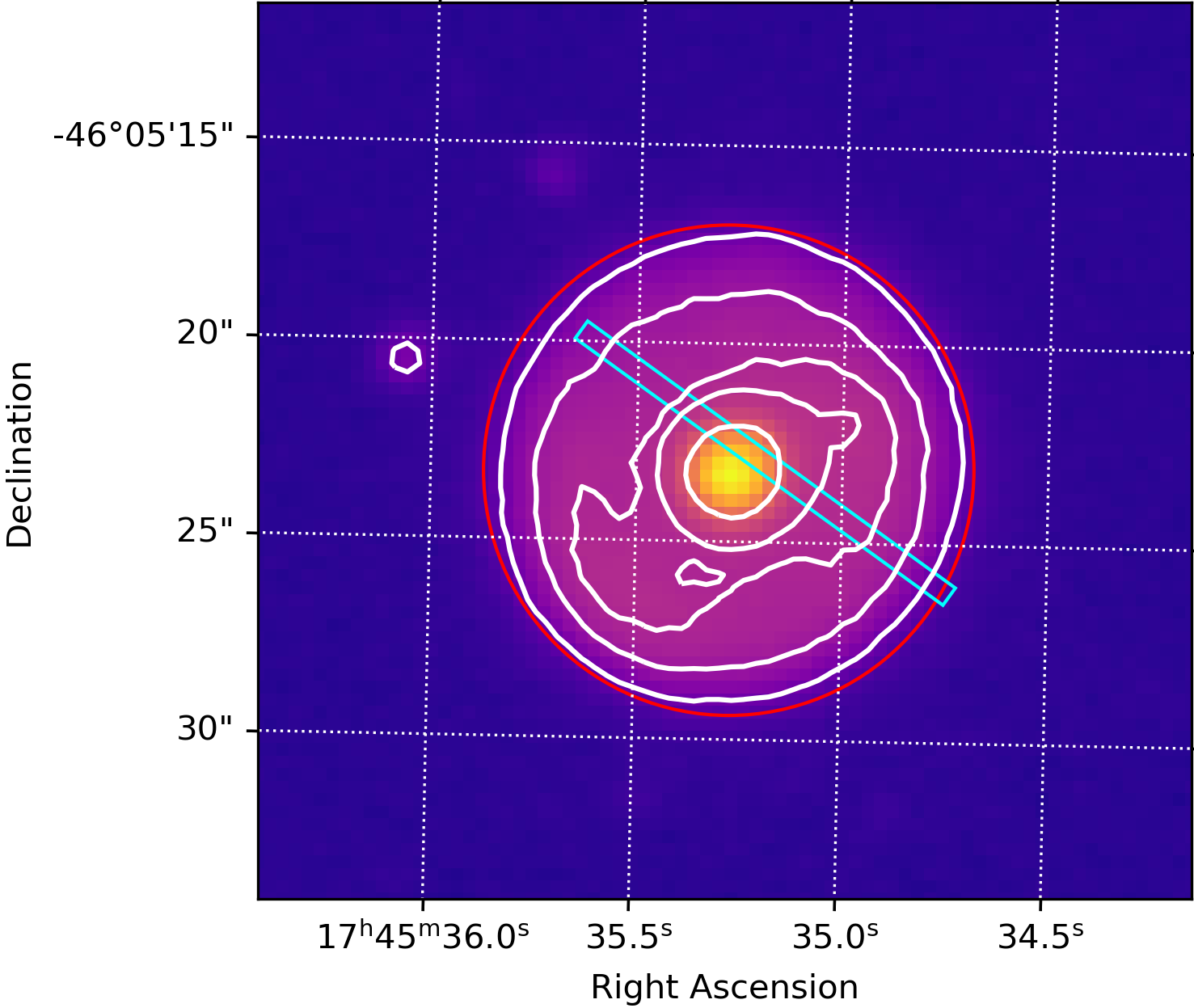}
  \caption{EFOSC2 R band image of Tc~1. The flux is in logarithmic scale. White contours correspond to 1.0, 2.0, 2.8, and 3.3 per cent of the peak flux. Around 96 per cent of the flux is emitted within the central contour. The Tc~1 projected morphology is only slightly deviated from a perfect circle (see the red circle for comparison). The blue rectangle indicates the position of the X-Shooter slit.}
  \label{Tc1_Contours}
\end{center}
\end{figure}

It should be noted that the elliptical model produces a slightly elliptical 2D structure projected into the plane of the sky due to the small inclination angle of 5--7 degrees. By scrutinising the observed optical and IR images from EFOSC2 (Fig.~\ref{Tc1_Contours}), we find that Tc~1 indeed displays a slightly elliptical structure with the major axis rotated by -50/-55 degrees rather than a spherical one, which is consistent with our model. Such slight asymmetry was also noted by \citet{2003A&A...405..627T}, who provide the size of the Tc~1 main shell of 12.9$\times$12.2~arcsec.

Figure~\ref{Tc1_Contours} also displays asymmetries in the NW and SE directions (PA$\sim$140~degrees). The overall structure of Tc~1 strongly resembles that of NGC~2392 \citep{2012ApJ...761..172G}, which also has similar asymmetries seen in [N~{\sc ii}]. The main difference between the two nebulae is that NGC~2392 shows a spherical rather than an elliptical morphology. The pole-on view of the PN NGC~7009 from the {\sc Shape} model of \citet{2009RMxAA..45..143S} also displays some asymmetries like NGC~2392, and its structure is very similar to our elliptical model of Tc~1 -- it is known that NGC 7009 is an ellipsoidal nebula \citep{1998AJ....116..360B}. These asymmetries may be associated with small-scale structures, bright in [N~{\sc ii}] or [S~{\sc ii} emission lines \citep[e.g. LISs,][]{2001ApJ...547..302G,2016MNRAS.455..930A}. The presence of small-scale structures, e.g., clumps or bipolar features, with different degree of ionization has also been proposed for Tc~1 by \cite{2008ApJ...677.1100W}. Narrow-band optical images with better spatial resolution or integral field spectroscopic data are needed to investigate these asymmetries as well as the distribution of line-emission in both spatial directions.

\section{A New Photoionization Model for Tc~1} \label{sect_models}

\citet{Pottasch_etal_2011} and \citet{2014MNRAS.437.2577O} have published photoionization models for Tc~1. Both works use a similar data set as a basis to obtain their {\sc cloudy} photoionization models. Their approaches were also similar, calculating a spherical shell model whose emission best matches the observed fluxes. Not surprisingly, therefore, they obtain similar results (see Table~\ref{Tab_Tc1_Literature}). \citet{Pottasch_etal_2011} discuss details of their Tc~1 model and its limitations in explaining some of the observed line fluxes. In particular, they report a difficulty in reproducing the oxygen lines.

Here we present a new photoionization model based on the high-quality VLT X-Shooter observations, as well as previous observations \citep{1994MNRAS.271..257K,2008ApJ...677.1100W,Pottasch_etal_2011}. An improvement over the previous models is that we consider the slit apertures when comparing our model to the observations. The slit position may have a significant impact on the line ratios of spatially-resolved nebulae, as different ions emit in different regions \citep{Fernandes_etal_2005}.

As shown in Sect.~\ref{sect_morphology}, Tc~1 is an ellipsoid. Although not exactly determined, its elongation seems to be small. Since Tc~1 is roughly spherical, simulations using a one-dimensional photoionization code should produce good results. We used two photoionization codes, {\sc Aangaba} \citep{1992ApJS...78..153G,2012A&A...541A.112K} and {\sc cloudy} \citep[version 17.03; ][]{2017RMxAA..53..385F}, to find the best model for Tc~1.

As the fluxes were measured using only extractions of the slit, we also simulated such slit extractions when determining the model fluxes. To this end, we used the Python library {\sc pycloudy} \citep{2013ascl.soft04020M}. This library includes a pseudo-3D module that allows the simulation of the three-dimensional structure of any given nebular morphology from a set of one-dimensional models. The effect of slit apertures of any shape in the calculated emission can then be simulated by doing a pseudo-3D model and integrating the line emissivities along the line of sight. This produces a 2D image for each emission line, on which we apply a mask of the slit corresponding to the observation. The {\sc pycloudy} modules have also been adapted to work with {\sc Aangaba} model outputs.

To simulate the extraction of the X-Shooter slit over the simulated nebula it is necessary to assume a distance to Tc~1. As the distance to Tc~1 is still uncertain, we explored a range of distances found in the literature (see Appendix~\ref{Ap_param_lit}). Our modelling efforts revealed that distances in the range $\sim$2-3~kpc provide better results.

The inner and outer radii of the nebula can be derived from the angular sizes of the nebula once a distance is defined. The inner radius is poorly known, but it has no strong effect on the model provided it is small compared to the outer radius. The outer radius of the main shell is 6~arcsec \citep{2008ApJ...677.1100W}.

For our models, we assumed that the Tc~1 central star emits as a blackbody. We also tried the WMBasic atmosphere model \citep{Pauldrach_etal_2004} used by \citet{Pottasch_etal_2011} and a more realistic model with higher surface gravity and solar abundance, but no significant changes in the resulting parameters of `best' model was found.

Some of the characteristics of Tc~1 have been constrained in previous sections and in other works in the literature (see Table \ref{Tab_Tc1_Literature} and Appendix \ref{Ap_param_lit}). To obtain our photoionization model, we used these constraints to guide the range of model parameters we explored.

The Tc~1 low-excitation nebular spectrum indicates it has a low-temperature central star. According to previous determinations in the literature, the central star temperature $T_{eff}$ is in the range of 30\,000 to 35\,000~K (see Appendix~\ref{Ap_param_lit} for a compilation of values and references). Although we explored a much larger range of temperatures using the photoionization codes and searching the 3MdB photoionization model database \citep{2015RMxAA..51..103M}, values outside this range of $T_{eff}$ models cannot explain well the Tc~1 nebular emission. The best matches between observations and models indicate that the Tc~1 central star temperature must be in the range above.

We assume a spherical matter distribution, with the density and elemental abundances constant across the nebula. We initially assumed the nebular abundances we determined for the elements in Table \ref{abund_reg} (except for Ne) and values found in the literature (Table \ref{Tab_Tc1_Literature}) otherwise. We gave preference for CEL abundances if determined. We use the He abundance from the classical method (Section \ref{sect_tot_abundances}), although the method provides only a lower limit as the ICF for this ion is poorly known. To obtain a good match to the observables, we had to vary the initial abundances. This will be discussed in more detail further in the text.

\begin{figure}
\begin{center}
\includegraphics[width=8cm,trim={0.3cm 0.6cm 0.2cm 0.2cm},clip]{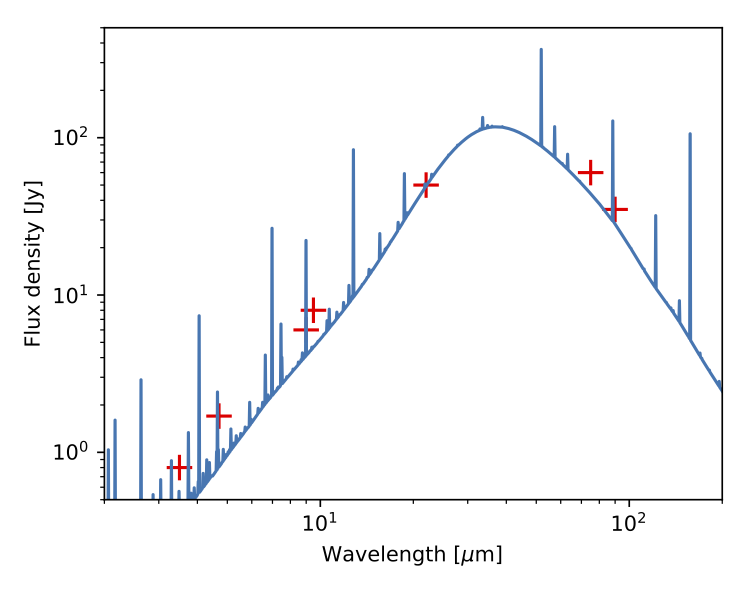} 
\caption{Infrared emission for the XS model (blue solid line) compared to the observations reported by \citet{2014MNRAS.437.2577O} (red crosses). The emission is dominated by the dust thermal emission.}
\label{fig:irsed}
\end{center}
\end{figure}

The dust model is defined by its composition, grain size distribution, and dust-to-gas ratio. As Tc~1 has C/O~=~1.4 and shows fullerene features (C-rich), we assumed the grains are composed of graphite. To define our dust model, we adjust the grain size range to reproduce the observed IR spectral distribution as given by \citet{2014MNRAS.437.2577O}. We use grain size distribution from 0.0005 to 1.5~$\mu$m, with the dust density following a power law of slope -3.5. We do not try here to justify this distribution or to explore a different one; we only require the energy budget to correctly take into account the presence of absorbing dust, which is done once a model fitting the IR emission is obtained. The resulting fit is shown in Fig.~\ref{fig:irsed}. 

The photoionization model for Tc~1 was determined by finding the best match to the observed emission properties. To quantify the quality of a model, we use the quality factor defined by \citet{2009A&A...507.1517M}:

\begin{equation}
\kappa(O) = \frac{\log(O_\mathrm{mod}) - \log(O_\mathrm{obs})}{\tau(O)},
\label{eq:QF}
\end{equation}

\noindent where $O_\mathrm{mod}$ and $O_\mathrm{obs}$ are the observed and modelled values of any observable and the acceptable tolerance in dex $\tau(O)$ for this observable is defined as $\tau(O) = log(1 + \frac{\Delta I}{I})$ for any line intensity I of uncertainty $\Delta I$. With this definition, values of $\kappa$(O) between -1 and 1 indicate a good fit for the line flux.

We define the relative uncertainties $\Delta I / I$ (used to derive the tolerance) to be 50, 30, and 20 per cent for lines lower than 10 per cent of H$\beta$, lower than H$\beta$ and higher than H$\beta$ respectively. We added 15 per cent to the UV and IR lines (observed by other instruments than LCO and X-Shooter). The values above take into account the observational uncertainties (including difficulties in the cross-calibration between to wavelength ranges, uncertainties in the slit positions, the effect of the seeing on the slit mask) as well as all the systematic effects that can cause deviations from the model for an observable: simplicity of the morphology, simplicity of the radial density law (constant in our case), uncertainties on the atomic data used to compute the model, etc.

In our search for the best model for Tc~1 we made use the 3MdB photoionization model database to browse a wide range of nebular and stellar parameters. As mentioned above, this helped us eliminate central stars with temperatures outside the range already published in the literature. To find a model that fit the observations, we run a grid of models with the codes {\sc aangaba} and {\sc pycloudy}, exploring the range of parameters found in the literature and in the previous sections.  To fine tune the model, we then use {\sc cloudy} and {\sc pycloudy}, to slightly change the parameters until we find a reasonable match to the observations. We use {\sc cloudy} since it allows to fit more observational features than {\sc aangaba}. We used {\sc aangaba} to obtain the krypton abundance, as this element is not included in {\sc cloudy} (see Sect.~\ref{kripton}).

The abundance of Si is not constrained by our models as no line observed for this element can be modelled with {\sc cloudy}. Also, apart from H and He lines, no effort has be done to fit the recombination lines, as the high value of the ADF cannot be reproduced by the classical model. This, however, would not cause a significant impact on the determination of other Tc~1 parameters.

\subsection{Best Model} \label{sec:bestmodel}

In the following, we discuss the best model we found to explain our X-Shooter integrated spectrum. We also imposed that the model should explain the UV and IR observations compiled by \citet{Pottasch_etal_2011}. We call this data set the XS model. For comparison, we also obtained a model to the full \citet{Pottasch_etal_2011} data set, which uses the optical data from \citet{2008ApJ...677.1100W} (obtained with the Las Campanas Observatory, LCO). The He~{\sc i} lines not reported by \citet{2008ApJ...677.1100W} were taken from \citet{1994MNRAS.271..257K} (data obtained with the Anglo Australian Telescope, AAT, at the Siding Spring Observatory). This model is hereafter referred as the LCO model. As we are using the same version of {\sc cloudy}, the same methodology and the same criteria for both data sets, this comparison can provide insights to the differences in the model caused by the optical data.

The parameters of the models that best explain each data set are given in Table~\ref{tab:models}. The resulting `best models' are very similar. The differences above can be seen as a measure of the uncertainties in the parameter determination and in the observations. Our values are also similar to values from the previous models by \citet{Pottasch_etal_2011} and \citet{2014MNRAS.437.2577O}. The parameters of those models are also given in Table~\ref{tab:models}.

\begin{figure}
\begin{center}
\includegraphics[width=8cm]{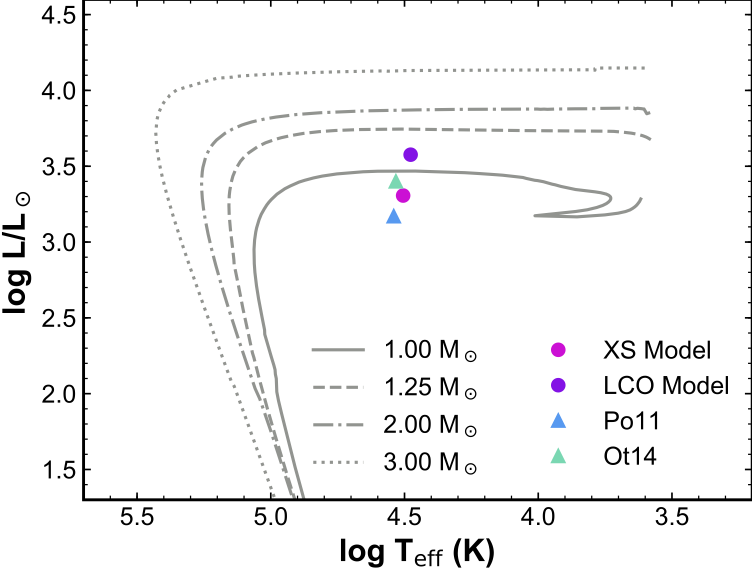}
\caption{Position of the Tc~1 central star in the H-R diagram. The curves are post-AGB evolutionary tracks for H-burning central stars with metalicity $Z$ = 0.01 obtained by \citet{2016A&A...588A..25M}.}
\label{hrdiagram}
\end{center}
\end{figure}

The LCO and XS models yield, respectively, $T_{eff}=$~30 and 32~kK and $L_\star=$~2000 and 3800~$L_\odot$ for the Tc~1 central star. Figure~\ref{hrdiagram} shows its position on the H-R diagram. Theoretical post-AGB evolutionary tracks \citep{2016A&A...588A..25M} were added for comparison. Despite the spread in the values of the luminosities, it seems clear that Tc~1 must have had a low-mass progenitor.

The gas density of the main shell can be very well constrained from the comparison of observed and simulated integrated line fluxes and spatial profiles. The best matches for the observations are found within the range $n_\mathrm{H}\sim$~1\,000-3\,000~cm$^{-3}$. Outside this range, the quality of the model fits to the observations deteriorates. A density of $n_\mathrm{H}=$~2\,500~cm$^{-3}$ provides good results for both the XS and LCO models.

To improve the match between models and observations, the initial abundances (empirical) had to be fine tuned. The differences between the abundances obtained with the XS and the LCO models, and and between those and literature values are within 0.4~dex. Such values are typical uncertainties. The largest difference between our models and the literature is seen for Mg. We determined the Mg abundance from the [Mg~{\sc i}]~4563~\AA\ line. However, no Mg line has been reported by \citet{Pottasch_etal_2011} or by the references therein, so it is not clear how they constrained their Mg abundance and hence the origin of the Mg discrepancy is not clear.

Table \ref{tab:models} includes the ICFs determined from the models. The calculation takes into account the slit position. The values for ICF(He$^{+}$) determined from the models should not be used, as the models do not fit well the He~{\sc i} lines (see discussion below). For C, N, O, S, and Ar, our modelled ICFs are similar to the corrections we used to obtain the abundances using the classical method. On the other hand, the correction we used for Ne seems to be an underestimate as we discussed before.

\begin{table*}
\centering
\caption{Tc~1 photoionization models results}
\label{tab:results}
\begin{tabular}{lcccccc}
\hline
& \multicolumn{3}{c}{LCO model} & \multicolumn{3}{c}{XS model} \\
\hline
Line & Observation & Model & $\kappa(O)$ & Observation & Model & $\kappa(O)$ \\
\hline
& \multicolumn{6}{c}{Whole Nebula Absolut Flux (in 10$^{-11}$ erg~s$^{-1}$cm$^{-2}$)$^{(a)}$}\\
~H$\beta$ 4861~\AA & 5.10 & 5.19 & --- & 5.10 & 5.10 & --- \\  
\hline
& \multicolumn{3}{c}{LCO Spectrum:} & \multicolumn{3}{c}{VLT/X-Shooter Spectrum:} \\
~H$\beta$ 4861~\AA & 100.0 & 100.0 &  --- &100.0 & 100.0 &  --- \\ 
~H$\gamma$ 4341~\AA & --- &  47.29 &    --- &46.66 &  47.12 &   0.05 \\ 
~He~{\sc i} 5876~\AA & --- &  12.17 &   --- &15.26 &   8.36 &  -3.30 \\ 
~He~{\sc i} 4471~\AA & --- &   4.05 &   --- &5.39 &   2.84 &  -2.45 \\ 
~[N~{\sc i}] 5199~\AA & 0.04 &   0.03 &  -0.75 &0.02 &   0.10 &   4.51 \\ 
~[N~{\sc ii}] 5755~\AA & 1.09 &   1.18 &   0.31 &1.20 &   1.02 &  -0.61 \\ 
~[N~{\sc ii}] 6583~\AA & 95.4 & 100.9 &   0.31 &111.3 & 108.8 &  -0.13 \\ 
~[O~{\sc i}] 6300~\AA & 0.12 &   0.07 &  -1.38 &0.09 &   0.70 &   5.16 \\ 
~[O~{\sc ii}] 3729~\AA & 86.0 &  84.9 &  -0.07 &148.9 & 137.8 &  -0.42 \\ 
~[O~{\sc ii}] 3726~\AA & 130.0 & 140.4 &   0.42 &234.1 & 228.2 &  -0.14 \\ 
~[O~{\sc ii}] 7323~\AA & 5.43 &   5.80 &   0.25 &6.11 &   7.66 &   0.86 \\ 
~[O~{\sc ii}] 7332~\AA & 4.57 &   4.75 &   0.15 &5.10 &   6.27 &   0.79 \\ 
~[O~{\sc iii}] 5007~\AA & 124.0 & 127.7 &   0.16 &109.1 & 110.8 &   0.08 \\ 
~[O~{\sc iii}] 4363~\AA & 0.55 &   0.67 &   0.47 &0.46 &   0.36 &  -0.64 \\ 
~[Ne~{\sc iii}] 3869~\AA & --- &   1.52 &    --- &0.60 &   0.59 &  -0.04 \\ 
~[Ne~{\sc iii}] 3967~\AA & --- &   0.46 &    --- &0.17 &   0.18 &   0.05 \\ 
~[Cl~{\sc iii}] 5518~\AA & 0.28 &   0.35 &   0.52 &0.28 &   0.29 &   0.10 \\ 
~[Cl~{\sc iii}] 5538~\AA & 0.30 &   0.40 &   0.69 &0.32 &   0.33 &   0.09 \\ 
~[S~{\sc ii}] 6731~\AA & 3.50 &   3.38 &  -0.14 &3.53 &   3.59 &   0.07 \\ 
~[S~{\sc ii}] 6716~\AA & 2.20 &   2.35 &   0.26 &2.31 &   2.51 &   0.32 \\ 
~[S~{\sc ii}] 4076~\AA & 0.18 &   0.21 &   0.36 &0.28 &   0.20 &  -0.85 \\ 
~[S~{\sc ii}] 4069~\AA & 0.62 &   0.67 &   0.18 &0.55 &   0.64 &   0.36 \\ 
~[S~{\sc iii}] 6312~\AA & 0.46 &   0.49 &   0.14 &0.57 &   0.18 &  -2.89 \\ 
~[S~{\sc iii}] 9069~\AA & 12.51 &   9.77 &  -1.36 &--- &   4.77 &    --- \\ 
~[Ar~{\sc iii}] 5192~\AA & 0.03 &   0.04 &   0.66 &0.03 &   0.04 &   0.43 \\ 
~[Ar~{\sc iii}] 7136~\AA & 5.65 &   6.30 &   0.42 &8.29 &   8.08 &  -0.10 \\ 
~[Ar~{\sc iii}] 7751~\AA & 1.66 &   1.50 &  -0.40 &1.96 &   1.92 &  -0.08 \\ 
~[Mg~{\sc i}] 4563~\AA & --- &   0.03 &    --- &0.06 &   0.06 &   0.07 \\ 
~[Fe~{\sc iii}] 4658~\AA & --- &   0.29 &    --- &0.27 &   0.22 &  -0.50 \\ 
~[Fe~{\sc iii}] 5270~\AA & --- &   0.24 &    --- &0.13 &   0.19 &   0.84 \\ 
~[Fe~{\sc iii}] 4702~\AA & --- &   0.10 &    --- &0.08 &   0.08 &  -0.13 \\ 
\hline 
& \multicolumn{6}{c}{IUE:} \\

~[C~{\sc ii}] 2326~\AA & 45.00 &  63.56 &   1.15 &45.00 &  69.46 &   1.45 \\ 
~[C~{\sc iii}] 1909~\AA & 27.00 &  35.68 &   0.93 &27.00 &  15.23 &  -1.91 \\ 
\hline 
& \multicolumn{6}{c}{ISO:}\\

~H~{\sc i} 12.37~$\mu$m & 0.92 &   0.98 &   0.12 &0.92 & 1.02 &   0.20 \\ 
~[Ne~{\sc ii}] 12.81~$\mu$m & 37.50 &  25.97 &  -1.22 & 37.50 &  25.02 &  -1.35 \\ 
~[Ne~{\sc iii}] 15.55~$\mu$m & 1.46 &   2.57 &   1.52 & 1.46 &   2.49 &   1.44 \\ 
~[S~{\sc iii}] 18.71~$\mu$m & 14.00 &  11.79 &  -0.57 & 14.00 &   7.12 &  -2.25 \\ 
~[S~{\sc iii}] 33.47~$\mu$m & 6.21 &   4.81 &  -0.69 & 6.21 &   2.89 &  -2.06 \\ 
~[S~{\sc iv}] 10.51~$\mu$m & 0.47 &   0.46 &  -0.05 &0.47 &   0.33 &  -0.71 \\ 
~H~{\sc i} 6.94~$\mu$m & 32.80$^{(b)}$ &   0.02 & -24.78 &32.80$^{(b)}$ &   0.02 & -24.67 \\ 
~[Ar~{\sc ii}] 6.98~$\mu$m & 32.80$^{(b)}$ &   4.93 &  -6.31 &32.80$^{(b)}$ &  14.06 &  -2.82 \\ 
~H~{\sc i} 7.09~$\mu$m & 32.80$^{(b)}$ &   0.02 & -24.26 &32.80$^{(b)}$ &   0.02 & -24.14 \\ 
~[Ar~{\sc iii}] 8.99~$\mu$m & 6.50 &   4.79 &  -0.82 &6.50 &   9.66 &   1.07 \\ 
~[Ar~{\sc iii}] 21.83~$\mu$m & 0.43 &   0.30 &  -0.70 &0.43 &   0.61 &   0.68 \\ 
\hline 
\multicolumn{7}{c}{$^{(a)}$The H$\beta$ absolute fluxes indicated here are integrated in the whole nebula. The observed}\\
\multicolumn{7}{c}{value is derived from radio observations by \citet{Pottasch_etal_2011}. $^{(b)}$Blended lines.}\\
\end{tabular}

\end{table*} 


\begin{figure*}
\begin{center}
\includegraphics[width=16.cm]{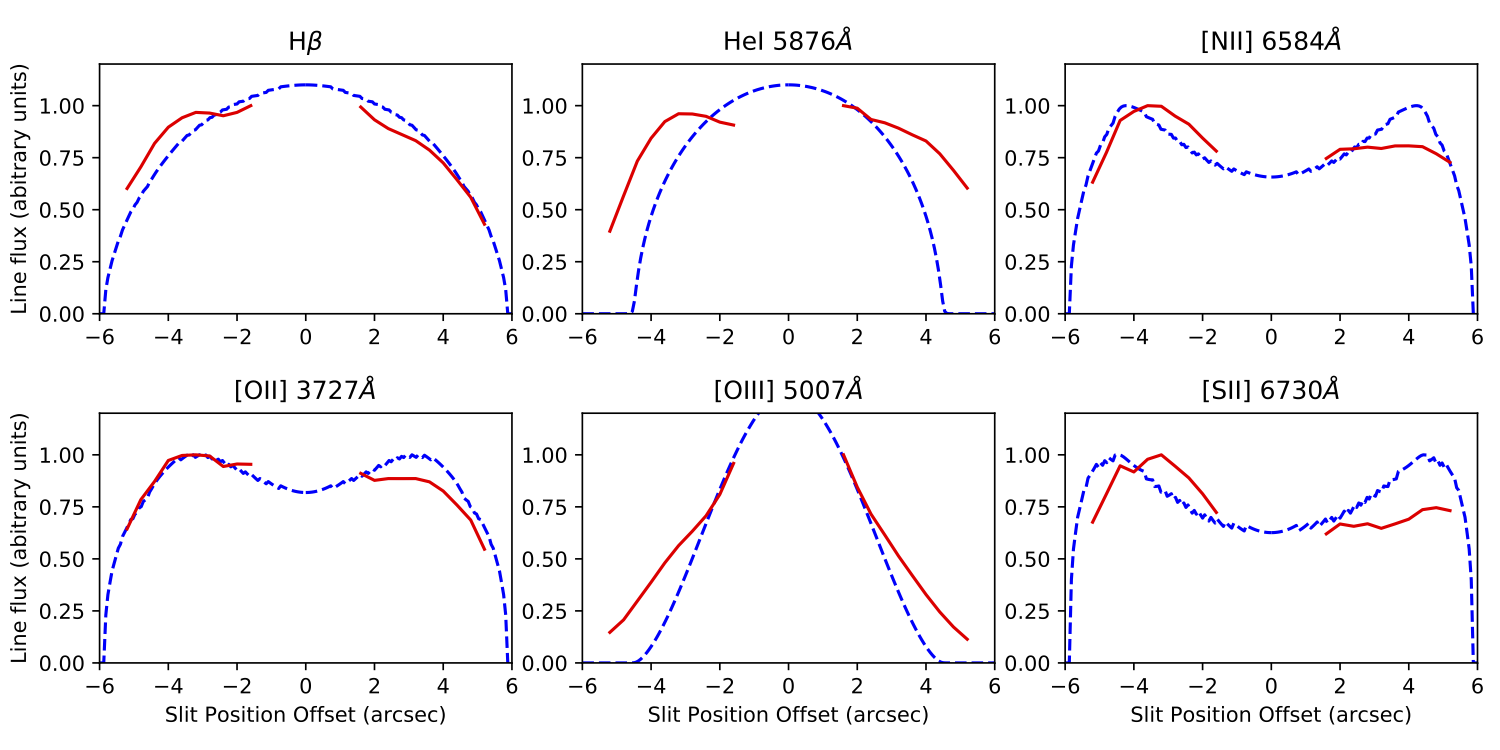}
\caption{Comparison of modelled (dashed) and observed (solid) line spatial profiles. The observed curves are normalised to the peak value. Dashed curves are the corresponding model curve, which were scaled to match the normalised observed profiles.}
\label{profiles_obs_model}
\end{center}
\end{figure*}

The observed and predicted line fluxes for both data sets, as well as the quality factor defined by Eq.~\ref{eq:QF} are presented in Table \ref{tab:results}. Differences in the fluxes of X-Shooter and LCO observations can be explained by the effect of different apertures. In comparison to the X-Shooter observations, LCO data indicates a higher ionization stage of the nebula: He~{\sc i}~5876~\AA/H$\beta$ changes from 0.09 to 0.15 between AAT and X-Shooter observations, while [O~{\sc ii}]~3727~\AA/[O~{\sc iii}]~5007~\AA changes from 1.7 to 3.5 between LCO and X-Shooter.

Most of the lines are well fitted by our models. Besides the line fluxes, our best model also reproduces the observed spatial profiles. The comparison between observed and modelled spatial profiles is shown in Fig.~\ref{profiles_obs_model}. The profiles of most of the lines, in particular H$\beta$ and [O~{\sc ii}] 3726~\AA\, are well reproduced. The observed He line profile shows a larger nebula (with a size similar to the size shown by the H$\beta$ line) than the simulated profile.

The total absolute H$\beta$ flux of Tc~1 is very well reproduced by both XS and LCO models. The total H$\beta$ fluxes for these model are $5.10 \times 10^{-11}$ and $5.19 \times 10^{-11}$~erg~cm$^{-2}$~s$^{-1}$, respectively. Such values are a very close match to the reddening corrected fluxes provided by \citep{1999A&A...352..297S} and \citep{Pottasch_etal_2011}, $4.1 \times 10^{-11}$~erg~cm$^{-2}$~s$^{-1}$ and $5.1 \times 10^{-11}$~erg~cm$^{-2}$~s$^{-1}$, respectively. Those authors determined the H$\beta$ fluxes from radio observations, using the relation between these two fluxes.

\citet{Pottasch_etal_2011} mention a difficulty in modelling the [O~{\sc iii}] lines. Our models fit well the [O~{\sc iii}] and [O~{\sc ii}] lines, but at the expense of fitting the He~{\sc i} lines; in the case of the new X-Shooter observations, they are underestimated by a factor of $\sim$2. From the models we studied, we observed that there is no parameter combination that can explain the O and He lines simultaneously. While the [O~{\sc ii}]/[O~{\sc iii}] line ratio points to a star with $T_{eff}$ in the 30-32~kK range, the He lines indicate that the star temperature should be a little higher (temperature closer to 40~kK, yielding to somewhat more ionized nebula). The integrated fluxes of the He lines can be better fitted if we increase the He abundances significantly, however this does not improve the match to their spatial profiles. Unfortunately, \citet{Pottasch_etal_2011} did not discuss how their model fitted the He lines.

The presence of high-density clumps, as we inferred from the hydrogen recombination lines in Section~\ref{sect_Hdecrement}, could perhaps be a factor in explaining the different ionization  between the 2 sets of observations.

Another possible solution is to consider matter bounded models. Matter bounded models can, in fact, better reproduce the ionization structure of Tc~1. They provide a better fit to the integrated line fluxes and to the bright line spatial profiles. The Tc~1 line intensities are better reproduced by models with $T_\mathrm{eff} <$ 35\,000~K. However, the sizes of the H~{\sc i} and He~{\sc i} emitting regions are practically equal, which, for such star temperatures, indicates the nebula (main shell) should be matter bounded \citep{2006agna.book.....O}. According to our model, the size of the main shell, however, should not be much different from the Str\"{o}mgren sphere.

A matter bounded main shell is also supported by the presence of the previously mentioned Tc~1 extended halo, which is a remnant from the progenitor AGB mass loss \citep{2003MNRAS.340..417C}. The faint halo is seen in deep image of H recombination line emission \citep{1992A&AS...96...23S,2003MNRAS.340..417C}. The halo has a very low surface brightness and should be composed of a very low-density gas. Its ionization may be due to ionizing photons escaping from the main shell. For the XS model, assuming an ionization bounded halo, we estimate that its density should be around 240~cm$^{-3}$. This value may be considered an upper limit, as for higher densities the halo would have to be smaller than observed. In case of a less dense halo, the nebula would then be matter bounded.

[Ar~{\sc ii}] 7~$\mu$m is strongly underpredicted by the model. This may be caused by blended lines. The IR observations reported by \citet{Pottasch_etal_2011} have not been corrected from contamination by H~{\sc i} lines and a C$_{60}$ feature, as pointed out by \citet{2019MNRAS.482.2354O}. The H~{\sc i} lines close to [Ar~{\sc ii}] are the transitions 19-8 and 20-8 at 7.09 and 6.95~$\mu$m respectively, and are predicted by our model to be less than one per cent of the [Ar~{\sc ii}] 7~$\mu$m intensity. Unfortunately, we can not estimate the C$_{60}$ contribution to the [Ar~{\sc ii}] 7~$\mu$m line with the current photoionization models.

\section{Detection and Abundance of Krypton} \label{kripton}

We detected the [Kr~{\sc iii}]~6827~\AA~ line in the Tc~1 X-Shooter spectrum. This is the first detection of Kr in this nebula. [Kr~{\sc iv}] lines within the X-Shooter spectral range were not detected, which is natural, as such lines are associated with high-excitation PNe. The [Kr~{\sc iii}] ion should dominate the emission in low excitation PNe as Tc~1, while [Kr~{\sc iv}] lines are produced very close to the centre of the nebula and therefore would not be detected in our slit extraction. 

We determine the krypton abundance using the photoionization code {\sc Aangaba}. Assuming an abundance of log(Kr/H) = -7.9 and adopting the XS model parameters, we match the observed flux of the [Kr~{\sc iii}]~6827~\AA~line. Such an abundance is in excess of 0.8~dex from the solar abundance \citep{2005ASPC..336...25A}. According to \citet{2008ApJS..174..158S}, Kr enrichment in excess of 0.3~dex relative to the solar abundance as found in Tc~1 can be produced by s-process nucleosynthesis in the progenitor star. 

All the PNe with Kr detections reported by \citet{2008ApJS..174..158S} are C-rich or mixed chemistry objects. There is a correlation between the abundance ratios Kr/O and the C/O ratio \citep[][and references therein]{2008ApJS..174..158S,2015MNRAS.452.2606G}. This correlation is expected, as during the third dredge up carbon is transported to the surface of the AGB stars together with the s-process elements. Using the Eq.~2 from \citet{2015MNRAS.452.2606G}\footnote{For this calculation we use the solar abundances from \citet{2005ASPC..336...25A}.}, the C/O ratio corresponding to the Tc~1 Kr/O value is C/O~=~1.9. The uncertainty in this value should be around 20-40\% \citep{2015MNRAS.452.2606G}, which makes the C/O~=~1.4 value from our best model consistent with the C/O determined from the Kr to C/O correlation.

\section{Summary \& Final Discussion}
\label{sect_conclusions}

Spatially resolved (11~arcsec slit) VLT/X-Shooter spectra at moderate spectral resolution ($R \sim$~9\,100--17\,000) covering a broad spectral range (3\,200-10\,100~\AA) are presented for the planetary nebula Tc~1. The spectra show a very rich set of atomic lines. From the spectra we derive physical conditions, ionization structure, elemental abundances, morphologies and central star characteristics for Tc~1.

\begin{enumerate}

\item The foreground$+$intrinsic visual extinction, $A_V$, associated with Tc~1 is about 0.96~mag. The extinction varies across the nebula between 0.36 and 1.31~mag.

\item Electron densities are derived from forbidden line diagnostics and the Balmer and Paschen decrement. While the forbidden line diagnostics provide $n_\mathrm{e} \sim 2\,000$~cm$^{-3}$, typical for ionized nebulae, the Balmer and Paschen lines indicate that much higher densities are present $n_\mathrm{e} > 10^5$~cm$^{-3}$. Such discrepancy could be explained by the presence of high-density clumps embedded in the ionized gas such as cometary knots, but this needs to be further investigated.

\item Elemental abundances are derived from colisionally excited lines (CELs) for N, O, Ne, Ar, and S, and from optical recombination lines (ORLs) for He, C, N, and O. The abundance discrepancy factors (ADF) for O and N are 2.2 and 9.4 respectively.

\item The clear double peaked structure (and variations thereof) seen for the lines of [N~{\sc ii}], [S~{\sc ii}], and [Cl~{\sc ii}] is consistent with a homologously expanding nebula. The lack of double peak profile for other emission lines is fully consistent with either lower expected expansion velocities (i.e. originating closer to the central star) or insufficient spectral resolving power at that wavelength.

\item A 3-D morpho-kinematical model was constructed to reproduce the [S~{\sc ii}] emission characteristics. The model shows that Tc~1 is a slightly elongated spheroid with an equatorial density enhancement seen almost pole-on (5--7 degrees inclination with respect to the line-of-sight).

\item In addition to the low-velocity (15--22~km\,s$^{-1}$) expanding gas there is also evidence for a faster `wind' with velocities up to several hundreds of km\,s$^{-1}$.

\item Based on detailed photoionization models for the VLT/X-Shooter observations, as well as observations available in the literature, we determined that the central star of Tc~1 has a temperature of 30 to 32~kK, a luminosity of 2\,000 to 4\,000~$L_\odot$, and is located at a distance of approximately 2.1-2.2~kpc. The nebular gas in the main shell has a density around 2\,000-3\,800~cm$^{-3}$.

\item Our models indicate that the main shell of Tc~1 is matter bounded. The leaking ionizing photons are enough to ionize the large and faint Tc~1 halo. The radiation field (in units of the Habing field) is $G_0\sim$~6\,000. The main shell density is $\sim$2000~cm$^{-3}$ and the halo density is $\sim$100~cm$^{-3}$. The density transition is somewhat abrupt. Temperatures in the frontier between the main shell and the halo are around 8\,000-9\,000~K. Hydrogen is fully ionized in both regions.

\item Comparison of the observed intensity to photoionization models provides a krypton abundance of log(Kr/H)~=~-7.9. The derived krypton abundance, which relates to the carbon to oxygen ratio, yields C/O~=~1.9, reasonably close to the value of 1.4 obtained from the empirical analysis and to the values derived from the photoionization models (C/O~=~1.4--2.2), confirming the carbon-rich nature of Tc~1.

\end{enumerate}

Tc~1 is the source with the strongest and cleanest C$_{60}$ emission in the IR. Narrow-band images reveal that the C$_{60}$ emission originates from a ring about 5~arcsec away from the central star \citep{2018Galax...6..101C}, which indicates that the fullerene emission is in or close to the interface of the denser material that forms the main shell and the fainter remnant of the AGB gas.

The absence of PAH and H$_2$ emission in the fullerene-emitting zone of Tc~1 in addition to our result that the nebula is matter bounded are strong evidence that there is no photodissociation region (PDR) surrounding the main shell (note that a PDR may exist in the region surrounding the halo). H$_2$ line have not been detected in Tc~1. H$_2$ lines are absent in both in our near infrared X-Shooter spectrum and in the mid infrared Spitzer spectrum \citep{2014MNRAS.437.2577O}. In fact, \citet{2014MNRAS.437.2577O} did not detect H$_2$ emission in the any of the C$_{60}$ PNe they studied. In Tc~1, PAH emission in seen in its central region \citep{2018Galax...6..101C}. PAHs are associated with photodissociation regions (PDRs) and, although fullerenes have also been detected in classical PDRs around H~{\sc ii} regions, the emission of PAH and fullerenes do not come from the same region \citep{2012PNAS..109..401B}. The absence of a PDR surrounding the main shell in this young PN indicates that this shell may be the star fast wind, which is overtaking the AGB remnant halo, as suggested by \citet{2003MNRAS.340..417C}.

From the temperature and radiation field we determined, there is no obvious constrains to the mechanisms of C$_{60}$ formation \citep[e.g.][]{2012PNAS..109..401B}. However, our results are consistent with the suggestion of \citet{2018Galax...6..101C} that the C$_{60}$ emission is related to the fast wind overtaking the slow wind, or the development of an ionization front can be the likely direct causes. They further argue that the dust destruction process must be related to the transition of AGB to PN, since C$_{60}$ bands are only detected in young PNe. \citet{2018Galax...6..101C} suggest a formation process where the dust condensation mechanism in an earlier phase (AGB) produce fullerenes, which are latter released to the gas when dust is destructed by the ionization front. Although this is a possible mechanism, further studies are needed to determine the C$_{60}$ formation route in Tc~1.

\section*{Acknowledgements}

We thank the anonymous referee for the suggestions to improved this paper. We also thanks Lex Kaper for the useful information provided regarding the X-shooter instrument. Studies of interstellar chemistry at Leiden Observatory are supported through the advanced-ERC grant 246976 from the European Research Council, through a grant by the Dutch Science Agency, NWO, as part of the Dutch Astrochemistry Network, and through the Spinoza prize from the Dutch Science Agency, NWO. I.A. also acknowledges the partial support of CNPq, Conselho Nacional de Desenvolvimento Cient\'ifico e Tecnol\'ogico - Brazil (Senior Postdoctoral Fellowship number 157806/2015-4) and CAPES, Ministry of Education, Brazil, through a PNPD fellowship. M.L.L.-F. and S.A. acknowledge the support of CNPq, Conselho Nacional de Desenvolvimento Cient\'ifico e Tecnol\'ogico - Brazil, processes numbers 248503/2013-8 and 300336/2016-0 respectively. CM thanks the CONACyT project CB2015/254132 and the UNAM-PAPIIT project 107215. RW is supported by ERC Grant 694520 SNDUST. Based on observations collected at the European Organisation for Astronomical Research in the Southern Hemisphere under ESO programme 385.C-0720. This work has made use of the computing facilities of the Laboratory of Astroinformatics (IAG/USP, NAT/Unicsul), whose purchase was made possible by the Brazilian agency FAPESP (grant 2009/54006-4) and the INCT-A. This research has made use of NASA's Astrophysics Data System.

\bibliographystyle{mnras} 
\bibliography{references}   

\begin{thebibliography}{}
\makeatletter
\relax
\def\mn@urlcharsother{\let\do\@makeother \do\$\do\&\do\#\do\^\do\_\do\%\do\~}
\def\mn@doi{\begingroup\mn@urlcharsother \@ifnextchar [ {\mn@doi@}
  {\mn@doi@[]}}
\def\mn@doi@[#1]#2{\def\@tempa{#1}\ifx\@tempa\@empty \href
  {http://dx.doi.org/#2} {doi:#2}\else \href {http://dx.doi.org/#2} {#1}\fi
  \endgroup}
\def\mn@eprint#1#2{\mn@eprint@#1:#2::\@nil}
\def\mn@eprint@arXiv#1{\href {http://arxiv.org/abs/#1} {{\tt arXiv:#1}}}
\def\mn@eprint@dblp#1{\href {http://dblp.uni-trier.de/rec/bibtex/#1.xml}
  {dblp:#1}}
\def\mn@eprint@#1:#2:#3:#4\@nil{\def\@tempa {#1}\def\@tempb {#2}\def\@tempc
  {#3}\ifx \@tempc \@empty \let \@tempc \@tempb \let \@tempb \@tempa \fi \ifx
  \@tempb \@empty \def\@tempb {arXiv}\fi \@ifundefined
  {mn@eprint@\@tempb}{\@tempb:\@tempc}{\expandafter \expandafter \csname
  mn@eprint@\@tempb\endcsname \expandafter{\@tempc}}}

\bibitem[\protect\citeauthoryear{{Acker}}{{Acker}}{1978}]{Acker1978}
{Acker} A.,  1978, \aaps, \href
  {http://adsabs.harvard.edu/abs/1978A%26AS...33..367A} {33, 367}

\bibitem[\protect\citeauthoryear{{Akras} \& {Gon{\c c}alves}}{{Akras} \&
  {Gon{\c c}alves}}{2016}]{2016MNRAS.455..930A}
{Akras} S.,  {Gon{\c c}alves} D.~R.,  2016, \mn@doi [\mnras]
  {10.1093/mnras/stv2139}, \href
  {http://adsabs.harvard.edu/abs/2016MNRAS.455..930A} {455, 930}

\bibitem[\protect\citeauthoryear{{Akras} \& {L{\'o}pez}}{{Akras} \&
  {L{\'o}pez}}{2012}]{2012MNRAS.425.2197A}
{Akras} S.,  {L{\'o}pez} J.~A.,  2012, \mn@doi [\mnras]
  {10.1111/j.1365-2966.2012.21578.x}, \href
  {http://adsabs.harvard.edu/abs/2012MNRAS.425.2197A} {425, 2197}

\bibitem[\protect\citeauthoryear{{Akras}, {Boumis}, {Meaburn}, {Alikakos},
  {L{\'o}pez}  \& {Gon{\c c}alves}}{{Akras} et~al.}{2015}]{2015MNRAS.452.2911A}
{Akras} S.,  {Boumis} P.,  {Meaburn} J.,  {Alikakos} J.,  {L{\'o}pez} J.~A.,
  {Gon{\c c}alves} D.~R.,  2015, \mn@doi [\mnras] {10.1093/mnras/stv1468},
  \href {http://adsabs.harvard.edu/abs/2015MNRAS.452.2911A} {452, 2911}

\bibitem[\protect\citeauthoryear{{Aleman} et~al.,}{{Aleman}
  et~al.}{2018}]{aleman2018}
{Aleman} I.,  et~al., 2018, \mn@doi [\mnras] {10.1093/mnras/sty966}, \href
  {http://adsabs.harvard.edu/abs/2018MNRAS.477.4499A} {477, 4499}

\bibitem[\protect\citeauthoryear{{Amnuel}, {Guseinov}, {Novruzova}  \&
  {Rustamov}}{{Amnuel} et~al.}{1984}]{Amnuel_etal_1984}
{Amnuel} P.~R.,  {Guseinov} O.~K.,  {Novruzova} K.~I.,   {Rustamov} I.~S.,
  1984, \mn@doi [\apss] {10.1007/BF00649612}, \href
  {http://adsabs.harvard.edu/abs/1984Ap%26SS.107...19A} {107, 19}

\bibitem[\protect\citeauthoryear{{Arrieta} \& {Torres-Peimbert}}{{Arrieta} \&
  {Torres-Peimbert}}{2003}]{2003ApJS..147...97A}
{Arrieta} A.,  {Torres-Peimbert} S.,  2003, \mn@doi [\apjs] {10.1086/374922},
  \href {http://adsabs.harvard.edu/abs/2003ApJS..147...97A} {147, 97}

\bibitem[\protect\citeauthoryear{{Asplund}, {Grevesse}  \& {Sauval}}{{Asplund}
  et~al.}{2005}]{2005ASPC..336...25A}
{Asplund} M.,  {Grevesse} N.,   {Sauval} A.~J.,  2005, in {Barnes} III T.~G.,
  {Bash} F.~N.,  eds,  Astronomical Society of the Pacific Conference Series
  Vol. 336, Cosmic Abundances as Records of Stellar Evolution and
  Nucleosynthesis. p.~25

\bibitem[\protect\citeauthoryear{{Balick}}{{Balick}}{1989}]{1989AJ.....97..476B}
{Balick} B.,  1989, \mn@doi [\aj] {10.1086/114996}, \href
  {http://adsabs.harvard.edu/abs/1989AJ.....97..476B} {97, 476}

\bibitem[\protect\citeauthoryear{{Balick}, {Alexander}, {Hajian}, {Terzian},
  {Perinotto}  \& {Patriarchi}}{{Balick} et~al.}{1998}]{1998AJ....116..360B}
{Balick} B.,  {Alexander} J.,  {Hajian} A.~R.,  {Terzian} Y.,  {Perinotto} M.,
   {Patriarchi} P.,  1998, \mn@doi [\aj] {10.1086/300429}, \href
  {http://adsabs.harvard.edu/abs/1998AJ....116..360B} {116, 360}

\bibitem[\protect\citeauthoryear{{Bernard-Salas}, {Cami}, {Peeters}, {Jones},
  {Micelotta}  \& {Groenewegen}}{{Bernard-Salas}
  et~al.}{2012}]{Jero:C60excitation}
{Bernard-Salas} J.,  {Cami} J.,  {Peeters} E.,  {Jones} A.~P.,  {Micelotta}
  E.~R.,   {Groenewegen} M.~A.~T.,  2012, \mn@doi [\apj]
  {10.1088/0004-637X/757/1/41}, \href
  {http://adsabs.harvard.edu/abs/2012ApJ...757...41B} {757, 41}

\bibitem[\protect\citeauthoryear{{Bern{\'e}} \& {Tielens}}{{Bern{\'e}} \&
  {Tielens}}{2012}]{2012PNAS..109..401B}
{Bern{\'e}} O.,  {Tielens} A.~G.~G.~M.,  2012, \mn@doi [Proceedings of the
  National Academy of Science] {10.1073/pnas.1114207108}, \href
  {http://adsabs.harvard.edu/abs/2012PNAS..109..401B} {109, 401}

\bibitem[\protect\citeauthoryear{{Bianchi}}{{Bianchi}}{1992}]{1992A&A...260..314B}
{Bianchi} L.,  1992, \aap, \href
  {https://ui.adsabs.harvard.edu/abs/1992A&A...260..314B} {260, 314}

\bibitem[\protect\citeauthoryear{{Cahn}}{{Cahn}}{1976}]{1976AJ.....81..407C}
{Cahn} J.~H.,  1976, \mn@doi [\aj] {10.1086/111900}, \href
  {http://adsabs.harvard.edu/abs/1976AJ.....81..407C} {81, 407}

\bibitem[\protect\citeauthoryear{{Cahn} \& {Kaler}}{{Cahn} \&
  {Kaler}}{1971}]{1971ApJS...22..319C}
{Cahn} J.~H.,  {Kaler} J.~B.,  1971, \mn@doi [\apjs] {10.1086/190227}, \href
  {http://adsabs.harvard.edu/abs/1971ApJS...22..319C} {22, 319}

\bibitem[\protect\citeauthoryear{{Cahn}, {Kaler}  \& {Stanghellini}}{{Cahn}
  et~al.}{1992}]{Cahn_etal_1992}
{Cahn} J.~H.,  {Kaler} J.~B.,   {Stanghellini} L.,  1992, \aaps, \href
  {http://adsabs.harvard.edu/abs/1992A%26AS...94..399C} {94, 399}

\bibitem[\protect\citeauthoryear{{Cami}, {Bernard-Salas}, {Peeters}  \&
  {Malek}}{{Cami} et~al.}{2010}]{2010Sci...329.1180C}
{Cami} J.,  {Bernard-Salas} J.,  {Peeters} E.,   {Malek} S.~E.,  2010, \mn@doi
  [Science] {10.1126/science.1192035}, \href
  {http://adsabs.harvard.edu/abs/2010Sci...329.1180C} {329, 1180}

\bibitem[\protect\citeauthoryear{{Cami}, {Peeters}, {Bernard-Salas}, {Doppmann}
   \& {De Buizer}}{{Cami} et~al.}{2018}]{2018Galax...6..101C}
{Cami} J.,  {Peeters} E.,  {Bernard-Salas} J.,  {Doppmann} G.,   {De Buizer}
  J.,  2018, \mn@doi [Galaxies] {10.3390/galaxies6040101}, \href
  {https://ui.adsabs.harvard.edu/abs/2018Galax...6..101C} {6, 101}

\bibitem[\protect\citeauthoryear{{Cardelli}, {Clayton}  \& {Mathis}}{{Cardelli}
  et~al.}{1989}]{1989ApJ...345..245C}
{Cardelli} J.~A.,  {Clayton} G.~C.,   {Mathis} J.~S.,  1989, \mn@doi [\apj]
  {10.1086/167900}, \href {http://adsabs.harvard.edu/abs/1989ApJ...345..245C}
  {345, 245}

\bibitem[\protect\citeauthoryear{{Cazetta} \& {Maciel}}{{Cazetta} \&
  {Maciel}}{2001}]{Cazetta_Maciel_2001}
{Cazetta} J.~O.,  {Maciel} W.~J.,  2001, \mn@doi [\apss]
  {10.1023/A:1012593212157}, \href
  {http://adsabs.harvard.edu/abs/2001Ap%26SS.277..393C} {277, 393}

\bibitem[\protect\citeauthoryear{{Clyne}, {Akras}, {Steffen}, {Redman}, {Gon{\c
  c}alves}  \& {Harvey}}{{Clyne} et~al.}{2015}]{2015A&A...582A..60C}
{Clyne} N.,  {Akras} S.,  {Steffen} W.,  {Redman} M.~P.,  {Gon{\c c}alves}
  D.~R.,   {Harvey} E.,  2015, \mn@doi [\aap] {10.1051/0004-6361/201526585},
  \href {http://adsabs.harvard.edu/abs/2015A\%26A...582A..60C} {582, A60}

\bibitem[\protect\citeauthoryear{{Corradi}, {Sch{\"o}nberner}, {Steffen}  \&
  {Perinotto}}{{Corradi} et~al.}{2003}]{2003MNRAS.340..417C}
{Corradi} R.~L.~M.,  {Sch{\"o}nberner} D.,  {Steffen} M.,   {Perinotto} M.,
  2003, \mn@doi [\mnras] {10.1046/j.1365-8711.2003.06294.x}, \href
  {https://ui.adsabs.harvard.edu/abs/2003MNRAS.340..417C} {340, 417}

\bibitem[\protect\citeauthoryear{{Corradi}, {Garc{\'{\i}}a-Rojas}, {Jones}  \&
  {Rodr{\'{\i}}guez-Gil}}{{Corradi} et~al.}{2015}]{corradi2015}
{Corradi} R.~L.~M.,  {Garc{\'{\i}}a-Rojas} J.,  {Jones} D.,
  {Rodr{\'{\i}}guez-Gil} P.,  2015, \mn@doi [\apj]
  {10.1088/0004-637X/803/2/99}, \href
  {http://adsabs.harvard.edu/abs/2015ApJ...803...99C} {803, 99}

\bibitem[\protect\citeauthoryear{{Cosby}, {Sharpee}, {Slanger}, {Huestis}  \&
  {Hanuschik}}{{Cosby} et~al.}{2006}]{2006JGRA..11112307C}
{Cosby} P.~C.,  {Sharpee} B.~D.,  {Slanger} T.~G.,  {Huestis} D.~L.,
  {Hanuschik} R.~W.,  2006, \mn@doi [Journal of Geophysical Research (Space
  Physics)] {10.1029/2006JA012023}, \href
  {http://adsabs.harvard.edu/abs/2006JGRA..11112307C} {111, A12307}

\bibitem[\protect\citeauthoryear{{Daub}}{{Daub}}{1982}]{1982ApJ...260..612D}
{Daub} C.~T.,  1982, \mn@doi [\apj] {10.1086/160283}, \href
  {http://adsabs.harvard.edu/abs/1982ApJ...260..612D} {260, 612}

\bibitem[\protect\citeauthoryear{{Davey}, {Storey}  \& {Kisielius}}{{Davey}
  et~al.}{2000}]{davey2000}
{Davey} A.~R.,  {Storey} P.~J.,   {Kisielius} R.,  2000, \mn@doi [\aaps]
  {10.1051/aas:2000139}, \href
  {http://adsabs.harvard.edu/abs/2000A%26AS..142...85D} {142, 85}

\bibitem[\protect\citeauthoryear{{Delgado-Inglada}, {Morisset}  \&
  {Stasi{\'n}ska}}{{Delgado-Inglada} et~al.}{2014}]{2014MNRAS.440..536D}
{Delgado-Inglada} G.,  {Morisset} C.,   {Stasi{\'n}ska} G.,  2014, \mn@doi
  [\mnras] {10.1093/mnras/stu341}, \href
  {http://adsabs.harvard.edu/abs/2014MNRAS.440..536D} {440, 536}

\bibitem[\protect\citeauthoryear{{Dere}, {Landi}, {Mason}, {Monsignori Fossi}
  \& {Young}}{{Dere} et~al.}{1997}]{1997A&AS..125..149D}
{Dere} K.~P.,  {Landi} E.,  {Mason} H.~E.,  {Monsignori Fossi} B.~C.,   {Young}
  P.~R.,  1997, \mn@doi [\aaps] {10.1051/aas:1997368}, \href
  {http://adsabs.harvard.edu/abs/1997A%26AS..125..149D} {125, 149}

\bibitem[\protect\citeauthoryear{{Escalante} \& {Victor}}{{Escalante} \&
  {Victor}}{1990}]{escalante1990}
{Escalante} V.,  {Victor} G.~A.,  1990, \mn@doi [\apjs] {10.1086/191479}, \href
  {http://adsabs.harvard.edu/abs/1990ApJS...73..513E} {73, 513}

\bibitem[\protect\citeauthoryear{{Ferland} et~al.,}{{Ferland}
  et~al.}{2017}]{2017RMxAA..53..385F}
{Ferland} G.~J.,  et~al., 2017, \rmxaa, \href
  {http://adsabs.harvard.edu/abs/2017RMxAA..53..385F} {53, 385}

\bibitem[\protect\citeauthoryear{{Fernandes}, {Gruenwald}  \&
  {Viegas}}{{Fernandes} et~al.}{2005}]{Fernandes_etal_2005}
{Fernandes} I.~F.,  {Gruenwald} R.,   {Viegas} S.~M.,  2005, \mn@doi [\mnras]
  {10.1111/j.1365-2966.2005.09596.x}, \href
  {http://adsabs.harvard.edu/abs/2005MNRAS.364..674F} {364, 674}

\bibitem[\protect\citeauthoryear{{Frew}, {Boji{\v c}i{\'c}}  \&
  {Parker}}{{Frew} et~al.}{2013}]{2013MNRAS.431....2F}
{Frew} D.~J.,  {Boji{\v c}i{\'c}} I.~S.,   {Parker} Q.~A.,  2013, \mn@doi
  [\mnras] {10.1093/mnras/sts393}, \href
  {http://adsabs.harvard.edu/abs/2013MNRAS.431....2F} {431, 2}

\bibitem[\protect\citeauthoryear{{Garc{\'{\i}}a-D{\'{\i}}az}, {L{\'o}pez},
  {Steffen}  \& {Richer}}{{Garc{\'{\i}}a-D{\'{\i}}az}
  et~al.}{2012}]{2012ApJ...761..172G}
{Garc{\'{\i}}a-D{\'{\i}}az} M.~T.,  {L{\'o}pez} J.~A.,  {Steffen} W.,
  {Richer} M.~G.,  2012, \mn@doi [\apj] {10.1088/0004-637X/761/2/172}, \href
  {http://adsabs.harvard.edu/abs/2012ApJ...761..172G} {761, 172}

\bibitem[\protect\citeauthoryear{{Garc{\'{\i}}a-Rojas}, {Madonna}, {Luridiana},
  {Sterling}, {Morisset}, {Delgado-Inglada}  \& {Toribio San
  Cipriano}}{{Garc{\'{\i}}a-Rojas} et~al.}{2015}]{2015MNRAS.452.2606G}
{Garc{\'{\i}}a-Rojas} J.,  {Madonna} S.,  {Luridiana} V.,  {Sterling} N.~C.,
  {Morisset} C.,  {Delgado-Inglada} G.,   {Toribio San Cipriano} L.,  2015,
  \mn@doi [\mnras] {10.1093/mnras/stv1415}, \href
  {http://adsabs.harvard.edu/abs/2015MNRAS.452.2606G} {452, 2606}

\bibitem[\protect\citeauthoryear{{Gesicki} \& {Zijlstra}}{{Gesicki} \&
  {Zijlstra}}{2007}]{Gesicki_Zijlstra_2007}
{Gesicki} K.,  {Zijlstra} A.~A.,  2007, \mn@doi [\aap]
  {10.1051/0004-6361:20077250}, \href
  {http://adsabs.harvard.edu/abs/2007A%26A...467L..29G} {467, L29}

\bibitem[\protect\citeauthoryear{{Gon{\c c}alves}, {Corradi}  \&
  {Mampaso}}{{Gon{\c c}alves} et~al.}{2001}]{2001ApJ...547..302G}
{Gon{\c c}alves} D.~R.,  {Corradi} R.~L.~M.,   {Mampaso} A.,  2001, \mn@doi
  [\apj] {10.1086/318364}, \href
  {http://cdsads.u-strasbg.fr/abs/2001ApJ...547..302G} {547, 302}

\bibitem[\protect\citeauthoryear{{Gon{\c c}alves}, {Wesson}, {Morisset},
  {Barlow}  \& {Ercolano}}{{Gon{\c c}alves} et~al.}{2012}]{2012IAUS..283..144G}
{Gon{\c c}alves} D.~R.,  {Wesson} R.,  {Morisset} C.,  {Barlow} M.,
  {Ercolano} B.,  2012, in IAU Symposium. pp 144--147 (\mn@eprint {arXiv}
  {1110.2709}), \mn@doi{10.1017/S174392131201085X}

\bibitem[\protect\citeauthoryear{{Gorny}, {Stasi{\'n}ska}  \&
  {Tylenda}}{{Gorny} et~al.}{1997}]{Gorny_etal_1997}
{Gorny} S.~K.,  {Stasi{\'n}ska} G.,   {Tylenda} R.,  1997, \aap, \href
  {http://adsabs.harvard.edu/abs/1997A%26A...318..256G} {318, 256}

\bibitem[\protect\citeauthoryear{{Green} et~al.,}{{Green}
  et~al.}{2015}]{2015ApJ...810...25G}
{Green} G.~M.,  et~al., 2015, \mn@doi [\apj] {10.1088/0004-637X/810/1/25},
  \href {http://adsabs.harvard.edu/abs/2015ApJ...810...25G} {810, 25}

\bibitem[\protect\citeauthoryear{{Gruenwald} \& {Viegas}}{{Gruenwald} \&
  {Viegas}}{1992}]{1992ApJS...78..153G}
{Gruenwald} R.~B.,  {Viegas} S.~M.,  1992, \mn@doi [\apjs] {10.1086/191623},
  \href {http://adsabs.harvard.edu/abs/1992ApJS...78..153G} {78, 153}

\bibitem[\protect\citeauthoryear{{Jones}, {Wesson}, {Garc{\'{\i}}a-Rojas},
  {Corradi}  \& {Boffin}}{{Jones} et~al.}{2016}]{jones2016}
{Jones} D.,  {Wesson} R.,  {Garc{\'{\i}}a-Rojas} J.,  {Corradi} R.~L.~M.,
  {Boffin} H.~M.~J.,  2016, \mn@doi [\mnras] {10.1093/mnras/stv2519}, \href
  {http://adsabs.harvard.edu/abs/2016MNRAS.455.3263J} {455, 3263}

\bibitem[\protect\citeauthoryear{{Kimura}, {Gruenwald}  \& {Aleman}}{{Kimura}
  et~al.}{2012}]{2012A&A...541A.112K}
{Kimura} R.~K.,  {Gruenwald} R.,   {Aleman} I.,  2012, \mn@doi [\aap]
  {10.1051/0004-6361/201118429}, \href
  {http://adsabs.harvard.edu/abs/2012A%26A...541A.112K} {541, A112}

\bibitem[\protect\citeauthoryear{{Kingsburgh} \& {Barlow}}{{Kingsburgh} \&
  {Barlow}}{1994}]{1994MNRAS.271..257K}
{Kingsburgh} R.~L.,  {Barlow} M.~J.,  1994, \mn@doi [\mnras]
  {10.1093/mnras/271.2.257}, \href
  {http://adsabs.harvard.edu/abs/1994MNRAS.271..257K} {271, 257}

\bibitem[\protect\citeauthoryear{{Kisielius}, {Storey}, {Davey}  \&
  {Neale}}{{Kisielius} et~al.}{1998}]{kisielius1998}
{Kisielius} R.,  {Storey} P.~J.,  {Davey} A.~R.,   {Neale} L.~T.,  1998,
  \mn@doi [\aaps] {10.1051/aas:1998319}, \href
  {http://adsabs.harvard.edu/abs/1998A%26AS..133..257K} {133, 257}

\bibitem[\protect\citeauthoryear{{Kisielius}, {Storey}, {Ferland}  \&
  {Keenan}}{{Kisielius} et~al.}{2009}]{kisielius2009}
{Kisielius} R.,  {Storey} P.~J.,  {Ferland} G.~J.,   {Keenan} F.~P.,  2009,
  \mn@doi [\mnras] {10.1111/j.1365-2966.2009.14989.x}, \href
  {http://adsabs.harvard.edu/abs/2009MNRAS.397..903K} {397, 903}

\bibitem[\protect\citeauthoryear{{Kroto}, {Heath}, {Obrien}, {Curl}  \&
  {Smalley}}{{Kroto} et~al.}{1985}]{1985Natur.318..162K}
{Kroto} H.~W.,  {Heath} J.~R.,  {Obrien} S.~C.,  {Curl} R.~F.,   {Smalley}
  R.~E.,  1985, \mn@doi [\nat] {10.1038/318162a0}, \href
  {http://adsabs.harvard.edu/abs/1985Natur.318..162K} {318, 162}

\bibitem[\protect\citeauthoryear{{Kudritzki}, {Mendez}, {Puls}  \&
  {McCarthy}}{{Kudritzki} et~al.}{1997}]{Kudritzki_etal_1997}
{Kudritzki} R.~P.,  {Mendez} R.~H.,  {Puls} J.,   {McCarthy} J.~K.,  1997, in
  {Habing} H.~J.,  {Lamers} H.~J.~G.~L.~M.,  eds,  IAU Symposium Vol. 180,
  Planetary Nebulae. p.~64.K

\bibitem[\protect\citeauthoryear{{Kudritzki}, {Urbaneja}  \&
  {Puls}}{{Kudritzki} et~al.}{2006}]{Kudritzki_etal_2006}
{Kudritzki} R.~P.,  {Urbaneja} M.~A.,   {Puls} J.,  2006, in {Barlow} M.~J.,
  {M{\'e}ndez} R.~H.,  eds,  IAU Symposium Vol. 234, Planetary Nebulae in our
  Galaxy and Beyond. pp 119--126 (\mn@eprint {} {astro-ph/0608094}),
  \mn@doi{10.1017/S1743921306002857}

\bibitem[\protect\citeauthoryear{{Landi}, {Del Zanna}, {Young}, {Dere}  \&
  {Mason}}{{Landi} et~al.}{2012}]{landi2012}
{Landi} E.,  {Del Zanna} G.,  {Young} P.~R.,  {Dere} K.~P.,   {Mason} H.~E.,
  2012, \mn@doi [\apj] {10.1088/0004-637X/744/2/99}, \href
  {http://adsabs.harvard.edu/abs/2012ApJ...744...99L} {744, 99}

\bibitem[\protect\citeauthoryear{{Leal-Ferreira}, {Gon{\c c}alves}, {Monteiro}
  \& {Richards}}{{Leal-Ferreira} et~al.}{2011}]{2011MNRAS.411.1395L}
{Leal-Ferreira} M.~L.,  {Gon{\c c}alves} D.~R.,  {Monteiro} H.,   {Richards}
  J.~W.,  2011, \mn@doi [\mnras] {10.1111/j.1365-2966.2010.17776.x}, \href
  {http://adsabs.harvard.edu/abs/2011MNRAS.411.1395L} {411, 1395}

\bibitem[\protect\citeauthoryear{{Lee} \& {Hyung}}{{Lee} \&
  {Hyung}}{2000}]{2000ApJ...530L..49L}
{Lee} H.-W.,  {Hyung} S.,  2000, \mn@doi [\apjl] {10.1086/312479}, \href
  {http://adsabs.harvard.edu/abs/2000ApJ...530L..49L} {530, L49}

\bibitem[\protect\citeauthoryear{{Liu}, {Storey}, {Barlow}  \& {Clegg}}{{Liu}
  et~al.}{1995}]{liu1995}
{Liu} X.-W.,  {Storey} P.~J.,  {Barlow} M.~J.,   {Clegg} R.~E.~S.,  1995,
  \mn@doi [\mnras] {10.1093/mnras/272.2.369}, \href
  {http://adsabs.harvard.edu/abs/1995MNRAS.272..369L} {272, 369}

\bibitem[\protect\citeauthoryear{{Liu}, {Storey}, {Barlow}, {Danziger}, {Cohen}
   \& {Bryce}}{{Liu} et~al.}{2000}]{2000MNRAS.312..585L}
{Liu} X.-W.,  {Storey} P.~J.,  {Barlow} M.~J.,  {Danziger} I.~J.,  {Cohen} M.,
   {Bryce} M.,  2000, \mn@doi [\mnras] {10.1046/j.1365-8711.2000.03167.x},
  \href {http://adsabs.harvard.edu/abs/2000MNRAS.312..585L} {312, 585}

\bibitem[\protect\citeauthoryear{{Maciel}}{{Maciel}}{1984}]{Maciel1984}
{Maciel} W.~J.,  1984, \aaps, \href
  {http://adsabs.harvard.edu/abs/1984A%26AS...55..253M} {55, 253}

\bibitem[\protect\citeauthoryear{{Maciel}, {Keller}  \& {Costa}}{{Maciel}
  et~al.}{2008}]{Maciel_etal_2008}
{Maciel} W.~J.,  {Keller} G.~R.,   {Costa} R.~D.~D.,  2008, \rmxaa, \href
  {http://adsabs.harvard.edu/abs/2008RMxAA..44..221M} {44, 221}

\bibitem[\protect\citeauthoryear{{Markwardt}}{{Markwardt}}{2009}]{2009ASPC..411..251M}
{Markwardt} C.~B.,  2009, in {Bohlender} D.~A.,  {Durand} D.,   {Dowler} P.,
  eds,  Astronomical Society of the Pacific Conference Series Vol. 411,
  Astronomical Data Analysis Software and Systems XVIII. p.~251 (\mn@eprint
  {arXiv} {0902.2850})

\bibitem[\protect\citeauthoryear{{Martin}}{{Martin}}{1994}]{Martin1994}
{Martin} W.,  1994, \aap, \href
  {http://adsabs.harvard.edu/abs/1994A%26A...281..526M} {281, 526}

\bibitem[\protect\citeauthoryear{{Mathis}, {Rumpl}  \& {Nordsieck}}{{Mathis}
  et~al.}{1977}]{1977ApJ...217..425M}
{Mathis} J.~S.,  {Rumpl} W.,   {Nordsieck} K.~H.,  1977, \mn@doi [\apj]
  {10.1086/155591}, \href
  {https://ui.adsabs.harvard.edu/abs/1977ApJ...217..425M} {217, 425}

\bibitem[\protect\citeauthoryear{{Mendez}, {Kudritzki}, {Herrero}, {Husfeld}
  \& {Groth}}{{Mendez} et~al.}{1988}]{Mendez1988}
{Mendez} R.~H.,  {Kudritzki} R.~P.,  {Herrero} A.,  {Husfeld} D.,   {Groth}
  H.~G.,  1988, \aap, \href
  {http://adsabs.harvard.edu/abs/1988A%26A...190..113M} {190, 113}

\bibitem[\protect\citeauthoryear{{Mendoza} \& {Zeippen}}{{Mendoza} \&
  {Zeippen}}{1982}]{mendoza1982}
{Mendoza} C.,  {Zeippen} C.~J.,  1982, \mn@doi [\mnras]
  {10.1093/mnras/198.1.127}, \href
  {http://adsabs.harvard.edu/abs/1982MNRAS.198..127M} {198, 127}

\bibitem[\protect\citeauthoryear{{Mendoza} \& {Zeippen}}{{Mendoza} \&
  {Zeippen}}{1983}]{mendoza1983}
{Mendoza} C.,  {Zeippen} C.~J.,  1983, \mn@doi [\mnras]
  {10.1093/mnras/202.4.981}, \href
  {http://adsabs.harvard.edu/abs/1983MNRAS.202..981M} {202, 981}

\bibitem[\protect\citeauthoryear{{Micelotta}, {Jones}, {Cami}, {Peeters},
  {Bernard-Salas}  \& {Fanchini}}{{Micelotta}
  et~al.}{2012}]{Elisabetta:arophatics}
{Micelotta} E.~R.,  {Jones} A.~P.,  {Cami} J.,  {Peeters} E.,  {Bernard-Salas}
  J.,   {Fanchini} G.,  2012, \mn@doi [\apj] {10.1088/0004-637X/761/1/35},
  \href {http://adsabs.harvard.edu/abs/2012ApJ...761...35M} {761, 35}

\bibitem[\protect\citeauthoryear{{Miller Bertolami}}{{Miller
  Bertolami}}{2016}]{2016A&A...588A..25M}
{Miller Bertolami} M.~M.,  2016, \mn@doi [\aap] {10.1051/0004-6361/201526577},
  \href {https://ui.adsabs.harvard.edu/abs/2016A&A...588A..25M} {588, A25}

\bibitem[\protect\citeauthoryear{{Milne} \& {Aller}}{{Milne} \&
  {Aller}}{1975}]{Milne_Aller_1975}
{Milne} D.~K.,  {Aller} L.~H.,  1975, \aap, \href
  {http://adsabs.harvard.edu/abs/1975A%26A....38..183M} {38, 183}

\bibitem[\protect\citeauthoryear{{Monteiro}, {Gon{\c c}alves}, {Leal-Ferreira}
  \& {Corradi}}{{Monteiro} et~al.}{2013}]{2013A&A...560A.102M}
{Monteiro} H.,  {Gon{\c c}alves} D.~R.,  {Leal-Ferreira} M.~L.,   {Corradi}
  R.~L.~M.,  2013, \mn@doi [\aap] {10.1051/0004-6361/201322220}, \href
  {http://adsabs.harvard.edu/abs/2013A%26A...560A.102M} {560, A102}

\bibitem[\protect\citeauthoryear{{Mor\'{e}}}{{Mor\'{e}}}{1978}]{More1978}
{Mor\'{e}} J.~J.,  1978, in {Watson} G.~A.,  ed., Numerical Analysis:
  Proceedings of the Biennial Conference Held at Dundee, June 28-July 1, 1977.
  Springer Berlin Heidelberg, Berlin, Heidelberg, pp 105--116,
  \mn@doi{10.1007/BFb0067700}, \url {http://dx.doi.org/10.1007/BFb0067700}

\bibitem[\protect\citeauthoryear{{Morisset}}{{Morisset}}{2013}]{2013ascl.soft04020M}
{Morisset} C.,  2013, {pyCloudy: Tools to manage astronomical Cloudy
  photoionization code}, Astrophysics Source Code Library (\mn@eprint {ascl}
  {1304.020})

\bibitem[\protect\citeauthoryear{{Morisset} \& {Georgiev}}{{Morisset} \&
  {Georgiev}}{2009}]{2009A&A...507.1517M}
{Morisset} C.,  {Georgiev} L.,  2009, \mn@doi [\aap]
  {10.1051/0004-6361/200912413}, \href
  {https://ui.adsabs.harvard.edu/#abs/2009A&A...507.1517M} {507, 1517}

\bibitem[\protect\citeauthoryear{{Morisset}, {Delgado-Inglada}  \&
  {Flores-Fajardo}}{{Morisset} et~al.}{2015}]{2015RMxAA..51..103M}
{Morisset} C.,  {Delgado-Inglada} G.,   {Flores-Fajardo} N.,  2015, \rmxaa,
  \href {https://ui.adsabs.harvard.edu/#abs/2015RMxAA..51..103M} {51, 103}

\bibitem[\protect\citeauthoryear{{O'Donnell}}{{O'Donnell}}{1994}]{1994ApJ...422..158O}
{O'Donnell} J.~E.,  1994, \mn@doi [\apj] {10.1086/173713}, \href
  {http://adsabs.harvard.edu/abs/1994ApJ...422..158O} {422, 158}

\bibitem[\protect\citeauthoryear{{Osterbrock} \& {Ferland}}{{Osterbrock} \&
  {Ferland}}{2006}]{2006agna.book.....O}
{Osterbrock} D.~E.,  {Ferland} G.~J.,  2006, {Astrophysics of gaseous nebulae
  and active galactic nuclei}.
University Science Books

\bibitem[\protect\citeauthoryear{{Otsuka}}{{Otsuka}}{2019}]{2019MNRAS.482.2354O}
{Otsuka} M.,  2019, \mn@doi [\mnras] {10.1093/mnras/sty2733}, \href
  {https://ui.adsabs.harvard.edu/\#abs/2019MNRAS.482.2354O} {482, 2354}

\bibitem[\protect\citeauthoryear{{Otsuka}, {Kemper}, {Cami}, {Peeters}  \&
  {Bernard-Salas}}{{Otsuka} et~al.}{2014}]{2014MNRAS.437.2577O}
{Otsuka} M.,  {Kemper} F.,  {Cami} J.,  {Peeters} E.,   {Bernard-Salas} J.,
  2014, \mn@doi [\mnras] {10.1093/mnras/stt2070}, \href
  {http://adsabs.harvard.edu/abs/2014MNRAS.437.2577O} {437, 2577}

\bibitem[\protect\citeauthoryear{{Pauldrach}, {Hoffmann}  \&
  {M{\'e}ndez}}{{Pauldrach} et~al.}{2004}]{Pauldrach_etal_2004}
{Pauldrach} A.~W.~A.,  {Hoffmann} T.~L.,   {M{\'e}ndez} R.~H.,  2004, \mn@doi
  [\aap] {10.1051/0004-6361:20034040}, \href
  {http://adsabs.harvard.edu/abs/2004A%26A...419.1111P} {419, 1111}

\bibitem[\protect\citeauthoryear{{Pequignot}, {Petitjean}  \&
  {Boisson}}{{Pequignot} et~al.}{1991}]{pequignot1991}
{Pequignot} D.,  {Petitjean} P.,   {Boisson} C.,  1991, \aap, \href
  {http://adsabs.harvard.edu/abs/1991A%26A...251..680P} {251, 680}

\bibitem[\protect\citeauthoryear{{Phillips} \& {Ramos-Larios}}{{Phillips} \&
  {Ramos-Larios}}{2005}]{Phillips_Ramos_Larios_2005}
{Phillips} J.~P.,  {Ramos-Larios} G.,  2005, \mn@doi [\mnras]
  {10.1111/j.1365-2966.2005.09611.x}, \href
  {http://adsabs.harvard.edu/abs/2005MNRAS.364..849P} {364, 849}

\bibitem[\protect\citeauthoryear{{Podobedova}, {Kelleher}  \&
  {Wiese}}{{Podobedova} et~al.}{2009}]{podobedova2009}
{Podobedova} L.~I.,  {Kelleher} D.~E.,   {Wiese} W.~L.,  2009, \mn@doi [Journal
  of Physical and Chemical Reference Data] {10.1063/1.3032939}, \href
  {http://adsabs.harvard.edu/abs/2009JPCRD..38..171P} {38, 171}

\bibitem[\protect\citeauthoryear{{Porter}, {Ferland}, {Storey}  \&
  {Detisch}}{{Porter} et~al.}{2013}]{porter2013}
{Porter} R.~L.,  {Ferland} G.~J.,  {Storey} P.~J.,   {Detisch} M.~J.,  2013,
  \mn@doi [\mnras] {10.1093/mnrasl/slt049}, \href
  {http://adsabs.harvard.edu/abs/2013MNRAS.433L..89P} {433, L89}

\bibitem[\protect\citeauthoryear{{Pottasch}, {Surendiranath}  \&
  {Bernard-Salas}}{{Pottasch} et~al.}{2011}]{Pottasch_etal_2011}
{Pottasch} S.~R.,  {Surendiranath} R.,   {Bernard-Salas} J.,  2011, \mn@doi
  [\aap] {10.1051/0004-6361/201116669}, \href
  {http://adsabs.harvard.edu/abs/2011A%26A...531A..23P} {531, A23}

\bibitem[\protect\citeauthoryear{{Schwarz}, {Corradi}  \& {Melnick}}{{Schwarz}
  et~al.}{1992}]{1992A&AS...96...23S}
{Schwarz} H.~E.,  {Corradi} R.~L.~M.,   {Melnick} J.,  1992, \aaps, \href
  {https://ui.adsabs.harvard.edu/abs/1992A%26AS...96...23S} {96, 23}

\bibitem[\protect\citeauthoryear{{Stanghellini} \& {Pasquali}}{{Stanghellini}
  \& {Pasquali}}{1995}]{1995ApJ...452..286S}
{Stanghellini} L.,  {Pasquali} A.,  1995, \mn@doi [\apj] {10.1086/176300},
  \href {http://adsabs.harvard.edu/abs/1995ApJ...452..286S} {452, 286}

\bibitem[\protect\citeauthoryear{{Stanghellini}, {Shaw}  \&
  {Villaver}}{{Stanghellini} et~al.}{2008}]{2008ApJ...689..194S}
{Stanghellini} L.,  {Shaw} R.~A.,   {Villaver} E.,  2008, \mn@doi [\apj]
  {10.1086/592395}, \href {http://adsabs.harvard.edu/abs/2008ApJ...689..194S}
  {689, 194}

\bibitem[\protect\citeauthoryear{{Stasi{\'n}ska} \& {Szczerba}}{{Stasi{\'n}ska}
  \& {Szczerba}}{1999}]{1999A&A...352..297S}
{Stasi{\'n}ska} G.,  {Szczerba} R.,  1999, \aap, \href
  {http://adsabs.harvard.edu/abs/1999A%26A...352..297S} {352, 297}

\bibitem[\protect\citeauthoryear{{Steffen} \& {L{\'o}pez}}{{Steffen} \&
  {L{\'o}pez}}{2006}]{2006RMxAA..42...99S}
{Steffen} W.,  {L{\'o}pez} J.~A.,  2006, \rmxaa, \href
  {http://adsabs.harvard.edu/abs/2006RMxAA..42...99S} {42, 99}

\bibitem[\protect\citeauthoryear{{Steffen}, {Esp{\'{\i}}ndola},
  {Mart{\'{\i}}nez}  \& {Koning}}{{Steffen}
  et~al.}{2009a}]{2009RMxAA..45..143S}
{Steffen} W.,  {Esp{\'{\i}}ndola} M.,  {Mart{\'{\i}}nez} S.,   {Koning} N.,
  2009a, \rmxaa, \href {http://adsabs.harvard.edu/abs/2009RMxAA..45..143S} {45,
  143}

\bibitem[\protect\citeauthoryear{{Steffen}, {Garc{\'{\i}}a-Segura}  \&
  {Koning}}{{Steffen} et~al.}{2009b}]{2009ApJ...691..696S}
{Steffen} W.,  {Garc{\'{\i}}a-Segura} G.,   {Koning} N.,  2009b, \mn@doi [\apj]
  {10.1088/0004-637X/691/1/696}, \href
  {http://adsabs.harvard.edu/abs/2009ApJ...691..696S} {691, 696}

\bibitem[\protect\citeauthoryear{{Steffen}, {Koning}, {Wenger}, {Morisset}  \&
  {Magnor}}{{Steffen} et~al.}{2011}]{2011ITVCG..17..454S}
{Steffen} W.,  {Koning} N.,  {Wenger} S.,  {Morisset} C.,   {Magnor} M.,  2011,
  \mn@doi [IEEE Transactions on Visualization and Computer Graphics, Volume 17,
  Issue 4, p.454-465] {10.1109/TVCG.2010.62}, \href
  {http://adsabs.harvard.edu/abs/2011ITVCG..17..454S} {17, 454}

\bibitem[\protect\citeauthoryear{{Sterling} \& {Dinerstein}}{{Sterling} \&
  {Dinerstein}}{2008}]{2008ApJS..174..158S}
{Sterling} N.~C.,  {Dinerstein} H.~L.,  2008, \mn@doi [\apjs] {10.1086/520845},
  \href {http://adsabs.harvard.edu/abs/2008ApJS..174..158S} {174, 158}

\bibitem[\protect\citeauthoryear{{Storey}}{{Storey}}{1994}]{storey1994}
{Storey} P.~J.,  1994, \aap, \href
  {http://adsabs.harvard.edu/abs/1994A%26A...282..999S} {282, 999}

\bibitem[\protect\citeauthoryear{{Storey} \& {Hummer}}{{Storey} \&
  {Hummer}}{1995}]{Storey_Hummer_1995}
{Storey} P.~J.,  {Hummer} D.~G.,  1995, \mn@doi [\mnras]
  {10.1093/mnras/272.1.41}, \href
  {http://adsabs.harvard.edu/abs/1995MNRAS.272...41S} {272, 41}

\bibitem[\protect\citeauthoryear{{Storey}, {Sochi}  \& {Bastin}}{{Storey}
  et~al.}{2017}]{storey2017}
{Storey} P.~J.,  {Sochi} T.,   {Bastin} R.,  2017, \mn@doi [\mnras]
  {10.1093/mnras/stx1189}, \href
  {http://adsabs.harvard.edu/abs/2017MNRAS.470..379S} {470, 379}

\bibitem[\protect\citeauthoryear{{Strelnitski}, {Smith}  \&
  {Ponomarev}}{{Strelnitski} et~al.}{1996}]{strel1996}
{Strelnitski} V.~S.,  {Smith} H.~A.,   {Ponomarev} V.~O.,  1996, \mn@doi [\apj]
  {10.1086/177937}, \href {http://adsabs.harvard.edu/abs/1996ApJ...470.1134S}
  {470, 1134}

\bibitem[\protect\citeauthoryear{{Tajitsu} \& {Tamura}}{{Tajitsu} \&
  {Tamura}}{1998}]{1998AJ....115.1989T}
{Tajitsu} A.,  {Tamura} S.,  1998, \mn@doi [\aj] {10.1086/300315}, \href
  {http://adsabs.harvard.edu/abs/1998AJ....115.1989T} {115, 1989}

\bibitem[\protect\citeauthoryear{{Tayal} \& {Zatsarinny}}{{Tayal} \&
  {Zatsarinny}}{2010}]{tayal2010}
{Tayal} S.~S.,  {Zatsarinny} O.,  2010, \mn@doi [\apjs]
  {10.1088/0067-0049/188/1/32}, \href
  {http://adsabs.harvard.edu/abs/2010ApJS..188...32T} {188, 32}

\bibitem[\protect\citeauthoryear{{Thum} \& {Greve}}{{Thum} \&
  {Greve}}{1997}]{tg1997}
{Thum} C.,  {Greve} A.,  1997, \aap, \href
  {http://adsabs.harvard.edu/abs/1997A%26A...324..699T} {324, 699}

\bibitem[\protect\citeauthoryear{{Tsamis}, {Walsh}, {P{\'e}quignot}, {Barlow},
  {Danziger}  \& {Liu}}{{Tsamis} et~al.}{2008}]{2008MNRAS.386...22T}
{Tsamis} Y.~G.,  {Walsh} J.~R.,  {P{\'e}quignot} D.,  {Barlow} M.~J.,
  {Danziger} I.~J.,   {Liu} X.-W.,  2008, \mn@doi [\mnras]
  {10.1111/j.1365-2966.2008.13051.x}, \href
  {http://adsabs.harvard.edu/abs/2008MNRAS.386...22T} {386, 22}

\bibitem[\protect\citeauthoryear{{Tylenda}, {Si{\'o}dmiak}, {G{\'o}rny},
  {Corradi}  \& {Schwarz}}{{Tylenda} et~al.}{2003}]{2003A&A...405..627T}
{Tylenda} R.,  {Si{\'o}dmiak} N.,  {G{\'o}rny} S.~K.,  {Corradi} R.~L.~M.,
  {Schwarz} H.~E.,  2003, \mn@doi [\aap] {10.1051/0004-6361:20030645}, \href
  {https://ui.adsabs.harvard.edu/abs/2003A&A...405..627T} {405, 627}

\bibitem[\protect\citeauthoryear{{Vernet} et~al.,}{{Vernet}
  et~al.}{2011}]{2011A&A...536A.105V}
{Vernet} J.,  et~al., 2011, \mn@doi [\aap] {10.1051/0004-6361/201117752}, \href
  {http://adsabs.harvard.edu/abs/2011A%26A...536A.105V} {536, A105}

\bibitem[\protect\citeauthoryear{{Walsh}, {Monreal-Ibero}, {Barlow}, {Ueta},
  {Wesson}  \& {Zijlstra}}{{Walsh} et~al.}{2016}]{2016A&A...588A.106W}
{Walsh} J.~R.,  {Monreal-Ibero} A.,  {Barlow} M.~J.,  {Ueta} T.,  {Wesson} R.,
   {Zijlstra} A.~A.,  2016, \mn@doi [\aap] {10.1051/0004-6361/201527988}, \href
  {http://adsabs.harvard.edu/abs/2016A%26A...588A.106W} {588, A106}

\bibitem[\protect\citeauthoryear{{Weinberger}}{{Weinberger}}{1989}]{1989A&AS...78..301W}
{Weinberger} R.,  1989, \aaps, \href
  {https://ui.adsabs.harvard.edu/abs/1989A&AS...78..301W} {78, 301}

\bibitem[\protect\citeauthoryear{{Wesson}}{{Wesson}}{2016}]{2016MNRAS.456.3774W}
{Wesson} R.,  2016, \mn@doi [\mnras] {10.1093/mnras/stv2946}, \href
  {http://adsabs.harvard.edu/abs/2016MNRAS.456.3774W} {456, 3774}

\bibitem[\protect\citeauthoryear{{Wesson}, {Liu}  \& {Barlow}}{{Wesson}
  et~al.}{2005}]{wesson2005}
{Wesson} R.,  {Liu} X.-W.,   {Barlow} M.~J.,  2005, \mn@doi [\mnras]
  {10.1111/j.1365-2966.2005.09325.x}, \href
  {http://adsabs.harvard.edu/abs/2005MNRAS.362..424W} {362, 424}

\bibitem[\protect\citeauthoryear{{Wesson}, {Stock}  \& {Scicluna}}{{Wesson}
  et~al.}{2012}]{2012MNRAS.422.3516W}
{Wesson} R.,  {Stock} D.~J.,   {Scicluna} P.,  2012, \mn@doi [\mnras]
  {10.1111/j.1365-2966.2012.20863.x}, \href
  {http://adsabs.harvard.edu/abs/2012MNRAS.422.3516W} {422, 3516}

\bibitem[\protect\citeauthoryear{{Wesson}, {Jones}, {Garc{\'{\i}}a-Rojas},
  {Boffin}  \& {Corradi}}{{Wesson} et~al.}{2018}]{wesson2018}
{Wesson} R.,  {Jones} D.,  {Garc{\'{\i}}a-Rojas} J.,  {Boffin} H.~M.~J.,
  {Corradi} R.~L.~M.,  2018, \mn@doi [\mnras] {10.1093/mnras/sty1871}, \href
  {http://adsabs.harvard.edu/abs/2018MNRAS.tmp.1786W} {}

\bibitem[\protect\citeauthoryear{{Williams}, {Jenkins}, {Baldwin}, {Zhang},
  {Sharpee}, {Pellegrini}  \& {Phillips}}{{Williams}
  et~al.}{2008}]{2008ApJ...677.1100W}
{Williams} R.,  {Jenkins} E.~B.,  {Baldwin} J.~A.,  {Zhang} Y.,  {Sharpee} B.,
  {Pellegrini} E.,   {Phillips} M.,  2008, \mn@doi [\apj] {10.1086/529065},
  \href {http://adsabs.harvard.edu/abs/2008ApJ...677.1100W} {677, 1100}

\bibitem[\protect\citeauthoryear{{Zeippen}}{{Zeippen}}{1982}]{zeippen1982}
{Zeippen} C.~J.,  1982, \mn@doi [\mnras] {10.1093/mnras/198.1.111}, \href
  {http://adsabs.harvard.edu/abs/1982MNRAS.198..111Z} {198, 111}

\bibitem[\protect\citeauthoryear{{Zhang}}{{Zhang}}{1995}]{1995ApJS...98..659Z}
{Zhang} C.~Y.,  1995, \mn@doi [\apjs] {10.1086/192173}, \href
  {http://adsabs.harvard.edu/abs/1995ApJS...98..659Z} {98, 659}

\bibitem[\protect\citeauthoryear{{Zhang}, {Liu}, {Wesson}, {Storey}, {Liu}  \&
  {Danziger}}{{Zhang} et~al.}{2004}]{2004MNRAS.351..935Z}
{Zhang} Y.,  {Liu} X.-W.,  {Wesson} R.,  {Storey} P.~J.,  {Liu} Y.,
  {Danziger} I.~J.,  2004, \mn@doi [\mnras] {10.1111/j.1365-2966.2004.07838.x},
  \href {http://adsabs.harvard.edu/abs/2004MNRAS.351..935Z} {351, 935}

\bibitem[\protect\citeauthoryear{{Zhen}, {Castellanos}, {Paardekooper},
  {Linnartz}  \& {Tielens}}{{Zhen} et~al.}{2014}]{Zhen:formation_lab}
{Zhen} J.,  {Castellanos} P.,  {Paardekooper} D.~M.,  {Linnartz} H.,
  {Tielens} A.~G.~G.~M.,  2014, \mn@doi [\apjl] {10.1088/2041-8205/797/2/L30},
  \href {http://adsabs.harvard.edu/abs/2014ApJ...797L..30Z} {797, L30}

\makeatother
\end{thebibliography}

\bsp	


\appendix

\section{Line Fluxes} \label{ap_linetable}

To measure and identify the lines in the Tc~1 VLT X-Shooter spectrum, we use two numerical codes and available literature. The main tool to measure line fluxes and identify their carriers was {\sc alfa} \citep[Automated Line Fitting Algorithm;][]{2016MNRAS.456.3774W}. Given an observed spectrum, the code first fits a continuum by sorting the flux in windows of 100 data points and taking the 25$^{th}$ ranked point as an estimator of the continuum flux at the central wavelength of the window. Then, it breaks the continuum-subtracted spectrum into chunks of 400 data points, and creates synthetic spectra based on a representative catalogue of emission lines. The line of sight velocity and resolution, and the peak flux of each emission line are initially free parameters that are optimised by means of a genetic algorithm, which uses random mutations and breeding of the best fitting solutions to attain a good fit to the observations. Blends are subsequently flagged, and the signal to noise ratio of each line estimated from the root mean square of the nearby residuals. Lines with signal-to-noise greater than 3 are considered to be detections and are included in the final output line list.

We first performed the measurement and identification for the numerous lines in the Tc~1 X-Shooter spectrum with \textsc{alfa}. The code detected and identified correctly dozens of lines automatically. It also detected and identified telluric lines in the spectra. Several lines were however either not detected or not well fitted by \textsc{alfa}. The reasons for this include blended lines and lines with profiles that differ too much from Gaussians, as well as lines not in the original \textsc{alfa} line catalogue. To solve this last case, we extended the list with identifications based on the general literature. \textsc{alfa} detected and identified 180 nebular lines, with 110 of them very well fitted.

The remaining lines were then measured using the code {\sc herfit}\footnote{The code was developed by I. Aleman and is available upon request.}. This code is based on the {\sc mpfit} algorithm \citep{2009ASPC..411..251M, More1978} and fits the line with a Gaussian profile and the underlying continuum emission with a polynomial. {\sc mpfit} is a widely used algorithm that uses the Levenberg-Marquardt technique to solve the least-squares problem. Differently from {\sc alfa}, {\sc herfit} needs the user to tailor the extractions for the continuum around each line and does not identify the line carriers. On the other hand, the user has more control over the choice of the continuum interval, which in many cases was important to produce a better fit to the line. {\sc herfit} also provides the line flux calculated by the integral under line, which is a better value in the case of lines that deviate significantly from a Gaussian profile.

\begin{figure}
   \centering
   \includegraphics[width=\columnwidth]{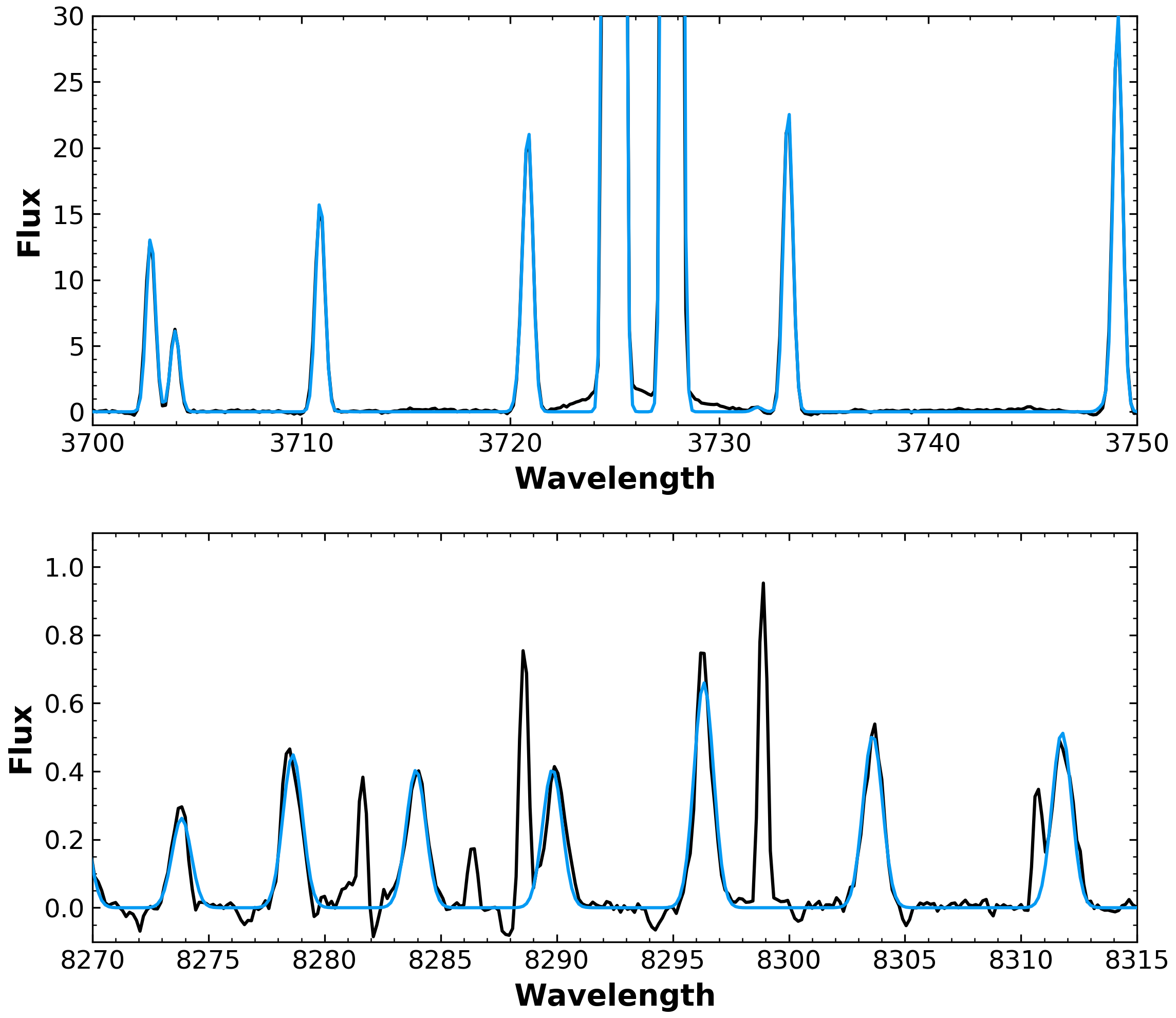}
   \caption{Two sections of the Tc~1 X-shooter spectrum for night 2, region 1 (black) fitted with the code {\sc Alfa} (blue). In this figure, the narrow lines absent in the blue spectrum were identified by {\sc Alfa} as telluric lines. Wavelength is given in Angstrons and flux in 10$^{-16}$~erg~s$^{-1}$~cm$^{-2}$~\AA$^{-1}$.}
   \label{alfafit}
\end{figure}

\begin{figure}
   \centering
   \includegraphics[width=\columnwidth]{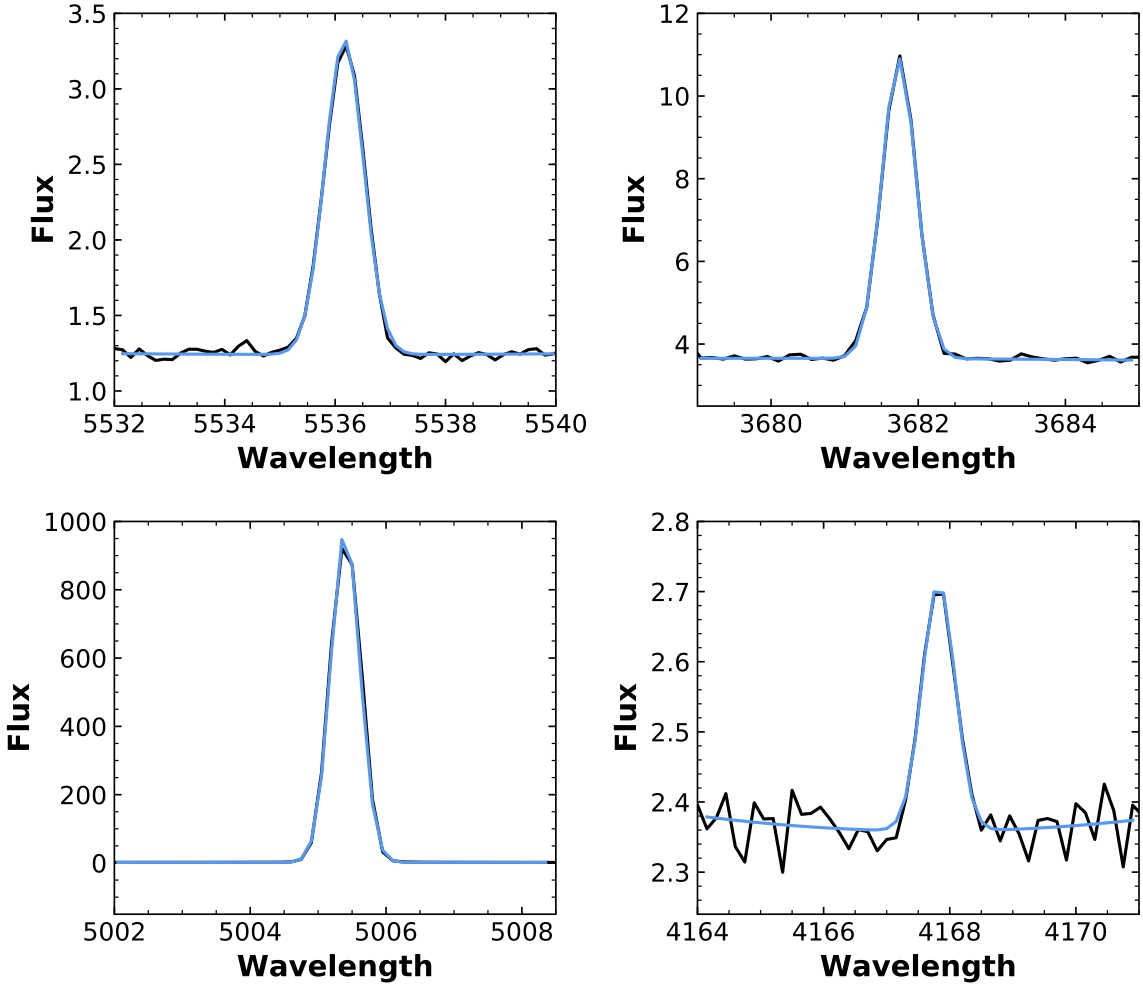}
   \caption{Examples of fits performed with the {\sc HerFit} code (blue curve). The black curve is the observed spectrum and the blue curve is the fitted continuum. Wavelength is given in Angstrons and flux in 10$^{-16}$~erg~s$^{-1}$~cm$^{-2}$~\AA$^{-1}$.}
   \label{herfit}
\end{figure}

\begin{figure}
   \centering
   \includegraphics[width=\columnwidth]{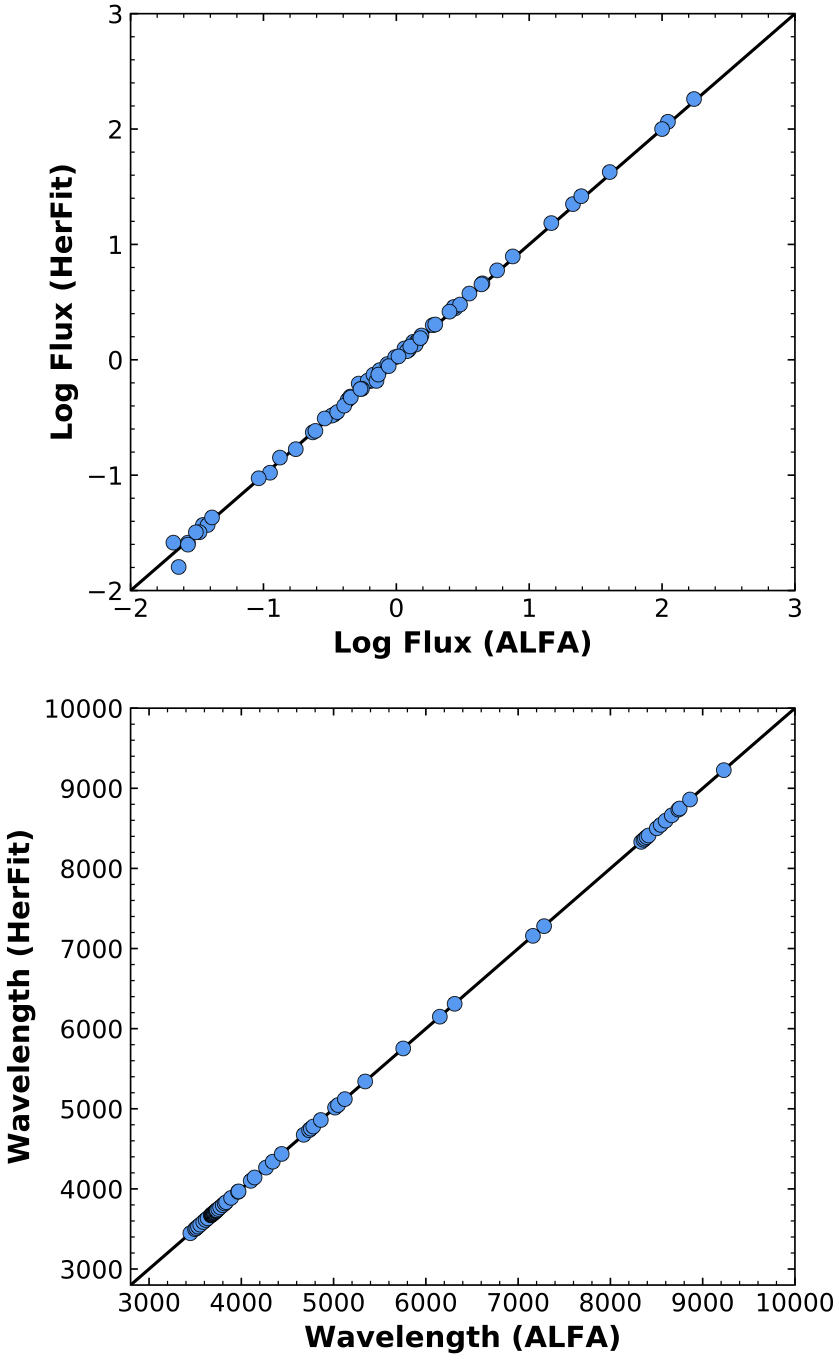}
   \caption{Comparison of {\sc HerFit} and {\sc Alfa} measurements. Dots indicate the measurements and the solid line the 1:1 curve. The top plot shows the measured central wavelength (in \AA) and the bottom plot shows the measured flux (H$\beta$~=~100).}
   \label{compherfit}
\end{figure}

Figures~\ref{alfafit} to \ref{compherfit} show the excellent quality and agreement of both {\sc alfa} and {\sc herfit} measurements. Examples of Gaussian line fits performed by the codes {\sc Alfa} and {\sc HerFit} are shown in Figs. \ref{alfafit} and \ref{herfit}, respectively. As different methods of measurements may produce different results \citep[see ][]{2016MNRAS.456.3774W}, we checked the consistence between the {\sc HerFit} and {\sc Alfa} measurements. In Fig.~\ref{compherfit}, we compare line fluxes of well fitted lines obtained with both codes. The central wavelengths and fluxes determined with both codes are very similar. Central wavelengths differ in less than 0.04 per cent in all cases and fluxes differs in less than 10 per cent in the majority of cases. Only three cases of higher differences in fluxes are seen (between 10 and 50 per cent) and all in lines with fluxes less than 1 per cent of H$\beta$.

As mentioned above, we could not extract a sky spectrum from the observations to properly eliminate the telluric lines. However, many of the intense telluric emission features were easily distinguished from the nebular emission lines. The telluric lines have different widths (typically much narrower) and are at the laboratory wavelength. To be sure we identified all the telluric lines and the nebular lines potentially affected by them, in addition to the {\sc alfa} identification, we also used the sky line models of \citet{2006JGRA..11112307C}. Nebular lines severely contaminated with sky emission features were eliminated from our analysis.

Tables~\ref{lines_uvb} and \ref{lines_vis} list the line fluxes measured in the Tc~1 X-Shooter UVB and VIS arms integrated spectra, respectively. We identified and measured 230 different lines from 14 elements (20 ions). The observed wavelengths have been corrected from the relative velocity of -87~km~s$^{-1}$ (determined by the \textsc{alfa} code). Note that in X-Shooter the fluxes in each arm are obtained with slits of different widths, 0.5~arcsec for UVB and 0.4~arcsec for VIS.


\begin{table}
\centering
\caption{X-Shooter UVB arm line fluxes.}
\label{lines_uvb}
\begin{tabular}{lrrrrl}

\hline
Ion & $\lambda _{rest}$ & $\lambda _{obs}$ & F$_\lambda^{(a)}$& I$_\lambda^{(a)}$& Comment\\
\hline

He~\textsc{I}  & 3258.27 & 3258.30 & 0.149 & 0.223 & \\ 
He~\textsc{I}  & 3296.77 & 3296.76 & 0.281 & 0.415 & \\ 
He~\textsc{I}  & 3354.56 & 3354.54 & 0.256 & 0.371 & \\ 
He~\textsc{I}  & 3447.59 & 3447.59 & 0.430 & 0.610 & \\ 
He~\textsc{I}  & 3478.97 & 3478.92 & 0.062 & 0.087 & \\ 
He~\textsc{I}  & 3487.77 & 3487.72 & 0.100 & 0.141 & \\ 
He~\textsc{I}  & 3498.69 & 3498.63 & 0.112 & 0.157 & \\ 
He~\textsc{I}  & 3512.52 & 3512.51 & 0.142 & 0.199 & \\ 
He~\textsc{I}  & 3530.50 & 3530.51 & 0.175 & 0.244 & \\ 
He~\textsc{I}  & 3554.42 & 3554.43 & 0.235 & 0.326 & \\ 
He~\textsc{I}  & 3587.28 & 3587.28 & 0.341 & 0.470 & \\ 
He~\textsc{I}  & 3613.64 & 3613.65 & 0.637 & 0.875 & \\ 
He~\textsc{I}  & 3634.25 & 3634.25 & 0.480 & 0.657 & \\ 
H27  & 3666.10 & 3666.10 & 0.433 & 0.589 & \\ 
H26  & 3667.68 & 3667.67 & 0.487 & 0.662 & \\ 
H25  & 3669.46 & 3669.45 & 0.466 & 0.634 & \\ 
H24  & 3671.48 & 3671.47 & 0.622 & 0.845 & \\ 
H23  & 3673.74 & 3673.76 & 0.660 & 0.897 & \\ 
H22  & 3676.36 & 3676.36 & 0.743 & 1.009 & \\ 
H21  & 3679.36 & 3679.35 & 0.814 & 1.105 & \\ 
H20  & 3682.81 & 3682.81 & 0.918 & 1.245 & \\ 
H19  & 3686.83 & 3686.83 & 1.050 & 1.423 & \\ 
H18  & 3691.56 & 3691.55 & 1.250 & 1.693 & \\ 
H17  & 3697.15 & 3697.15 & 1.430 & 1.935 & \\ 
H16  & 3703.86 & 3703.87 & 1.550 & 2.094 & \\ 
He~\textsc{I}  & 3705.02 & 3705.03 & 0.709 & 0.958 & \\ 
H15  & 3711.97 & 3711.97 & 1.990 & 2.685 & \\ 
H14  & 3721.94 & 3721.95 & 2.770 & 3.730 &  +[S~\textsc{III}]?\\

[O~\textsc{II}]  & 3726.03 & 3726.04 & 174.00 & 234.14 & (b)\\ 

[O~\textsc{II}]  & 3728.82 & 3728.83 & 110.70 & 148.89 & (b)\\ 
H13  & 3734.37 & 3734.38 & 2.700 & 3.628 & \\ 
H12  & 3750.15 & 3750.16 & 3.550 & 4.755 & \\ 
H11  & 3770.63 & 3770.63 & 4.450 & 5.937 & \\ 
He~\textsc{I}  & 3784.89 & 3784.89 & 0.050 & 0.067 & \\ 
H10  & 3797.90 & 3797.90 & 5.730 & 7.604 & \\ 
He~\textsc{I}  & 3805.74 & 3805.72 & 0.053 & 0.070 & \\ 
He~\textsc{I}  & 3819.62 & 3819.63 & 1.220 & 1.612 & \\ 
Mg~\textsc{I}  & 3829.35 & 3829.68 & 0.020 & 0.026 & \\ 
Mg~\textsc{I}  & 3831.68 & 3831.64 & 0.056 & 0.074 & \\ 
He~\textsc{I}  & 3833.55 & 3833.55 & 0.068 & 0.090 & \\ 
H9  & 3835.39 & 3835.38 & 7.530 & 9.916 & \\ 
He~\textsc{I}  & 3838.10 & 3838.32 & 0.067 & 0.088 & \\ 

~+Mg~\textsc{I}&&&&&\\

Si~\textsc{II}  & 3856.02 & 3856.01 & 0.093 & 0.122 & \\ 
Si~\textsc{II}  & 3862.60 & 3862.59 & 0.047 & 0.062 & \\ 
He~\textsc{I}  & 3867.53 & 3867.48 & 0.078 & 0.102 & \\ 

[Ne~\textsc{III}]  & 3868.75 & 3868.76 & 0.456 & 0.596 & \\ 
He~\textsc{I}  & 3871.83 & 3871.81 & 0.103 & 0.135 & \\ 
H8   & 3888.65 & 3888.71 & 21.400 & 27.860 & \\ 

~+He~\textsc{I}&&&&&\\

C~\textsc{II}  & 3918.98 & 3918.94 & 0.293 & 0.379 & \\ 
C~\textsc{II}  & 3920.69 & 3920.66 & 0.553 & 0.715 & \\ 
He~\textsc{I}  & 3926.54 & 3926.56 & 0.150 & 0.194 & \\ 
He~\textsc{I}  & 3964.73 & 3964.73 & 1.190 & 1.522 & \\ 

[Ne~\textsc{III}]  & 3967.46 & 3967.47 & 0.137 & 0.175 & \\ 
H7  & 3970.07 & 3970.09 & 15.300 & 19.547 & \\ 
He~\textsc{I}  & 4009.26 & 4009.25 & 0.199 & 0.252 & \\ 
He~\textsc{I}  & 4026.08 & 4026.21 & 2.120 & 2.671 & \\ 

~+N~\textsc{II}&&&&&\\

[S~\textsc{II}]  & 4068.60 & 4068.60 & 0.443 & 0.552 & \\ 
O~\textsc{II}  & 4072.16 & 4072.16 & 0.025 & 0.031 & \\ 

[S~\textsc{II}]  & 4076.35 & 4076.30 & 0.226 & 0.281 & +O II?\\ 
H6  & 4101.74 & 4101.74 & 24.700 & 30.515 & \\ 
\hline
\end{tabular}
\end{table}

\begin{table}
\centering
\contcaption{UVB X-Shooter Arm Line Fluxes.}
\begin{tabular}{lrrrrl}

\hline
Ion & $\lambda _{rest}$ & $\lambda _{obs}$ & F$_\lambda^{(a)}$& I$_\lambda^{(a)}$& Comment\\
\hline
O~\textsc{II}  & 4110.78 & 4110.75 & 0.014 & 0.017 & \\ 
He~\textsc{I}  & 4120.84 & 4120.85 & 0.204 & 0.251 & \\ 

~+O~\textsc{II}&&&&&\\

Si~\textsc{II}  & 4128.53 & 4128.55 & 0.034 & 0.042 & \\ 
He~\textsc{I}  & 4143.76 & 4143.77 & 0.326 & 0.398 & \\ 
O~\textsc{II}  & 4153.30 & 4153.40 & 0.017 & 0.021 & \\ 
He~\textsc{I}   & 4168.97 & 4168.99 & 0.052 & 0.063 & \\ 

~+O~\textsc{II}&&&&&\\

C~\textsc{II}  & 4267.15 & 4267.16 & 0.551 & 0.650 & \\ 
O~\textsc{II}  & 4317.14 & 4317.19 & 0.013 & 0.015 & \\ 
H5  & 4340.47 & 4340.47 & 40.400 & 46.665 & \\ 
O~\textsc{II}  & 4345.55 & 4345.55 & 0.026 & 0.030 & \\ 
O~\textsc{II}  & 4349.43 & 4349.42 & 0.019 & 0.022 & \\ 

[O~\textsc{III}]  & 4363.21 & 4363.20 & 0.403 & 0.462 & \\ 
N~\textsc{III}  & 4366.89 & 4366.96 & 0.013 & 0.015 & \\ 
He~\textsc{I}  & 4387.93 & 4387.94 & 0.606 & 0.690 & \\ 
O~\textsc{II}  & 4414.90 & 4414.92 & 0.036 & 0.041 & \\ 
O~\textsc{II}  & 4416.97 & 4416.99 & 0.027 & 0.031 & \\ 
He~\textsc{I}  & 4437.55 & 4437.58 & 0.092 & 0.103 & \\ 
Ne~\textsc{II}  & 4457.05 & 4457.07 & 0.025 & 0.028 & \\ 

~+N~\textsc{II}&&&&&\\

He~\textsc{I}  & 4471.50 & 4471.51 & 4.850 & 5.393 & \\ 
O~\textsc{II}  & 4477.90 & 4477.90 & 0.012 & 0.013 & \\ 
Mg~\textsc{II}  & 4481.21 & 4481.21 & 0.012 & 0.013 & \\ 
O~\textsc{II}   & 4491.23 & 4491.07 & 0.019 & 0.021 & \\ 

~+C~\textsc{II}&&&&&\\

Mg~\textsc{I}]  & 4562.60 & 4562.61 & 0.034 & 0.037 & \\ 
Mg~\textsc{I}]  & 4571.10 & 4571.12 & 0.068 & 0.073 & \\ 
O~\textsc{II}  & 4590.97 & 4590.98 & 0.015 & 0.016 & \\ 
N~\textsc{II}  & 4601.48 & 4601.46 & 0.030 & 0.032 & \\ 
N~\textsc{II}  & 4607.16 & 4607.02 & 0.050 & 0.053 &  +[Fe~\textsc{III}]?\\ 
N~\textsc{II}  & 4613.68 & 4613.67 & 0.026 & 0.028 &  +[O~\textsc{II}]?\\ 
C~\textsc{II}  & 4620.26 & 4620.25 & 0.013 & 0.014 & \\ 
N~\textsc{II}   & 4621.39 & 4621.38 & 0.035 & 0.037 & \\

~+O~\textsc{II}&&&&&\\

N~\textsc{II}  & 4630.54 & 4630.53 & 0.101 & 0.107 & \\ 
N~\textsc{III}  & 4634.14 & 4633.98 & 0.068 & 0.072 & \\ 
O~\textsc{II}  & 4647.80 & 4647.83 & 0.007 & 0.007 & \\ 
O~\textsc{II}  & 4649.13 & 4649.16 & 0.037 & 0.039 & \\ 
O~\textsc{II}  & 4650.84 & 4650.85 & 0.021 & 0.022 & \\ 

[Fe~\textsc{III}]  & 4658.10 & 4658.10 & 0.257 & 0.271 & \\ 
O~\textsc{II}  & 4661.63 & 4661.66 & 0.024 & 0.025 & \\ 
O~\textsc{II}  & 4676.24 & 4676.29 & 0.016 & 0.017 & \\ 
O~\textsc{II}  & 4699.22 & 4699.05 & 0.014 & 0.015 & \\ 

[Fe~\textsc{III}]  & 4701.62 & 4701.58 & 0.080 & 0.083 & \\ 
O~\textsc{II}  & 4705.35 & 4705.33 & 0.019 & 0.020 & \\ 
He~\textsc{I}  & 4713.17 & 4713.17 & 0.544 & 0.565 & \\ 

[Fe~\textsc{III}]  & 4733.91 & 4733.91 & 0.027 & 0.028 & \\ 
O~\textsc{II}  & 4752.96 & 4752.96 & 0.021 & 0.022 & \\ 

[Fe~\textsc{III}]  & 4754.69 & 4754.69 & 0.046 & 0.047 & \\ 
O~\textsc{I}  & 4756.70 & 4756.70 & 0.026 & 0.027 & \\ 

[Fe~\textsc{III}]  & 4769.40 & 4769.40 & 0.025 & 0.026 & \\ 
N~\textsc{II}  & 4779.72 & 4779.71 & 0.035 & 0.036 & \\ 
N~\textsc{II}  & 4788.13 & 4788.12 & 0.034 & 0.035 & \\ 
N~\textsc{II}  & 4789.57 & 4789.56 & 0.030 & 0.031 & \\ 
N~\textsc{II}  & 4803.29 & 4803.28 & 0.048 & 0.049 & (c)\\ 
S~\textsc{II}  & 4815.55 & 4815.54 & 0.018 & 0.018 & \\ 
H4  & 4861.33 & 4861.31 & 100.000 & 100.000 & (b)\\ 

[Fe~\textsc{III}]  & 4881.11 & 4881.05 & 0.101 & 0.101 & \\ 
O~\textsc{II}  & 4906.83 & 4906.82 & 0.012 & 0.012 & \\ 
He~\textsc{I}  & 4921.93 & 4921.93 & 1.402 & 1.381 & \\ 
O~\textsc{II}  & 4924.53 & 4924.46 & 0.013 & 0.013 & \\ 

[O~\textsc{III}]  & 4958.91 & 4958.92 & 37.400 & 36.527 & (b)\\ 

\hline
\end{tabular}
\end{table}

\begin{table}
\centering
\contcaption{UVB X-Shooter Arm Line Fluxes.}

\begin{tabular}{lrrrrl}

\hline
Ion & $\lambda _{rest}$ & $\lambda _{obs}$ & F$_\lambda^{(a)}$& I$_\lambda^{(a)}$& Comment\\
\hline

[Fe~\textsc{III}]  & 4985.46 & 4985.83 & 0.023 & 0.022 & \\ 

[Fe~\textsc{III}]  & 4987.38 & 4987.30 & 0.033 & 0.032 & \\ 
N~\textsc{II}  & 4994.35 & 4994.36 & 0.044 & 0.043 & \\ 

[O~\textsc{III}]  & 5006.84 & 5006.85 & 113.000 & 109.143 & (b)\\ 
He~\textsc{I}  & 5015.68 & 5015.67 & 3.020 & 2.911 & \\ 
C~\textsc{II}   & 5032.00 & 5031.99 & 0.064 & 0.061 & \\ 

~+S~\textsc{II}&&&&&\\

O~\textsc{II}  & 5035.78 & 5035.77 & 0.061 & 0.059 & \\ 

Si~\textsc{II}  & 5041.02 & 5040.93 & 0.073 & 0.070 & \\ 

N~\textsc{II}  & 5045.00 & 5045.11 & 0.041 & 0.039 & \\ 
He~\textsc{I}  & 5047.74 & 5047.73 & 0.246 & 0.235 & \\ 
Si~\textsc{II}  & 5055.98 & 5055.98 & 0.136 & 0.130 & (d) 0.138\\ 
C~\textsc{II}  & 5121.82 & 5121.80 & 0.038 & 0.036 & \\ 

[Ar~\textsc{III}]  & 5191.82 & 5191.71 & 0.032 & 0.030 & \\ 

[N~\textsc{I}]  & 5198.26 & 5197.83 & 0.017 & 0.016 & \\ 

[Fe~\textsc{III}]  & 5270.40 & 5270.52 & 0.147 & 0.134 & \\ 
C~\textsc{II}  & 5342.38 & 5342.37 & 0.033 & 0.030 & \\ 
S~\textsc{II}  & 5453.83 & 5454.03 & 0.019 & 0.017 & \\ 
C~\textsc{II}  & 5480.08 & 5480.16 & 0.011 & 0.010 & \\ 
N~\textsc{II}  & 5495.66 & 5495.61 & 0.019 & 0.017 & \\ 
\hline
\multicolumn{6}{l}{$^{a}$ Fluxes normalised to H$\beta$ = 100. 
$^{b}$ Lines with broad wings.}\\
\multicolumn{6}{l}{$^{c}$ Blended line. $^{d}$ Line with a profile that deviates}\\
\multicolumn{6}{l}{significantly from a Gaussian curve. The number that}\\
\multicolumn{6}{l}{follows is the flux obtained by the integration of the}\\
\multicolumn{6}{l}{(continuum subtracted) spectrum line in the 3-sigma range.}\\
\multicolumn{6}{l}{See text for discussion.}\\

\end{tabular}
\end{table}



\begin{table}
\centering
\caption{X-Shooter VIS arm line fluxes.} 
\label{lines_vis}

\begin{tabular}{lrrrrl}

\hline
Ion & $\lambda _{rest}$ & $\lambda _{obs}$ & F$_\lambda^{(a)}$& I$_\lambda^{(a)}$& Comment\\

\hline

N~\textsc{II}  & 5666.63 & 5666.60 & 0.070 & 0.060 & \\ 
N~\textsc{II}  & 5676.02 & 5676.02 & 0.031 & 0.026 & \\ 
N~\textsc{II}  & 5679.56 & 5679.55 & 0.095 & 0.081 & \\ 
N~\textsc{II}  & 5686.21 & 5686.20 & 0.023 & 0.020 & \\ 
C~\textsc{III}?  & 5695.92 & 5695.76 & 0.251 & 0.214 & \\ 
N~\textsc{II}  & 5710.77 & 5710.78 & 0.015 & 0.013 & \\ 

[N~\textsc{II}]  & 5754.60 & 5754.56 & 1.420 & 1.198 & \\ 
He~\textsc{I}  & 5875.66 & 5875.65 & 18.400 & 15.262 & \\ 
Na~\textsc{I}  & 5889.95 & 5889.86 & 0.152 & 0.126 & \\ 
Na~\textsc{I}  & 5895.92 & 5896.04 & 0.041 & 0.034 & \\ 
N~\textsc{II}  & 5927.81 & 5927.81 & 0.035 & 0.029 & \\ 
N~\textsc{II}  & 5931.78 & 5931.77 & 0.047 & 0.039 & \\ 
N~\textsc{II}  & 5941.65 & 5941.66 & 0.086 & 0.071 & \\ 
S~\textsc{III}  & 5978.97 & 5978.75 & 0.033 & 0.027 & \\ 
C~\textsc{II}  & 6151.43 & 6151.28 & 0.031 & 0.025 & \\ 

[O~\textsc{I}]  & 6300.30 & 6300.23 & 0.111 & 0.087 & (d) 0.140\\ 

[S~\textsc{III}]  & 6312.10 & 6312.03 & 0.732 & 0.574 & \\ 
Si~\textsc{II}  & 6347.10 & 6347.02 & 0.109 & 0.085 & (d) 0.106\\ 

[O~\textsc{I}]  & 6363.78 & 6363.79 & 0.042 & 0.033 & \\ 
S~\textsc{III}  & 6371.38 & 6371.31 & 0.049 & 0.038 & (d) 0.051\\ 
C~\textsc{II}  & 6461.95 & 6461.80 & 0.078 & 0.060 & \\ 

[N~\textsc{II}]  & 6527.11 & 6527.07 & 0.026 & 0.020 & \\ 

[N~\textsc{II}]  & 6548.10 & 6548.09 & 46.100 & 35.113 & (d)\\ 
H3  & 6562.77 & 6562.76 & 398.300 & 302.828 & (b); (d) 417.7\\ 
C~\textsc{II}  & 6578.05 & 6578.02 & 1.320 & 1.002 & \\ 

[N~\textsc{II}]  & 6583.50 & 6583.40 & 146.800 & 111.330 & (b); (d) 151.7\\ 
He~\textsc{I}  & 6678.16 & 6678.13 & 5.850 & 4.386 & \\ 

[S~\textsc{II}]  & 6716.44 & 6716.40 & 3.090 & 2.306 & (d) 3.45\\ 

[S~\textsc{II}]  & 6730.82 & 6730.78 & 4.740 & 3.531 & (d) 5.33\\ 

[Kr~\textsc{III}]  & 6826.70 & 6826.79 & 0.031 & 0.023 & \\ 

\hline
\end{tabular}
\end{table}

\begin{table}
\centering
\contcaption{VIS X-Shooter Arm Line Fluxes.}

\begin{tabular}{lrrrrl}

\hline
Ion & $\lambda _{rest}$ & $\lambda _{obs}$ & F$_\lambda^{(a)}$& I$_\lambda^{(a)}$& Comment\\

\hline

He~\textsc{I}  & 7062.26 & 7062.28 & 0.028 & 0.020 & \\ 
He~\textsc{I}  & 7065.25 & 7065.19 & 6.570 & 4.697 & (d) 6.43\\ 

[Ar~\textsc{III}]  & 7135.80 & 7135.74 & 11.700 & 8.292 & \\ 
He~\textsc{I}  & 7160.56 & 7160.51 & 0.041 & 0.029 & \\ 

C~\textsc{II}  & 7231.32 & 7231.31 & 0.706 & 0.495 & (d) 0.720\\ 
He~\textsc{I}  & 7281.35 & 7281.32 & 1.350 & 0.940 & \\ 
He~\textsc{I}  & 7298.04 & 7298.05 & 0.063 & 0.044 & \\ 

[O~\textsc{II}]  & 7318.92 & 7318.97 & 2.070 & 1.434 & \\ 

[O~\textsc{II}]  & 7319.99 & 7320.04 & 6.750 & 4.677 & \\ 

[O~\textsc{II}]  & 7329.67 & 7329.72 & 3.700 & 2.560 & \\ 

[O~\textsc{II}]  & 7330.73 & 7330.78 & 3.670 & 2.539 & \\ 
He~\textsc{I}  & 7499.84 & 7499.81 & 0.070 & 0.047 & \\ 
C~\textsc{II}  & 7519.87 & 7519.64 & 0.066 & 0.045 & (d) 0.068\\ 

[Ar~\textsc{III}]  & 7751.09 & 7751.07 & 2.980 & 1.959 & \\ 
He~\textsc{I}  & 7816.16 & 7816.09 & 0.097 & 0.063 & \\ 
P34  & 8267.94 & 8267.92 & 0.119 & 0.074 & \\ 
P33  & 8271.93 & 8271.93 & 0.125 & 0.077 & \\ 
P32  & 8276.31 & 8276.16 & 0.104 & 0.064 & \\ 
P31  & 8281.12 & 8280.96 & 0.205 & 0.127 & \\ 
P30  & 8286.43 & 8286.40 & 0.164 & 0.102 & \\ 
P29  & 8292.31 & 8292.38 & 0.159 & 0.098 & \\ 
P27  & 8306.11 & 8306.05 & 0.207 & 0.128 & \\ 
P26  & 8314.26 & 8314.24 & 0.213 & 0.131 & (c)\\ 
P25  & 8323.42 & 8323.26 & 0.208 & 0.128 & \\ 
P24  & 8333.78 & 8333.76 & 0.289 & 0.178 & \\ 
P22  & 8359.00 & 8358.97 & 0.359 & 0.221 & \\ 
He~\textsc{I}  & 8361.74 & 8361.71 & 0.140 & 0.086 & \\ 
P21  & 8374.48 & 8374.46 & 0.405 & 0.248 & \\ 
P20  & 8392.40 & 8392.37 & 0.454 & 0.278 & \\ 
P19  & 8413.32 & 8413.29 & 0.537 & 0.328 & \\ 
He~\textsc{I}  & 8421.99 & 8421.98 & 0.016 & 0.010 & \\ 
P18  & 8437.95 & 8437.93 & 0.647 & 0.394 & \\ 
He~\textsc{I}  & 8444.69 & 8444.47 & 0.040 & 0.024 & \\ 
He~\textsc{I}  & 8486.31 & 8486.24 & 0.031 & 0.019 & \\ 

P16  & 8502.48 & 8502.44 & 0.876 & 0.531 & \\ 
He~\textsc{I}  & 8518.04 & 8518.00 & 0.033 & 0.020 & \\ 
He~\textsc{I}  & 8528.99 & 8529.01 & 0.042 & 0.025 & \\ 
P15  & 8545.38 & 8545.35 & 1.038 & 0.626 & \\ 

[Cl~\textsc{II}]  & 8578.70 & 8578.67 & 0.169 & 0.102 & (d) 0.215\\ 
He~\textsc{I}  & 8582.24 & 8582.21 & 0.100 & 0.060 & \\ 
P14  & 8598.39 & 8598.35 & 1.274 & 0.765 & \\ 
He~\textsc{I}  & 8648.25 & 8648.21 & 0.055 & 0.033 & \\ 
P13  & 8665.02 & 8664.98 & 1.516 & 0.905 & \\ 
He~\textsc{I}  & 8733.43 & 8733.43 & 0.074 & 0.044 & \\ 
He~\textsc{I}  & 8736.04 & 8736.03 & 0.027 & 0.016 & \\ 

P12  & 8750.47 & 8750.46 & 1.962 & 1.163 & \\ 
He~\textsc{I}  & 8776.79 & 8776.73 & 0.088 & 0.052 & \\ 
He~\textsc{I}  & 8845.37 & 8845.34 & 0.092 & 0.054 & \\ 
P11  & 8862.78 & 8862.76 & 2.511 & 1.475 & \\ 
He~\textsc{I}  & 8914.77 & 8914.76 & 0.041 & 0.024 & \\ 
He~\textsc{I}  & 8930.77 & 8930.68 & 0.026 & 0.015 & \\ 
He~\textsc{I}  & 8997.02 & 8996.96 & 0.112 & 0.065 & \\ 
P10  & 9014.91 & 9014.98 & 2.627 & 1.526 & \\ 

[S~\textsc{III}]  & 9068.60 & 9068.88 & 20.340 & 11.773 & \\ 
He~\textsc{I}  & 9210.33 & 9210.31 & 0.166 & 0.095 & \\ 
He~\textsc{I}  & 9213.23 & 9213.18 & 0.066 & 0.038 & \\ 
P9  & 9229.01 & 9229.02 & 4.374 & 2.505 & (d) 4.65\\ 

?  & 9236.39 & 9236.26 & 0.051 & 0.029 & \\ 
P8  & 9545.97 & 9545.73 & 4.200 & 2.360 & (b); (d) 4.71\\ 
He~\textsc{I}  & 9603.44 & 9603.38 & 0.082 & 0.046 & \\ 
O~\textsc{II}  & 9903.39 & 9903.43 & 0.213 & 0.117 & \\ 
~+ C~\textsc{II}&&&&&\\

P7  & 10049.30 & 10049.32 & 8.600 & 4.701 & (b)\\ 

\hline
\end{tabular}
\end{table}

\section{Atomic data} \label{ap_atomicdata}

Table~\ref{atomicdatatable} lists the atomic data used in \textsc{Neat} and \textsc{2D\_Neb} codes for the calculations in this work.

\begin{table}
\caption{Atomic data used in {\sc Neat}.}
\begin{tabular}{ll}
\hline
\multicolumn{2}{c}{Collisionally Excited Lines}\\
\hline
Ion & Collision Strengths  \\
\hline
O$^{+}$ & \citet{kisielius2009}  \\
S$^{+}$ & \citet{tayal2010}  \\
S$^{2+}$ & \citet{mendoza1983}  \\

\hline
Ion &  Transition Probabilities \\
\hline
O$^{+}$ &  \citet{zeippen1982} \\
S$^{+}$ &  \citet{podobedova2009} \\
S$^{2+}$ &  \citet{mendoza1982} \\

\hline
\multicolumn{2}{c}{Recombination Lines}\\
\hline
Ion & Recombination Coefficients \\
\hline
H$^{+}$ & \citet{Storey_Hummer_1995} \\
He$^{+}$ & \citet{porter2013} \\
He$^{2+}$ & \citet{Storey_Hummer_1995} \\
C$^{2+}$ & \citet{davey2000} \\
C$^{3+}$ & \citet{pequignot1991} \\
N$^{2+}$ & \citet{escalante1990} \\
N$^{3+}$ & \citet{pequignot1991} \\
O$^{2+}$ (3s--3p) &  \citet{storey1994} \\
O$^{2+}$ (3p--3d, 3d--4f) & \citet{liu1995} \\
Ne$^{2+}$ (3s--3p) & \citet{kisielius1998} \\
Ne$^{2+}$ (3d-4f) & Storey (unpublished) \\
\hline
\multicolumn{2}{l}{Collisional data for heavy element ions not listed here is}\\
\multicolumn{2}{l}{from the {\sc Chianti} database version 7.0 (\citealt{1997A&AS..125..149D},}\\
\multicolumn{2}{l}{\citealt{landi2012})}\\
\label{atomicdatatable}
\end{tabular}
\end{table}

\section{Line Diagnostics} \label{ap_diag}

{\sc neat} and {\sc 2d\_neb} derive electronic densities and temperatures using typical line ratio diagnostics. Calculations are based on solving the statistical equilibrium equations for the level populations in an N-level atomic model with an iterative method.

To derive $n_\mathrm{e}$ or $T_\mathrm{e}$ from the emission line ratio of a given ionic species, {\sc neat} and {\sc 2d\_neb} both adopt the $T_\mathrm{e}$ or $n_\mathrm{e}$ given by the line diagnostic of an ion with similar ionization potential as input, with the exception of the input parameter used to derive $T_\mathrm{e}$~[O~{\sc iii}], for which we use $n_\mathrm{e} (Low)$ due to the lack of a better alternative.  Table \ref{diagnostic} lists the line ratio and the corresponding pair $n_\mathrm{e}$/$T_\mathrm{e}$ used as input in the calculation for each diagnostic.

Ratios of O~{\sc ii} recombination lines may also be used to estimate the temperature and density; the ratio 4649~\AA/4089~\AA\ is mainly sensitive to the electron temperature, while 4649~\AA/4662~\AA\ is mainly sensitive to the electron density (\citealt{storey2017}). Unfortunately, we do not detect the 4089~\AA\ line. The ratio of 4649~\AA/4662~\AA\ is 1.54, which, depending on the electron temperature in the gas emitting O~{\sc ii} recombination lines, is consistent with electron densities between 250 and 1\,500~cm$^{3}$. Given this weak constraint, and the insensitivity of recombination line abundance ratios to assumed physical conditions, we adopt the medium ionization zone temperatures and densities to calculate RL abundances.

\begin{table}
\centering
\caption{List of line ratio diagnostics, input electronic densities and temperatures, and emission lines used.} \label{diagnostic}
\begin{tabular}{lccc}
\hline
$n_\mathrm{e}$/$T_\mathrm{e}$ & Line Ratio & Input $n_\mathrm{e}$/$T_\mathrm{e}$ & Input $n_\mathrm{e}$/$T_\mathrm{e}$\\
Output & & (Integrated) & (Pixel by pixel)\\
\hline
$n_\mathrm{e}$[S~{\sc ii}]& $\dfrac{6717 \textrm{\AA}}{6731 \textrm{\AA}}$ & $T_\mathrm{e}$(Low) & $T_\mathrm{e}$[N~{\sc ii}] \\[9pt]

$n_\mathrm{e}$[O~{\sc ii}]& $\dfrac{3726 \textrm{\AA}}{3729 \textrm{\AA}}$ & $T_\mathrm{e}$(Low) & $T_\mathrm{e}$[O~{\sc ii}]\\[9pt] 

$T_\mathrm{e}$[N~{\sc ii}]& $\dfrac{6548 \textrm{\AA} + 6583 \textrm{\AA}}{5755 \textrm{\AA}}$ & $n_\mathrm{e}$(Low) & $n_\mathrm{e}$[S~{\sc ii}]\\[9pt]

$T_\mathrm{e}$[O~{\sc ii}]& $\dfrac{3726 \textrm{\AA} + 3729 \textrm{\AA}}{7319 \textrm{\AA} + 7330 \textrm{\AA}}$ & $n_\mathrm{e}$(Low) & $n_\mathrm{e}$[O~{\sc ii}]\\[9pt]

$T_\mathrm{e}$[S~{\sc ii}]& $\dfrac{6717 \textrm{\AA} + 6731 \textrm{\AA}}{4068 \textrm{\AA} + 4075 \textrm{\AA}}$ & $n_\mathrm{e}$(Low) & -- \\[9pt] 

$T_\mathrm{e}$[Ar~{\sc iii}]& $\dfrac{7135 \textrm{\AA} + 7751 \textrm{\AA}}{5192 \textrm{\AA}}$ & $n_\mathrm{e}$(Low) & -- \\[9pt] 

$T_\mathrm{e}$[S~{\sc iii}]& $\dfrac{9069 \textrm{\AA} + 9531 \textrm{\AA}}{6312 \textrm{\AA}}$ & $n_\mathrm{e}$(Low) & -- \\[9pt] 

$T_\mathrm{e}$[O~{\sc iii}]& $\dfrac{4959 \textrm{\AA} + 5007 \textrm{\AA}}{4363 \textrm{\AA}}$ & $n_\mathrm{e}$(Low) & $n_\mathrm{e}$[O~{\sc ii}]\\[9pt]

$T_\mathrm{e}$(He A) & $\dfrac{5876 \textrm{\AA}}{4471 \textrm{\AA}}$ & 5000~cm$^{-3}$ & --\\[9pt]

$T_\mathrm{e}$(He B) & $\dfrac{6678 \textrm{\AA}}{4471 \textrm{\AA}}$ & 5000~cm$^{-3}$ & --\\[6pt]

\hline 

Abundance & Lines & Input $n_\mathrm{e}$/$T_\mathrm{e}$ & Input $n_\mathrm{e}$/$T_\mathrm{e}$ \\
Output && (Integrated) & (Pixel by pixel)\\

\hline

N$^{+}$/H$^{+}$& 6548\AA, 6583\AA &$n_\mathrm{e}$(Low), & $n_\mathrm{e}$[S~\textsc{ii}], \\

&&$T_\mathrm{e}$(Low) & $T_\mathrm{e}$[N~\textsc{ii}]\\

O$^{0}$/H$^{+}$& 6300\AA, 6363\AA &$n_\mathrm{e}$(Low),& --\\

&&$T_\mathrm{e}$(Low)& --\\

O$^{+}$/H$^{+}$&3726\AA, 3729\AA& $n_\mathrm{e}$(Low),& $n_\mathrm{e}$[O~\textsc{ii}], \\

&&$T_\mathrm{e}$(Low)&$T_\mathrm{e}$[N~\textsc{ii}]\\

O$^{2+}$/H$^{+}$ & 4959\AA, 5007\AA, & $n_\mathrm{e}$(Low), & $n_\mathrm{e}$[O~{\sc ii}], \\

&(4363\AA)$^{(a)}$ &$T_\mathrm{e}$(Medium)&$T_\mathrm{e}$[O~{\sc iii}]\\

Ne$^{2+}$/H$^{+}$& 3868\AA, 3967\AA & $n_\mathrm{e}$(Low),& --\\

&&$T_\mathrm{e}$(Medium) & --\\

Ar$^{2+}$/H$^{+}$& 7135\AA, 7751\AA & $n_\mathrm{e}$(Low), & --\\

&&$T_\mathrm{e}$(Medium) & --\\

S$^{+}$/H$^{+}$&6717\AA, 6731\AA & $n_\mathrm{e}$(Low), & $n_\mathrm{e}$[S~\textsc{ii}], \\

&&$T_\mathrm{e}$(Low)&$T_\mathrm{e}$[N~\textsc{ii}]\\

S$^{2+}$/H$^{+}$& 9069\AA, 9531\AA & $n_\mathrm{e}$(Low), & --\\

&&$T_\mathrm{e}$(Medium)&--\\

He$^{+}$/H$^{+}$&4471\AA, 5876\AA, & $n_\mathrm{e}$(Low), & $n_\mathrm{e}$[S~\textsc{ii}],  \\

&6678\AA&$T_\mathrm{e}$(Low)&$T_\mathrm{e}$[N~\textsc{ii}]\\
\hline
\multicolumn{4}{l}{(a) This line was considered in the pixel-by-pixel analysis,}\\
\multicolumn{4}{l}{but was not considered by {\sc Neat} in the investigation of the}\\
\multicolumn{4}{l}{integrated spectrum.}\\

\end{tabular}
\end{table}

\section{Intrinsic Line Widths} \label{Ap_intrinsic_widt}

We calculated the intrinsic line width using the expression

\begin{equation}
\sigma^2_{\rm int} = \sigma^2_{\rm obs} - \sigma^2_{\rm instr} - \sigma^2_{\rm therm} - \sigma^2_{\rm fs},
\end{equation}

\noindent where $\sigma_{\rm instr}$ is the instrumental broadening (33~km~s$^{-1}$ for the UVB arm; 17.5~km~s$^{-1}$ for the VIS arm); $\sigma_{\rm therm}$ is the thermal broadening (21.4~km~s$^{-1}$ for H~\textsc{i}; 5.4~km~s$^{-1}$ for [O~\textsc{iii}] and [O~\textsc{ii}] and 5.8~km~s$^{-1}$ for [N~\textsc{ii}] -- all evaluated at 10\,000~K) and $\sigma_{\rm fs}$ is fine structure broadening (relevant for H only) of 6.4~km~s$^{-1}$. For more details, we refer to \citet{2015MNRAS.452.2911A}.

\section{Tc~1 Parameters in the Literature} \label{Ap_param_lit}

\begin{table}
\centering
\caption{Tc~1 distances from the literature}
\label{literature_distance}
\begin{tabular}{cl}
\hline
Distance (kpc) & Reference\\
\hline
2.1 & \citet{1971ApJS...22..319C} \\
2.5 & \citet{Milne_Aller_1975}\\
2.5 & \citet{1976AJ.....81..407C} \\
2.0 & \citet{Acker1978}\\
0.6 & \citet{1982ApJ...260..612D}\\
1.0 & \citet{Maciel1984}\\
1.4 & \citet{Amnuel_etal_1984}\\
3.8 & \citet{Mendez1988} \\
2.6 & \citet{Cahn_etal_1992}\\
2.8 & \citet{Martin1994}\\
3.0 & \citet{1995ApJ...452..286S}\\
1.5 & \citet{1995ApJS...98..659Z}\\
4.1 & \citet{Gorny_etal_1997} \\
3.1 & \citet{Gorny_etal_1997} \\
3.0 & \citet{1998AJ....115.1989T}\\
2.3 & \citet{Cazetta_Maciel_2001}\\
3.7 & \citet{Pauldrach_etal_2004}\\
2.7 & \citet{2008ApJ...689..194S}\\
1.8 & \citet{Pottasch_etal_2011}\\
2.1-2.2 & This Work \\
\hline
\end{tabular}
\end{table} 

\begin{table*}
\centering
\caption{Tc~1 central star data from the literature.}
\label{param_literature_star}
\begin{tabular}{ccll}
\hline
$T_{\star}$ (10$^3$~K) & $L_{\star}$ (10$^3 L_{\sun}$) & Reference & Method\\
\hline
28.2  & 6.6   & \citet{Gorny_etal_1997} & $T_{Z}$(H~\textsc{I})\\
33.0  & <25.1 & \citet{Kudritzki_etal_1997} & Atmosphere Model\\
35.0  & 12.6  & \citet{Pauldrach_etal_2004} & Atmosphere Model\\
29.0  & --    & \citet{Phillips_Ramos_Larios_2005} & $T_{Z}$(H~\textsc{I})\\
<48.2 & --    & \citet{Phillips_Ramos_Larios_2005} & $T_{Z}$(He~\textsc{II}) \\
34.0  & --    & \citet{Kudritzki_etal_2006} & Spectroscopic\\
32.0  & --    & \citet{Gesicki_Zijlstra_2007} & Dynamical Method\\
--    & 5.9   & \citet{Maciel_etal_2008}    & Atmosphere Model\\ 
34.7  & 1.4   & \citet{Pottasch_etal_2011} & Photoionization Model\\
32.0  & 4.5   & \citet{Pottasch_etal_2011} & Evolution Theory\\
34.1  & 2.5   & \citet{2014MNRAS.437.2577O} & Photoionization Model\\
30.0-32.0 & 2.0-3.8 & This work & Photoionization Model\\
\hline
\end{tabular}
\end{table*} 

Tc~1 distances and central star parameters available in the literature are listed in Tables~\ref{literature_distance} and \ref{param_literature_star}.

\label{lastpage}
\end{document}